\documentclass[]{llncs} 
\usepackage[utf8]{inputenc}
\usepackage{multirow}
\usepackage{tikz}
\usepackage{xspace}
\usepackage{geometry}

\usepackage[giveninits=true,
uniquename=false,
uniquelist=false,
hyperref=auto,
maxbibnames=1,
maxcitenames=2,
style=numeric,
citestyle=numeric,
backref=false,
natbib=true,
style=trad-abbrv,
doi=true,
sortcites,
backend=biber]{biblatex}
\usepackage{acro}
\usepackage[T1]{fontenc}
\usepackage{textcomp}
\usepackage{fancyvrb}
\usepackage{algpseudocode}
\usepackage{mathtools}
\usepackage{bbm}
\usepackage[flushleft]{threeparttable}
\usepackage{etoolbox}\AtBeginEnvironment{algorithmic}{\scriptsize} 
\usepackage{adjustbox}
\usepackage{pgfplots}
\pgfplotsset{compat=1.8}
\usepackage{pgfplotstable}
\usepgfplotslibrary{colorbrewer}
\usetikzlibrary{pgfplots.statistics, pgfplots.colorbrewer} \usepackage{forest}
\usetikzlibrary{arrows.meta, shapes.geometric, calc, shadows}
\usepackage{booktabs}
\usepackage{relsize}
\usepackage{siunitx}
\usepackage{verbatim}
\usepackage{subcaption}

\newcommand{\PCA}{Principal Component analysis\xspace}
\newcommand{\PC}{Principal Component\xspace}
\newcommand{\ML}{machine learning\xspace}
\newcommand{\LBM}{Lattice Boltzmann Method\xspace}
\newcommand{\spMVM}{sparse matrix-vector multiplication\xspace}
\newcommand{\LLC}{last-level cache\xspace}
\newcommand{\CRS}{Compressed Row Storage\xspace}

\newcommand{\SMcomm}[1]{{\color{purple} {#1} \color{black}} }

\newif\ifblind
\blindfalse
\begin{document}
%\title{Searching the Best-suited Techniques for the Analysis of Parallel Applications (due May 20)}
%\title{Exploring the Efficacy of Various Techniques for the Analysis of Parallel Applications on Clusters}
\title{Exploring Techniques for the Analysis of Spontaneous Asynchronicity in MPI-Parallel Applications}
\ifblind
\author{Authors omitted for double-blind review process}
\institute{\email{}}
\else
\author{Ayesha Afzal\inst{1}, Georg Hager\inst{1}, Gerhard Wellein\inst{1,2}, and Stefano Markidis\inst{3}}
\authorrunning{A.\ Afzal et al.} 
\institute{Erlangen National High Performance Computing Center (NHR@FAU), 91058 Erlangen, Germany,\\
% \email{ayesha.afzal@fau.de, georg.hager@fau.de}
\and
Department of Computer Science, University of Erlangen-N\"urnberg, 91058 Erlangen, Germany,\\
% \email{gerhard.wellein@fau.de}
\and
Department of Computer Science, KTH Royal Institute of Technology, 11428 Stockholm, Sweden,\\
% \email{markidis@kth.se}
\email{ayesha.afzal@fau.de, georg.hager@fau.de, gerhard.wellein@fau.de, markidis@kth.se}
}
\fi

\maketitle

\begin{abstract}
This paper studies the utility of using data analytics and machine learning techniques for identifying, classifying, and characterizing the dynamics of large-scale parallel (MPI) programs.
To this end, we run microbenchmarks and realistic proxy applications with the regular compute-communicate structure on two different supercomputing platforms and choose the per-process performance and MPI time per time step as relevant observables. Using principal component analysis, clustering techniques, correlation functions, and a new ``phase space plot,'' we show how desynchronization patterns (or lack thereof) can be readily identified from a data set that is much smaller than a full MPI trace. Our methods also lead the way towards a more general classification of parallel program dynamics. 
%perform temporal evolution and workload characterization via investigating various simple and advanced worthwhile metrics.
%The analysis has been performed for a suite of synthetic benchmarks and two proxy applications on two different supercomputers.
\end{abstract}

%%%%%%%%%%%%%%%%%%%%%%%%%%%%%%%%%%%%%%%%%%%%%%%%%%%%%%%%%%%%%%%%%%%%%
\section{Introduction and related work} \label{sec:introduction}
    % Related work: is it also used to extract information from very noisy signals?
    Highly parallel MPI programs with no or weak global synchronization points 
    show interesting dynamics that go beyond what is expected from their usually
    regular compute-communicate structure. Initiated by what is typically
    called ``noise,'' a plethora of patterns can emerge: Propagating delays 
    emanating from strong one-off disturbances, so-called \emph{idle waves}~\cite{markidis2015idle}, 
    can interact~\cite{AfzalHW19} and eventually decay~\cite{AfzalHW19,AfzalHW20,AfzalHW21} via
    various mechanisms. Caused by idle waves, but also under the natural, fine-grained system 
    noise, some 
    applications are unstable and leave their initial lock-step mode {(Figure~2 (left))} where all processes
    either compute or communicate. It was shown~\cite{AfzalHW20} that a hardware
    bottleneck such as main memory bandwidth is a prerequisite for this 
    \emph{bottleneck evasion} to occur. As a consequence, such programs settle in a
    metastable state, a \emph{computational wavefront}, where neighboring processes
    are shifted in time with respect to each other~{(Figure~2 (right))}. It was also shown~\cite{AfzalHW22role}
    that this \emph{desynchronization} can lead to substantial speedups via 
    automatic overlap of communication and code execution.
    
    Investigating these dynamics typically requires the analysis of MPI traces taken
    by tools such as Intel Trace Analyzer/Collector or VAMPIR. Apart from the 
    often prohibitive amount of data contained in such traces, the relevant patterns
    are often hidden in the data and not readily visible to the human eye. Furthermore, it is hard, if not impossible, to obtain this data in a production environment without adverse effects on the performance of applications. For applications
    that have natural regular compute-communicate cycles, we propose to use the 
    MPI waiting time per process and time step (i.e., the time spent in the MPI library, 
    regardless of whether communication takes place or not) as a starting point
    and input metric for data analysis methods that can identify the structural 
    processes described above. The performance per process and time step can
    serve as a supplemental metric to track the impact of automatic communication
    overlap.

    %\paragraph{Contribution}
    This paper makes the following relevant contributions:
    \begin{itemize}%[nosep]
        \item We demonstrate how to automatically characterize different flavors of synchronous versus non-synchronous execution of MPI-parallel codes without taking full MPI traces or in-depth application analysis.
        %\item We scan a wide spectrum of metrics beyond the timelines of MPI processes to manifest the analysis of a set of parallel programs.  
        \item We show that the MPI waiting time per process and time step provides a powerful input metric for principal component analysis (PCA) and clustering methods in order to spot these patterns.
        \item We introduce the \emph{MPI phase space plot} as a tool to visualize the long-term evolution and peculiar patterns of MPI waiting time in a parallel program. 
        %\item We perform the comparison of analysis for two observables, such as MPI times and performance, that enable us to explore the effectiveness of each metric for such analysis.
    \end{itemize}
    %\paragraph{Overview}
    This paper is organized as follows:
    We first provide details about our experimental environment and methodology in Sect.~\ref{sec:environment}.
    To investigate the dynamics of large-scale parallel programs,
    simple metrics such as the histogram and timelines for all or individual MPI processes are studied in Sect.~\ref{sec:metrics},
    while Sect.~\ref{sec:additionalmetrics} covers advanced methods like correlation coefficient matrices and phase space plots. Sect.~\ref{sec:MLTechniques} addresses \ML techniques such as \PCA and k-means clustering.
    Finally, Sect.~\ref{sec:conclusion} concludes the paper and gives an outlook to future work.

%%%%%%%%%%%%%%%%%%%%%%%%%%%%%%%%%%%%%%%%%%%%%%%%%%%%%%%%%%%%%%%%%%%%%
\section{Case studies, testbed and experimental methods} \label{sec:environment}
    \begin{figure}[tb]
	\centering
	\begin{minipage}[c]{0.48\textwidth}
		\resizebox{\columnwidth}{!}{%
	    \includegraphics{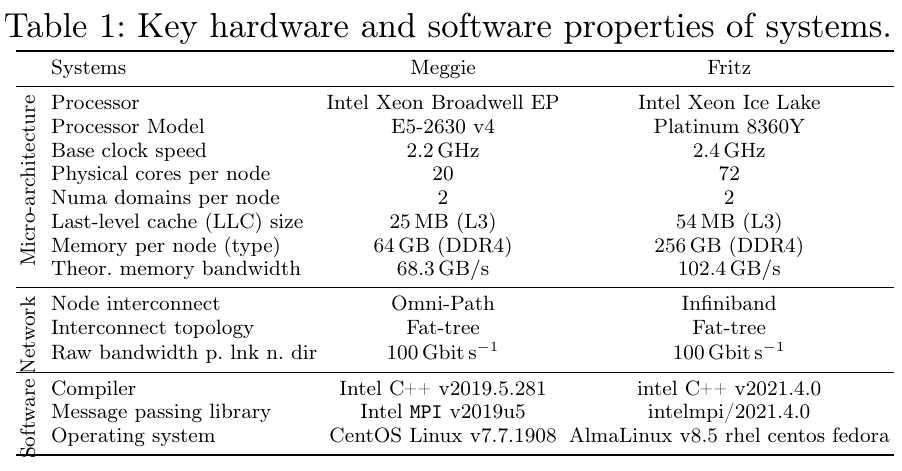}
	    }
		\label{tab:system}
	\end{minipage}\hfill
	\begin{minipage}[c]{0.27\textwidth}
	    \vspace{0.05em}
		\begin{tikzpicture}
		\begin{axis}[trim axis left, trim axis right, scale only axis,
		width=0.75\textwidth,height=0.55\textwidth,
		xlabel = {\textbf{Cores per Meggie socket}},
		ylabel = {\textbf{MEM bandwidth [\si{\mega \byte / \second}]}},
		xmin=0,
		ymin=0,
		ymax=90000,
		y label style={at={(-0.1,0.4)},font=\scriptsize},
		x label style={font=\scriptsize},
		x tick label style={font=\normalsize},
		y tick label style={font=\normalsize},
		xmajorgrids,
		ymajorgrids,
		tick style={thick},
		legend columns = 1, 
		legend style = {
			nodes={inner sep=0.117em},
			draw=none,
			font=\tiny,
			cells={align=left},
			anchor=east,
			at={(0.99,0.83)},
			/tikz/column 1/.style={column sep=5pt,},
		},
		]

		\addplot+[error bars/.cd, y dir=both, y explicit,]
		table
		[
		x expr=\thisrow{Cores}, 
		y error minus expr=\thisrow{Median}-\thisrow{Min},
		y error plus expr=\thisrow{Max}-\thisrow{Median},
		row sep=crcr
		]{
%cos DIV
		Cores	Median		Min		Max\\
1	8296.5011	8296.5011	8296.5011\\
2	15875.1178	15875.1178	15875.1178\\
3	22530.8013	22530.8013	22530.8013\\
4	29022.0039	29022.0039	29022.0039\\
5	34677.882	34677.882	34677.882\\
6	40004.6952	40004.6952	40004.6952\\
7	44520.1861	44520.1861	44520.1861\\
8	48264.4511	48264.4511	48264.4511\\
9	51905.5896	51905.5896	51905.5896\\
10	52738.2617	52738.2617	52738.2617\\
% DIV
% 1	15798.553	15798.553	15798.553\\
% 2	27668.0578	27668.0578	27668.0578\\
% 3	37574.0117	37574.0117	37574.0117\\
% 4	44959.4246	44959.4246	44959.4246\\
% 5	49132.1641	49132.1641	49132.1641\\
% 6	51752.9495	51752.9495	51752.9495\\
% 7	52156.8531	52156.8531	52156.8531\\
% 8	53223.1771	53223.1771	53223.1771\\
% 9	53799.3908	53799.3908	53799.3908\\
% 10	53273.828	53273.828	53273.828\\
        };
		\addlegendentry{\textbf{Sch\"onauer Triad}} %Slow 
		
		\addplot+[error bars/.cd, y dir=both, y explicit,]
		table
		[
		x expr=\thisrow{Cores}, 
		y error minus expr=\thisrow{Median}-\thisrow{Min},
		y error plus expr=\thisrow{Max}-\thisrow{Median},
		row sep=crcr]{
		Cores	Median		Min		Max\\
1	16081.349	16081.349	16081.349\\
2	28282.9129	28282.9129	28282.9129\\
3	38011.239	38011.239	38011.239\\
4	45320.711	45320.711	45320.711\\
5	49324.4899	49324.4899	49324.4899\\
6	51933.2305	51933.2305	51933.2305\\
7	52217.5162	52217.5162	52217.5162\\
8	53376.9859	53376.9859	53376.9859\\
9	52863.7581	52863.7581	52863.7581\\
10	53463.242	53463.242	53463.242\\
		};
		\addlegendentry{\textbf{STREAM Triad}}
 		\end{axis}
 	\end{tikzpicture}
 	\caption{\footnotesize Saturation attributes}
 	\label{fig:scalability}
	\end{minipage}\hfill
	\begin{minipage}[c]{0.23\textwidth}
	    \vspace{2em}
		\resizebox{\columnwidth}{!}{%
			\begin{tikzpicture}
			\put(-2.5,-0.8) {\includegraphics[width=0.65\textwidth,height=0.21 \textheight]{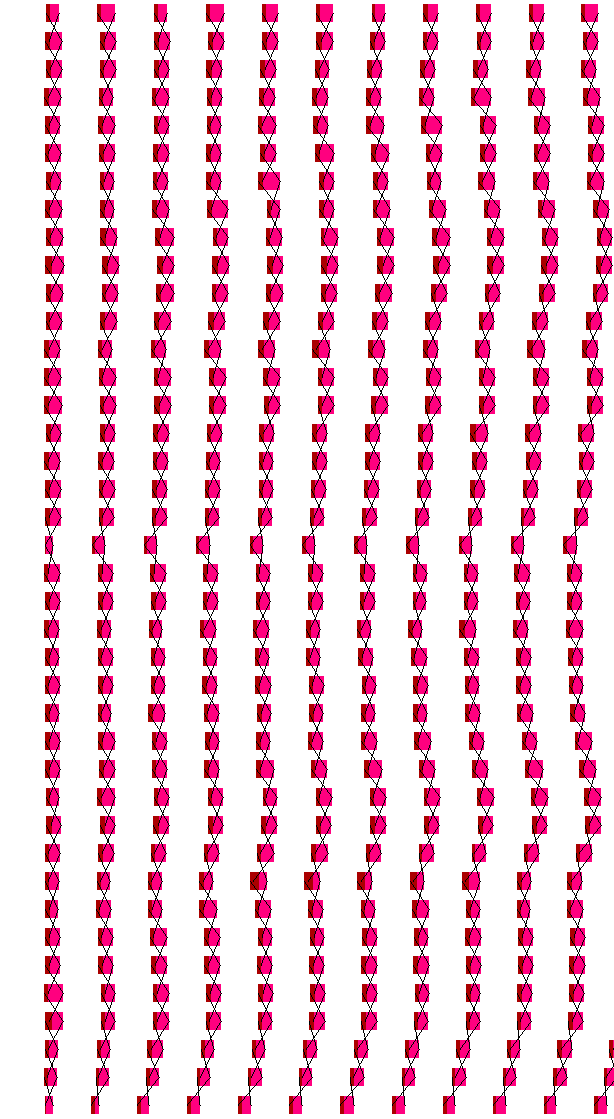}}
			\begin{axis}[
				trim axis left, trim axis right, scale only axis,
				width=0.65\textwidth,height=0.21\textheight,
				title style={at={(0.5,-0.5)}},
				ylabel = {\textbf{Rank}},
				y label style={font=\huge,at={(-0.8,0.5)}},
				xlabel = {\textbf{Walltime [s]}},
				x label style={font=\Large}, 		
				x tick label style={font=\Large},
				y tick label style={font=\Large}, 
				xmin=0, xmax=55,
				ymin=0, ymax=39,
				xtick={1,4,33.5,37.5,45,54.2},
				xticklabels={},
				ytick={0,10,20,30,39},
			    yticklabels={\textbf{39},\textbf{29},\textbf{19},\textbf{9},\textbf{0}},
				]
				\end{axis}
				\node at (-1.15,3.6) {\Large \textbf{\emph{S0}}}; 
				\node at (-1.15,2.6) {\Large \textbf{\emph{S1}}}; 
				\node at (-1.15,1.6) {\Large \textbf{\emph{S2}}}; 
				\node at (-1.15,0.6) {\Large \textbf{\emph{S3}}};
			\end{tikzpicture}
			\begin{tikzpicture}
			\put(1.5,-0.8) {
			\includegraphics[width=0.64\textwidth,height=0.21 \textheight]{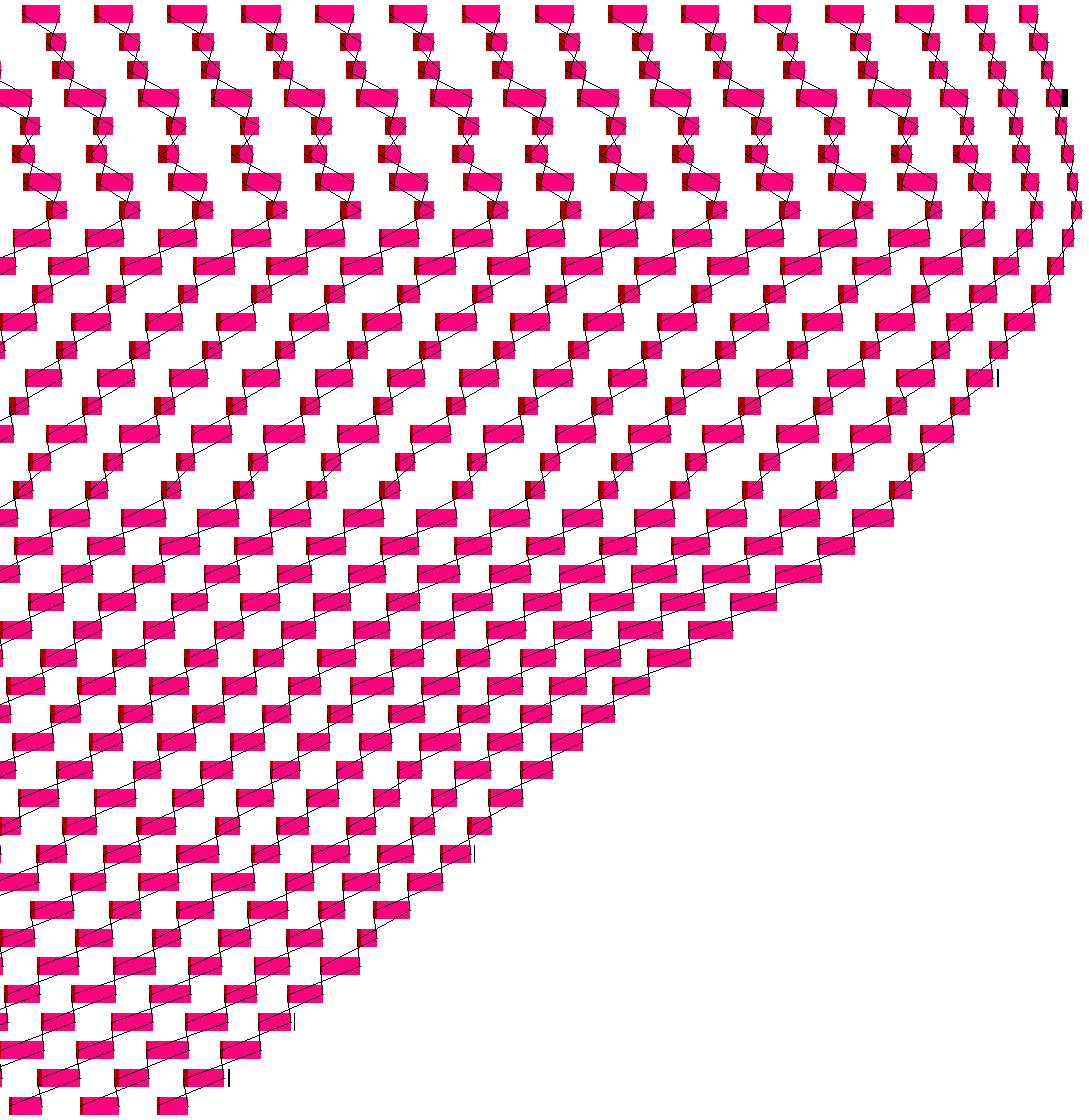}}
			\begin{axis}[
				trim axis left, trim axis right, scale only axis,
				width=0.65\textwidth,height=0.21\textheight,
				title style={at={(0.5,-0.5)}},
				y label style={font=\huge,at={(-0.8,0.5)}},
				xlabel = {\textbf{Walltime [s]}},
				x label style={font=\Large}, 			
				x tick label style={font=\Large},
				y tick label style={font=\Large}, 
				xmin=0, xmax=55,
				ymin=0, ymax=39,
				xtick={},
				xticklabels={},
				ytick={0,10,20,30,39},
				yticklabels={},
				]
				\end{axis}
				% \draw [thick,densely dashed,blue] (0.9,0) -- (2.65,1.8);
				% \draw [thick,densely dashed,blue] (2.65,1.8) -- (3.1,2.6);
				% \draw [thick,densely dashed,blue] (3.03,3.4) -- (3.15,2.65);
			\end{tikzpicture}
		}
		\caption{\footnotesize Timeline traces of synchronized (left) and desynchronized (right) MPI processes on sockets (\emph{Si}).
% 		\GHcomm{The sizes of these graphs are ridiculous}\AAcomm{made the fonts higher as far as possible}
		}
		\label{fig:sync}
	\end{minipage}
    \label{fig:intro}	
\end{figure}

    \subsection{Test systems and methodology}
    %We conduct various experiments on two different clusters.
    The details of the hardware and software environments on the ``Meggie''\footnote{\ifblind{URL redacted for double-blind review}\else\url{https://hpc.fau.de/systems-services/systems-documentation-instructions/clusters/meggie-cluster/}\fi} and ``Fritz''\footnote{\ifblind{URL redacted for double-blind review}\else\url{https://hpc.fau.de/systems-services/systems-documentation-instructions/clusters/fritz-cluster/}\fi} clusters can be found  in Table~1.
    By default, hyper-threading (SMT) is disabled on both systems and Sub-NUMA Clustering (SNC) is enabled on the Fritz system.
    The optimization flag \texttt{-O3} was utilized with the Intel compiler.
    Process-core affinity was enforced using the \texttt{I\_MPI\_PIN\_PROCESSOR\_LIST} environment variable.
    The clock frequency was always fixed to the base value of the respective CPUs.
    Working sets were chosen large enough to not fit into any cache, i.e., at least ten times the \LLC size on both systems. % (non-inclusive victim L3 cache in the Ice Lake processor of Fritz).
    All floating-point computations were done in double precision.
    All experiments were performed for \SI{500}{\kilo~iterations} (compute-communicate cycles), except for LBM where we used \SI{100}{\kilo~iterations}. Each experiment was repeated at least five times to ensure that the runtime analysis is stable.
    %The scripts and additional material are available online at \url{hptts://}.
    
    \subsection{Synthetic microbenchmarks}
    We ran pure-MPI versions of the McCalpin STREAM Triad~\cite{mccalpin1995memory} (\texttt{A(:)=B(:)+s*C(:))}) and a ``slow'' Schönauer vector Triad (\texttt{A(:)=B(:)+{cos(C(:)/D(:))}})\footnote{The low-throughput cosine and floating-point division shifts the bandwidth saturation point to a higher number of cores} with bidirectional next-neighbor communication. The saturation characteristics of these streaming kernels on one socket (ccNUMA domain) of Meggie are shown in Fig.~1.
    Each MPI rank $i$ sends and receives messages to and from each of its direct neighbors $i + 1$ and $i - 1$ after each full loop traversal.
    Each array had \SI{400}{\mega~elements}, which yields a \SI{9.6}{\giga\byte} working set for the STREAM Triad and a \SI{12.8}{\giga\byte} working set for the Schönauer Triad.
    To mimic scalable applications, we set up a PISOLVER code which calculates the value of $\pi$ by
    evaluating $\int_0^14/(1+x^2)\,\mathrm dx$ using the mid-point rule with
    \SI{500}{\mega~steps}. Overall, four microbenchmark variants were employed:
    %We consider the characteristics of four parallel benchmark programs (\emph{Benchi}, where i$ = 1, 2, 3, 4$)
    \begin{enumerate}%[nosep]
        \item MPI-parallel STREAM Triad with \SI{5}{\mega\byte} messages
        \item MPI-parallel STREAM Triad with \SI{8}{\byte} messages
        \item MPI-parallel ``slow'' Schönauer Triad with \SI{8}{\byte} messages 
        \item MPI-parallel PISOLVER with \SI{8}{\byte} messages
    \end{enumerate}
    These four cases were run in two scenarios on the Meggie system:   
    (A) open-chain process topology with $40$ MPI processes on four ccNUMA domains (sockets), and (B) closed-ring process topology with $400$ MPI processes on $40$ sockets. Later these scenarios will be denoted \textit{``Bench iA''} and \textit{``Bench iB''}, respectively, where $i$ is the label in the enumeration list above.

    \subsection{Proxy memory-bound parallel applications}
    We experiment with the following two MPI-parallel proxy applications and run them with $1440$ MPI processes on $40$ sockets of the Fritz system.
    \paragraph{MPI-parallel LBM solver}
        This is a prototype application based on a \LBM (LBM) from computational fluid dynamics using the Bhatnagar–Gross–Krook collision operator~\cite{bhatnagar1954model} and implementing a 3D lid-driven cavity scenario.
        It is purely memory bound on a single ccNUMA domain, but the halo exchange makes it communication dominated in strong scaling scenarios. 
        The double-precision implementation employs a three-dimensional \textsc{D3Q19} space discretization~\cite{qian1992lattice}.
        %These probabilities of the neighboring slice (contained in ghost layers for each slice) are used in the next iteration’s stream step and after each iteration are updated from neighboring slices.
        %A cell is updated by reading one particle distribution function (PDF) value each from its $19$ neighbors.
        The domain decomposition is performed by cutting slices in the $z$ direction.
        For halo exchange, five PDFs per boundary cell must be communicated. The MPI communication is done with non-blocking point-to-point calls, but no explicit overlapping of communication with useful work is implemented.
        We use an overall problem size of $n_x\times n_y\times n_z=1440^3$ lattice cells, which amounts to a working set of \SI{908}{\giga\byte} plus halo layers. 
        Due to the one-dimensional domain decomposition, the communication volume per halo depends on $n_x$ and $n_y$ only and is independent of the number of processes. 
        
    \paragraph{MPI-parallel spMVM solver} 
        The \spMVM (SpMVM) $ \vec{b}=A\vec{x}$ is a most relevant, time-consuming building block of numerous applications in science and engineering.
        Here, $A$ is an $n \times n$ sparse matrix, and $\vec{b}$, $\vec{x}$ are $n$-dimensional vectors.
        SpMVM plays a central role in the iterative solution of sparse linear systems, eigenvalue problems and Krylov subspace solvers.
        Due to its low computational intensity, the SpMVM kernel is mostly limited by the main memory bandwidth if the matrix does not fit into a cache.
        Our implementation uses non-blocking point-to-point communication calls, where the communication requests for reading the remote parts of $\vec{x}$ are issued and then collectively finished via \texttt{MPI\_Waitall}. After that, the whole SpMVM kernel is executed.
        The communication volume is crucially dependent on the structure of the matrix; the distribution of the nonzero entries plays a decisive role.
        In this paper, we use a matrix that arises from strongly correlated electron-phonon systems in solid state physics. 
        It describes a Holstein-Hubbard model~\cite{fehske2004quantum} comprising $3$ electrons on $8$ lattice sites coupled to $10$ phonons.
        The sparse matrix $A$ has $60,988,928$ rows and columns and $889,816,368$ non-zero entries, respectively, which leads to an average of 13 nonzeros per row and an overall data set size of \SI{10.9}{\giga\byte} using four-byte indices in the \CRS format (one-dimensional arrays for values, column indices, and row pointers).

    \subsection{Observables for analysis}
    We instrument all codes to collect the time stamps of entering and leaving MPI calls (MPI waiting time per process) at each iteration of each MPI process across the full run.
    From this data we construct a non-square matrix of size $N_p \times N_{it}$, where $N_p$ is the number of MPI processes and $N_{it}$ is the number of iterations.
    Each row (column) of the observable matrix represents the observable value, i.e., the time spent in MPI, for each process (iteration).
    There is a choice as to how this data can be used in analysis: One can either use the full timeline per process, which takes the end-to-end evolution of execution characteristics (such as desynchronization) into account, or cut out a number of consecutive iterations from different execution phases, which allows to investigate the development of interesting patterns in more detail. 
    In addition, for some experiments we collect performance per MPI process averaged over the 1000 time steps.

    %We use two observables for the analyses of parallel applications and ensure that  our finding of methods (such as, dot cloud, etc.) for both should correlate:
    %\begin{enumerate}%[nosep]
    %    \item MPI times [\si{\sec}]
    %    \item Performance [\si{iterations \per \sec}]
    %\end{enumerate}
    %Performance is averaging over a certain period of time.
    %We performed analysis in two views:
    %firstly, we consider the integrated behaviour of application over the whole run time that includes all changes in characteristics from the beginning till evolved shape encompassing in-between transitional regions.
    %Secondly, we show snippets over a few iterations for all data analytic metrics to investigate the evolution of asynchronous dynamics from the initial synchronized state to the final evolved state. 

%%%%%%%%%%%%%%%%%%%%%%%%%%%%%%%%%%%%%%%%%%%%%%%%%%%%%%%%%%%%%%%%%%%%%  
\section{Simple timeline metrics for analysis} \label{sec:metrics}
    %%%%%%%%%%%%%%%%%%%%%%%%%%%%%%%%%%%%%%%%%%%%%%%%%%%%%%%%%%%%%%%%%%%%% 
    % Adding figures
    \begin{figure*}[t]
	\centering
	\begin{minipage}{\textwidth}
	    \begin{subfigure}[t]{0.245\textwidth} 
			\includegraphics[scale=0.125]{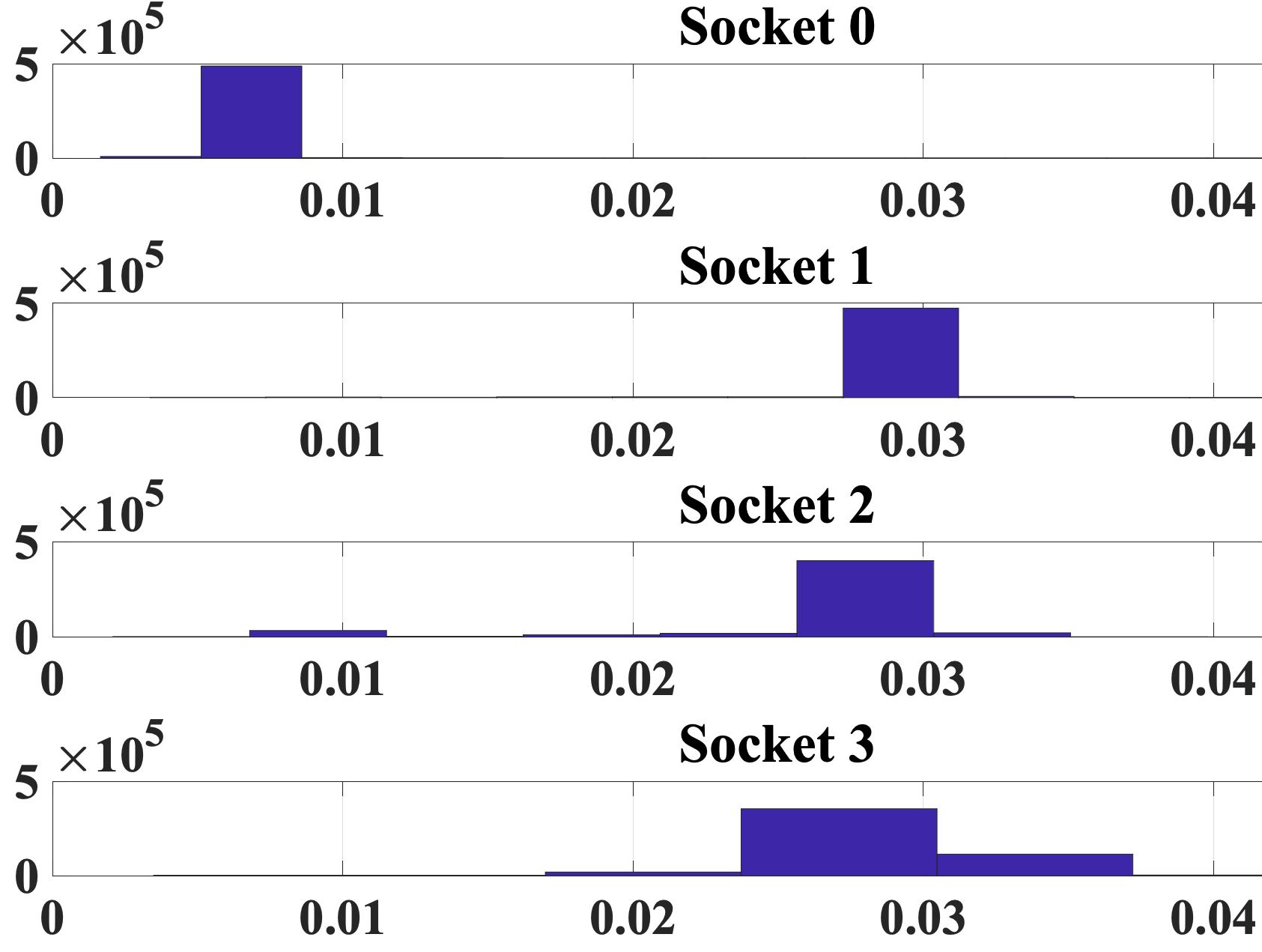}
		    \caption{Bench1A-500\si{\kilo~it}}
		\end{subfigure}
		\begin{subfigure}[t]{0.245\textwidth} 
			\includegraphics[scale=0.125]{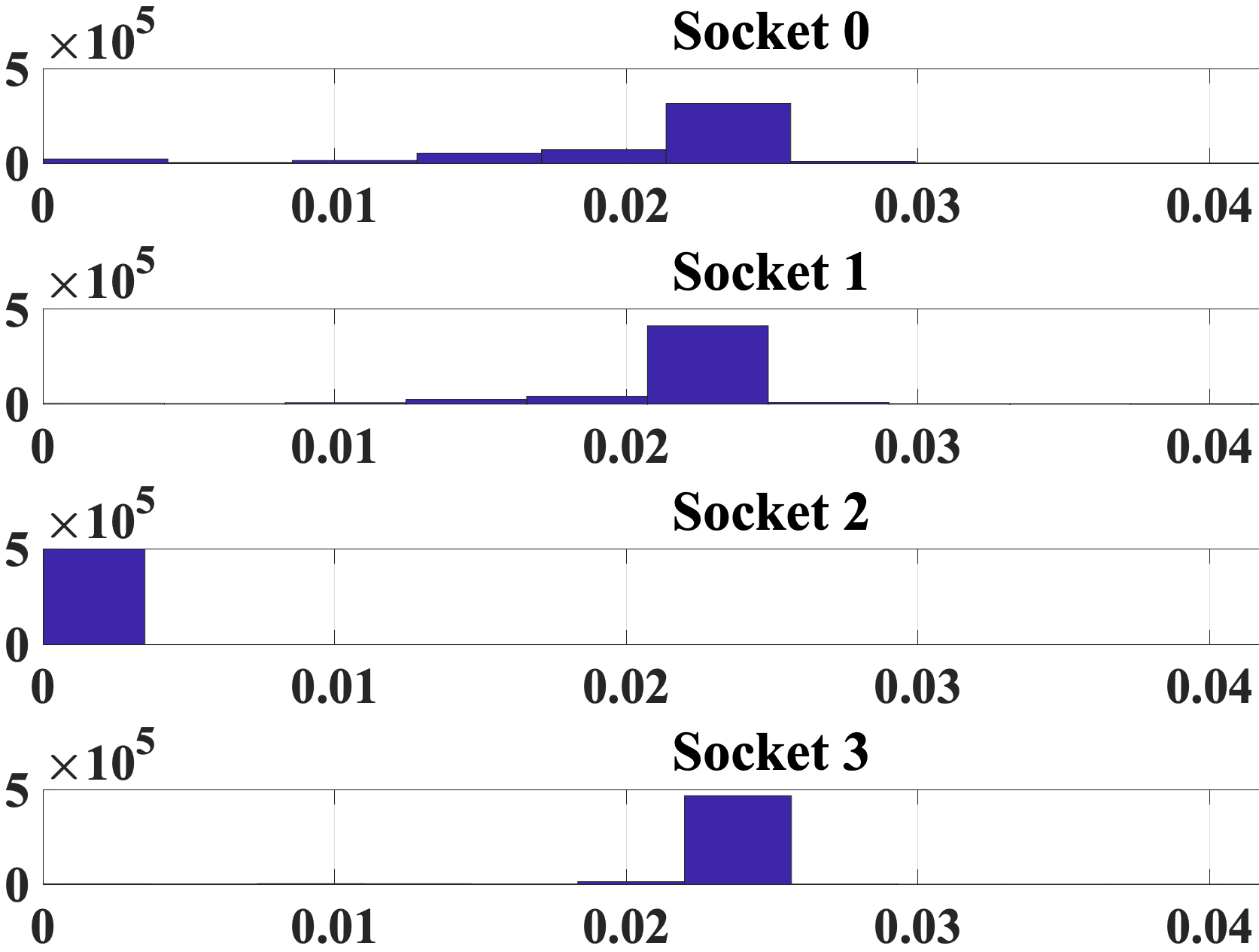}
		    \caption{Bench2A-500\si{\kilo~it}}
		\end{subfigure}
		\begin{subfigure}[t]{0.245\textwidth} 
			\includegraphics[scale=0.125]{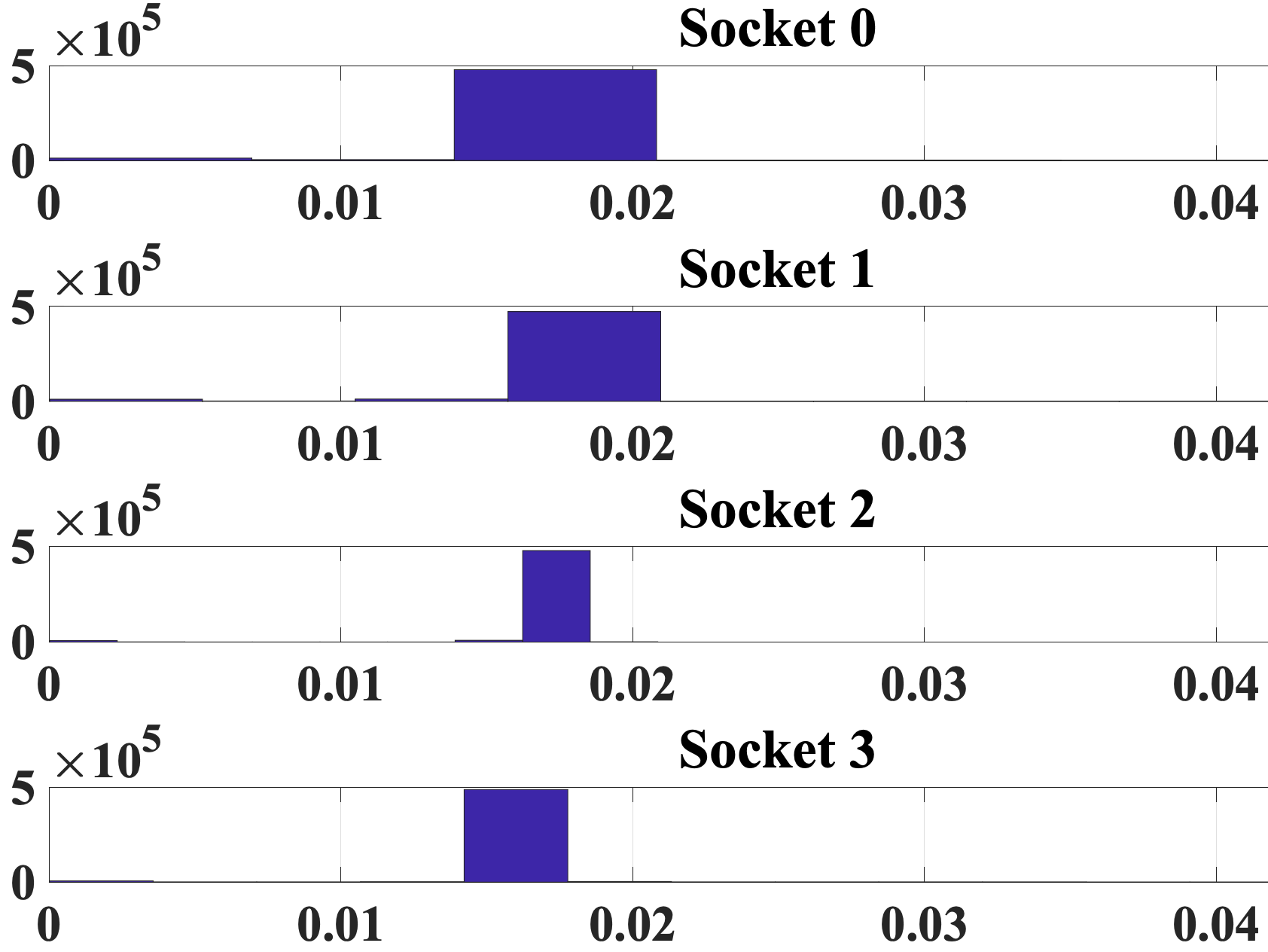}
		    \caption{Bench3A-500\si{\kilo~it}}
		\end{subfigure}
		\begin{subfigure}[t]{0.245\textwidth} 
			\includegraphics[scale=0.125]{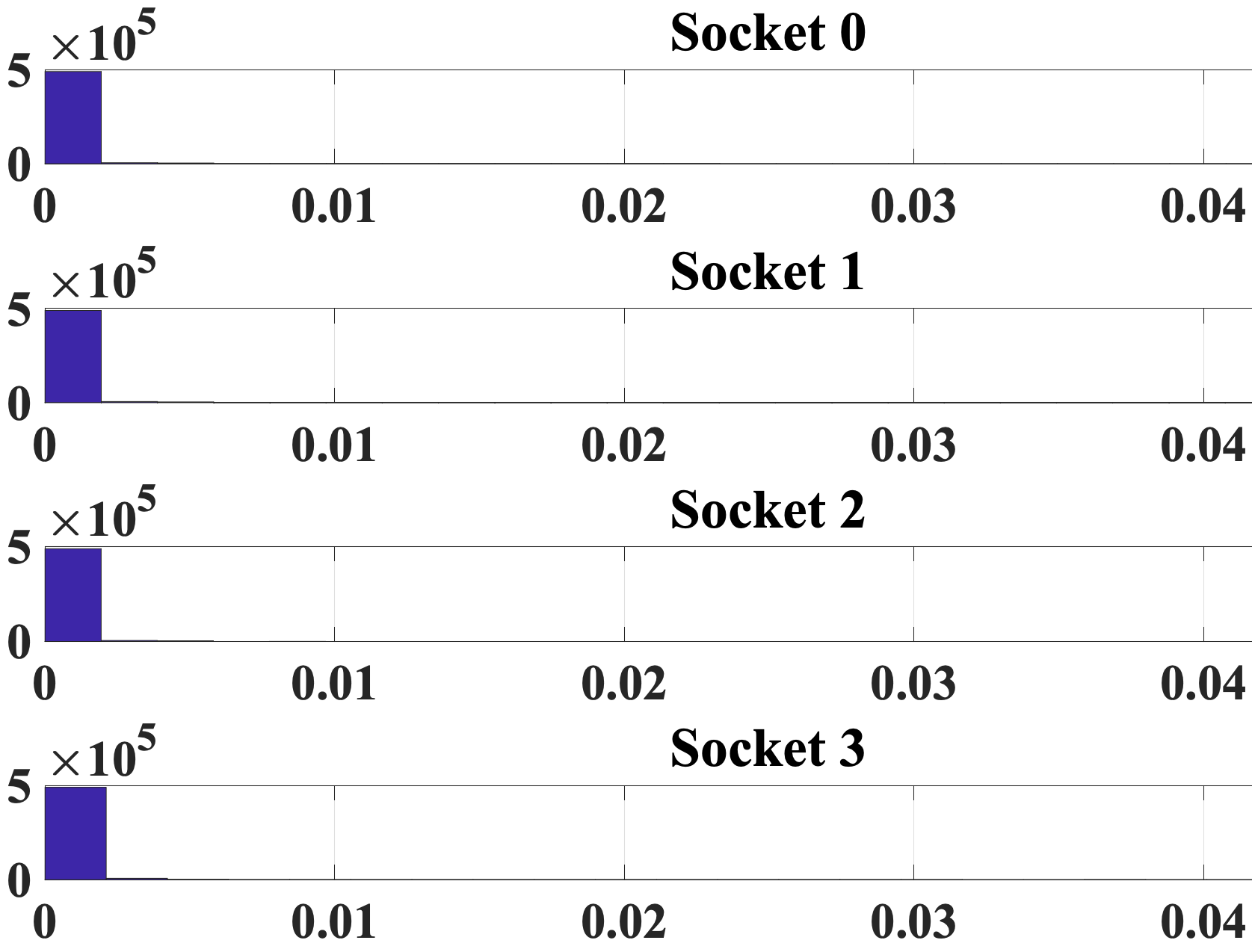}
		    \caption{Bench4A-500\si{\kilo~it}}
		\end{subfigure}
	\end{minipage}
% 	\begin{minipage}{\textwidth}
% 	    \begin{subfigure}[t]{0.33\textwidth} 
% 			\includegraphics[scale=0.142]{figures/histogramMPItimes/Case1b.png}
% 		    \caption{Bench1B-500\si{\kilo~it}-M}
% 		\end{subfigure}
% 		\begin{subfigure}[t]{0.33\textwidth} 
% 			\includegraphics[scale=0.142]{figures/histogramMPItimes/Case2b.png}
% 		    \caption{Bench2B-500\si{\kilo~it}-M}
% 		\end{subfigure}
% 		\begin{subfigure}[t]{0.33\textwidth} 
% 			\includegraphics[scale=0.142]{figures/histogramMPItimes/Case3b.png}
% 		    \caption{Bench3B-500\si{\kilo~it}-M}
% 		\end{subfigure}
% 	\end{minipage}
% 	\begin{minipage}{\textwidth}
% 	    \begin{subfigure}[t]{0.33\textwidth} 
% 			\includegraphics[scale=0.142]{figures/histogramMPItimes/Case4b.png}
% 		    \caption{Bench4B-500\si{\kilo~it}-M}
% 		\end{subfigure}
% 	    \begin{subfigure}[t]{0.33\textwidth} 
% 			\includegraphics[scale=0.142]{figures/histogramMPItimes/Case5.png}
% 		    \caption{LBM-100\si{\kilo~it}-F}
% 		\end{subfigure}
% 		\begin{subfigure}[t]{0.33\textwidth} 
% 			\includegraphics[scale=0.142]{figures/histogramMPItimes/Case6.png}
% 		    \caption{spMVM-500\si{\kilo~it}-F}
% 		\end{subfigure}
% 	\end{minipage}
	\caption{Histograms sorting MPI times [\si{\sec}] into bins for all benchmarks on the first process of each Meggie
% 	(M) and Fritz (F) 
	socket used in the run.
	The $x$-axes show the MPI times and the $y$-axes  indicate the number of MPI time values in each bin. 
%	\GHcomm{Delete all but (a) and (d) and just describe the rest in the text. It is impossible and nonsensical to show all this data that nobody can read anyway.}
	}
    \label{fig:histogramMPItimes}
\end{figure*}
    %%%%%%%%%%%%%%%%%%%%%%%%%%%%%%%%%%%%%%%%%%%%%%%%%%
    \subsection{Rank/ccNUMA-wise timelines and histogram of MPI time and performance} % plot(Performance(:,i),'.') % hist(Performance(:,i),35)
    The histograms in Figure~\ref{fig:histogramMPItimes} sort the MPI time values of end-to-end (\SI{500}{\kilo~iterations}) runs of Bench[1--4]A into 35 bins.
    %The bins are displayed as rectangles such that the height of each rectangle (y-axis) indicates the number of MPI time values in the bin and the x-axis is the MPI times.
    For memory-bound code, idle times are lower for desynchronized processes if the bandwidth saturation on a ccNUMA domain is weaker~\cite{AfzalHW20} %((b)--(d) for \SI{8}{\byte} messages and (f)--(h)) for \SI{5}{\mega\byte} messages).
    (see Figure~\ref{fig:histogramMPItimes}(c)).
    In the compute-bound PISOLVER case (Figure~\ref{fig:histogramMPItimes}(d)), all processes are synchronized because of the absence of any contention on the memory interface or on the network.
    Open-chain boundary conditions and strong memory contention ((a) and (b)) lead to a single synchronized socket.
    In the other cases, all sockets desynchronize gradually over \SI{500}{\kilo~iterations}, which causes a spread in the histogram because processes evolve from lower to higher idle times.
    We have observed that this spread is more prominent for codes with stronger saturation and higher communication (\emph{Bench1B}, LBM, and spMVM; data not shown for brevity).
    %in Figure~\ref{fig:histogramMPItimes}(e, i, j)).  
    
\begin{figure*}[t]
	\centering
	\begin{minipage}{\textwidth}
	    \begin{subfigure}[t]{0.365\textwidth} 
			\includegraphics[scale=0.068]{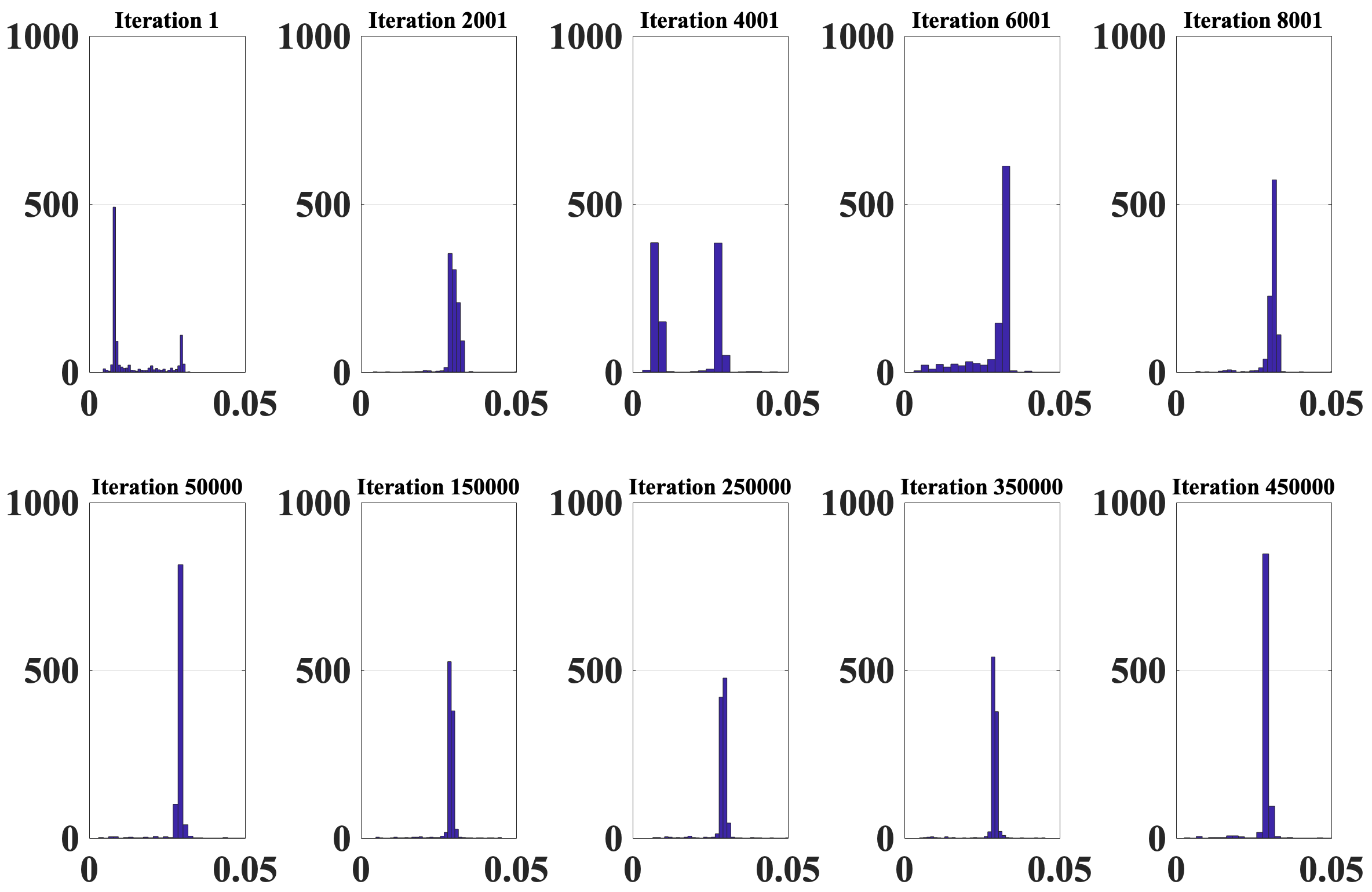}
		    \caption{Bench1A-500\si{\kilo~it}-Rank 20}
		\end{subfigure}
		\begin{subfigure}[t]{0.22\textwidth} 
			\includegraphics[scale=0.08]{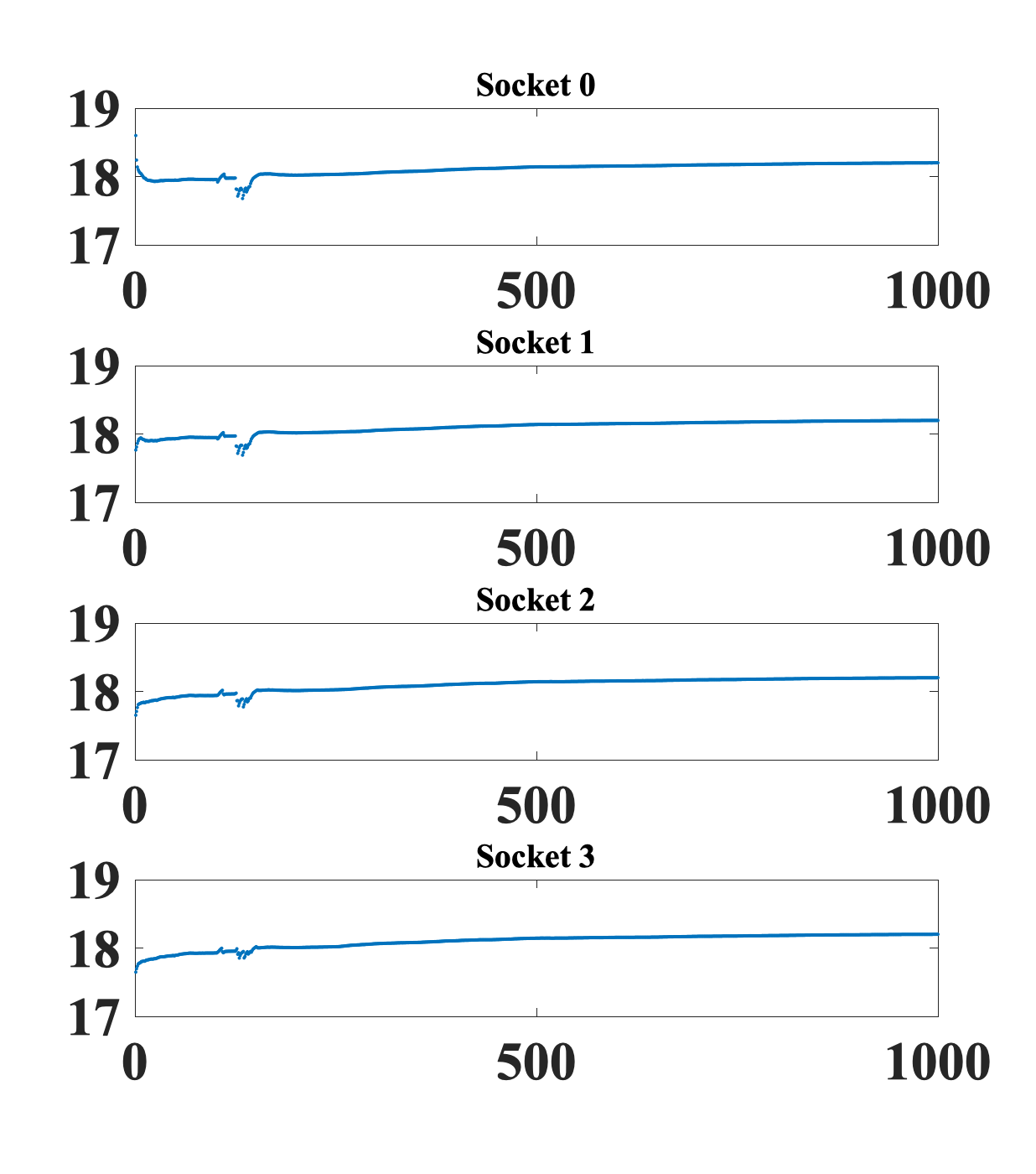}
			\caption{Bench1A-1\si{\kilo~it}}% of 500\si{\kilo~it} run}
		\end{subfigure}
		\begin{subfigure}[t]{0.372\textwidth} 
			\includegraphics[scale=0.178]{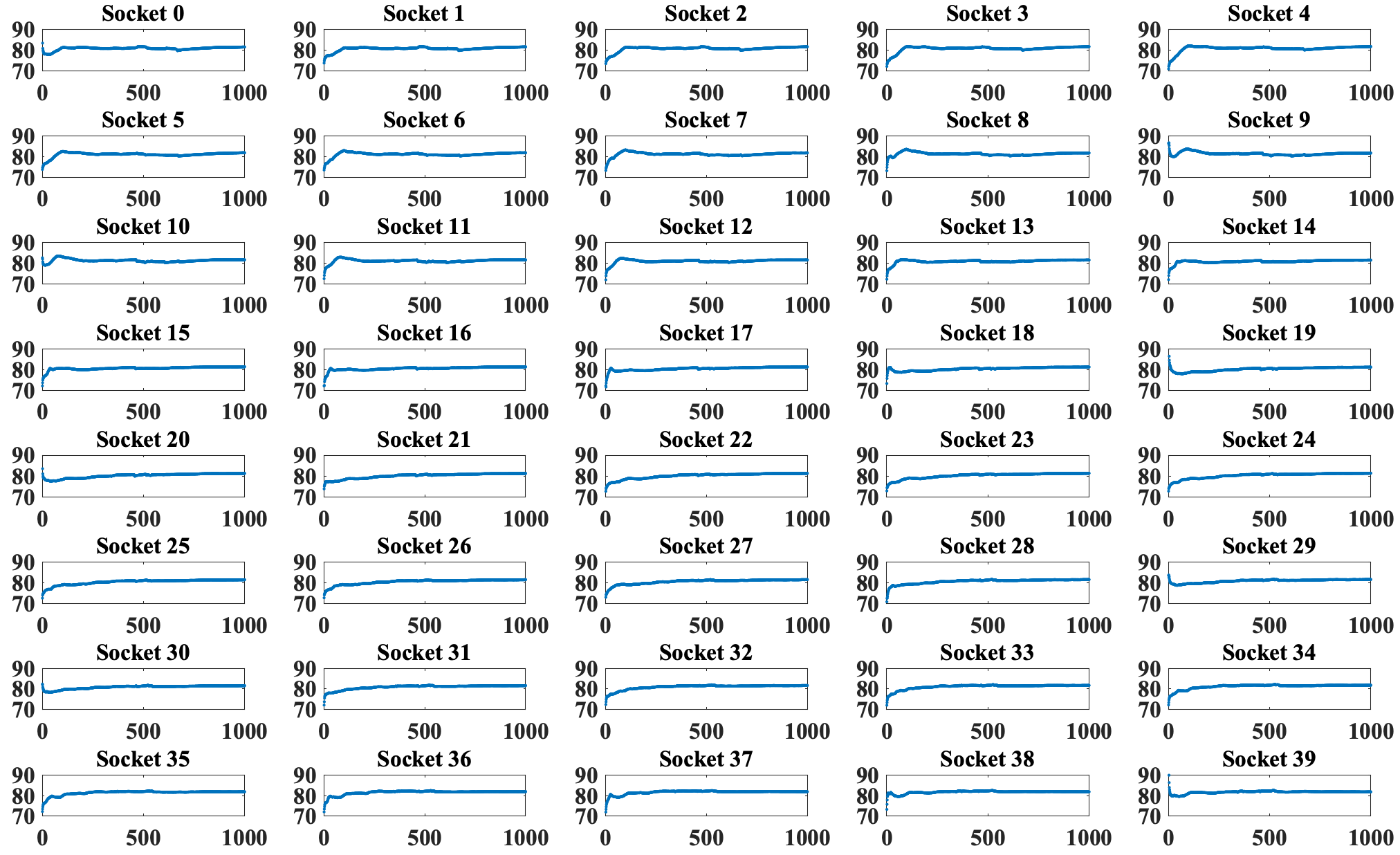}
		    \caption{Bench1B-1\si{\kilo~it} of 500\si{\kilo~it} run}
		\end{subfigure}
	\end{minipage}
	\caption{(a) Snippet view of histograms for MPI times [\si{\sec}] (x-axes) of rank 20 only and 
	(b-c) performance [\si{iterations \per \sec}] (y-axes) on every first MPI process of each Meggie socket for the initial-zoomed \SI{1}{\kilo~iterations} (x-axes) snapshot of Bench1A and Bench1B.
	Since the performance remains constant afterwards, we don't show the whole run of \SI{500}{\kilo~iterations}.}
    \label{fig:histogramPerformance}
\end{figure*}

    We first investigate the open chain high communication overhead benchmark mode (\emph{Bench1A}). Figure~\ref{fig:histogramPerformance}(a) shows the histograms at the different stages of evolution of a single MPI process (i.e., rank 20 on third ccNUMA domain) through the whole execution. Each histogram encompasses
    1000 iterations.
    Initially (e.g., till 50 \si{\kilo~iterations}), the distributions are multimodal, which indicates different phases.
    %effect of MPI times shows different peaks for synchronized and asynchronized ccNUMA domains on two middle sockets which got swamped in longer scale run of \SI{500}{\kilo~iterations} due to its smaller contribution.
    On closer inspection it can be observed that the peak snaps from left to right as the process goes out of sync with its neighbors.
    This corroborates that the MPI waiting time is a good observable in our context.
    Since desynchronization cannot yield significant speedup if communication is insignificant, we show plots of performance vs.\ time step for significant communication cases only
    (\emph{Bench1A} in Figure~\ref{fig:histogramPerformance}(b)
    and \emph{Bench1B} in Figure~\ref{fig:histogramPerformance}(c)). 
    These plots show the initial \SI{1}{\kilo~iterations}.
    %for full \SI{500}{\kilo~iterations} experiments. 
    With open boundary conditions (b), one observes fluctuating performance as processes get desynchronized on all but one socket.
    %\AAcomm{Fluctuations are there even with 1K iterations, though they are not prominent enough to be visible at the current y-axis scale. While if one zoom y-axes [17 19] then fluctuations are clearly visible.} \GHcomm{The y axis is zoomed anyway, so [17 19] would be OK IMO}
    However, this slow synchronized socket does not permit a global performance increase as desynchronized processes on other sockets cannot lag behind indefinitely.
    With closed boundary conditions (c), as the simulation progresses, performance (along with  MPI  waiting times) increases by about 15\% and stays constant at the higher level till the end of the run. 
    %Since bi-model effect is so small, it gets swamped and invisible on the long-scale runs and ultimately all sockets effectively achieve the similar performance as lagers pull the other behind.
    %Since all MPI processes are desynchronized in Figure~\ref{fig:histogramPerformance}(c), then one observes the performance gain of about 72-82~\si{iterations\per\second}.
    %However, there is no performance gain (constant at 18~\si{iterations\per\second}) in Figure~\ref{fig:histogramPerformance}(b) because of the presence of one sync ccNUMA domain.

    %%%%%%%%%%%%%%%%%%%   

    \begin{figure*}[t]
	\centering
	\begin{minipage}{\textwidth}
	    \begin{subfigure}[t]{0.19\textwidth} 
			\includegraphics[scale=0.153]{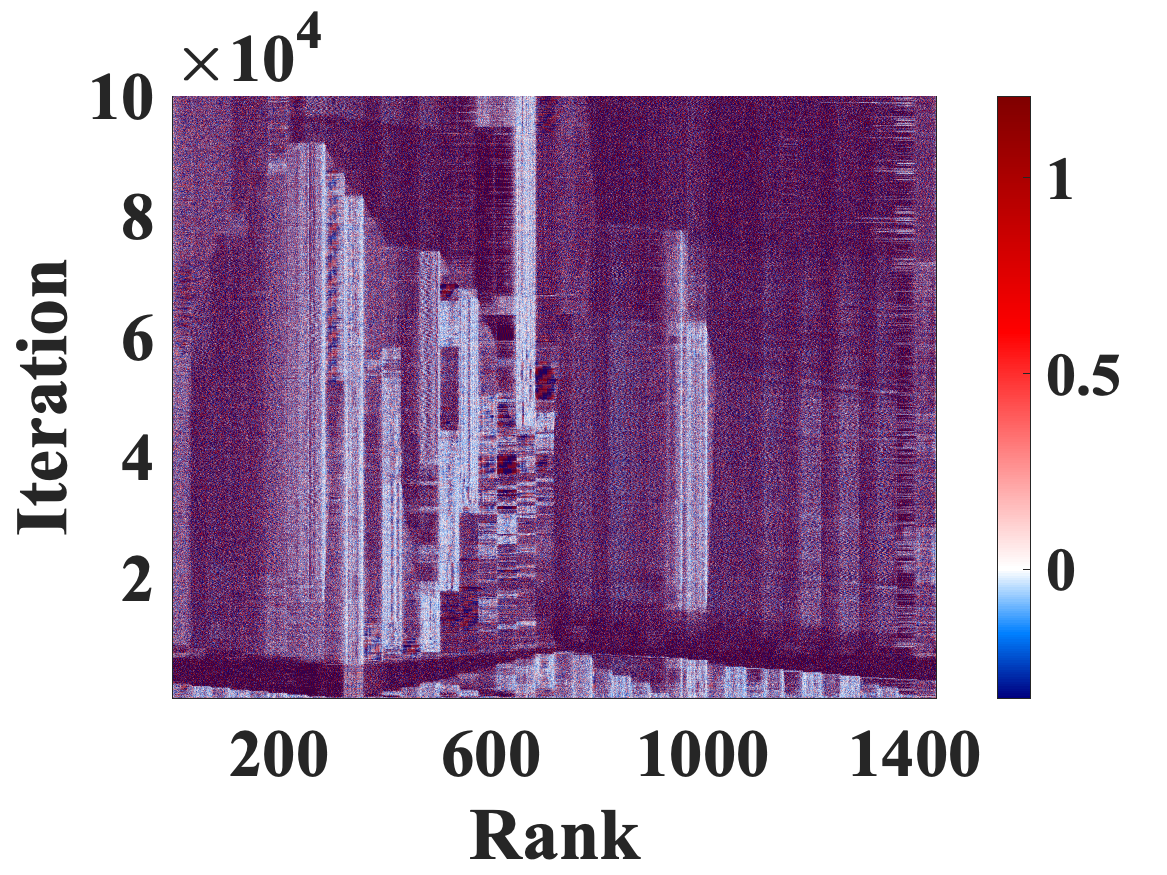}
		    \caption{LBM-100\si{\kilo~it}}
		\end{subfigure}
		\begin{subfigure}[t]{0.19\textwidth} 
			\includegraphics[scale=0.153]{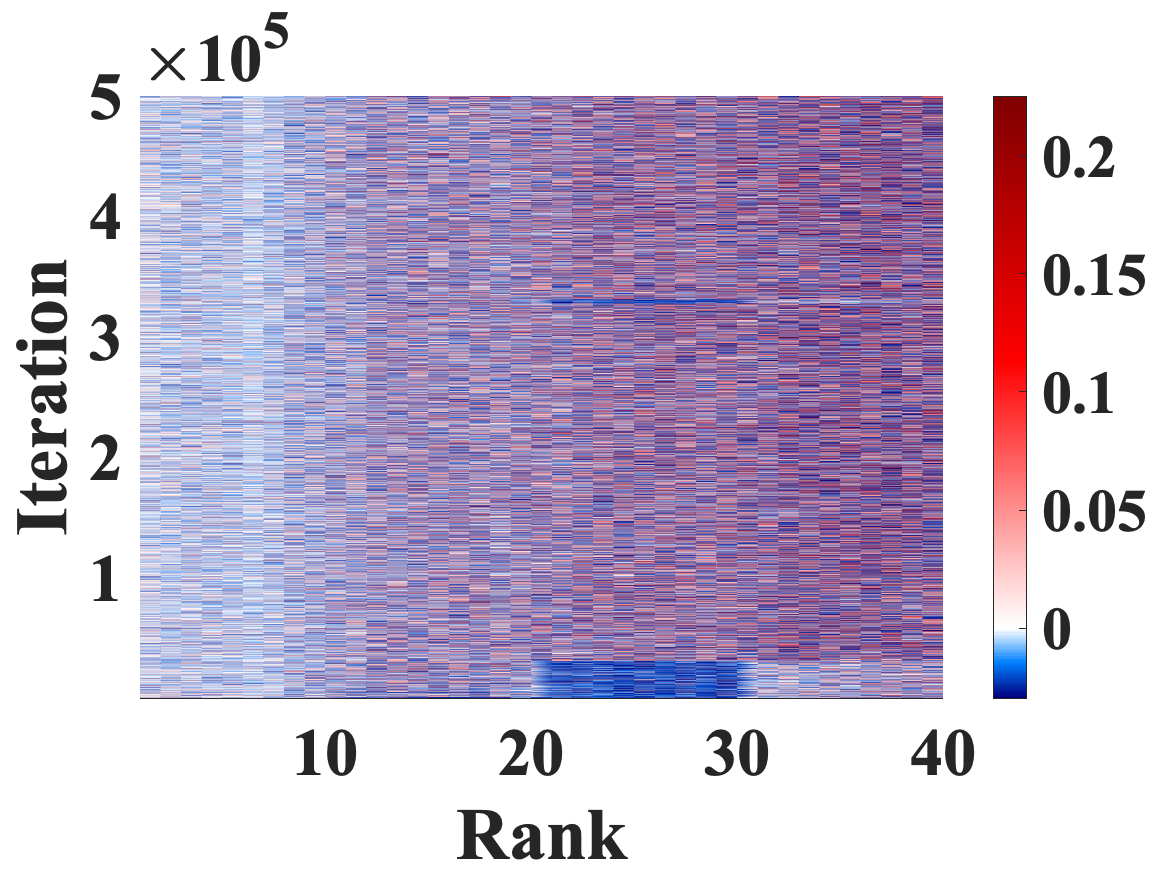}
			\caption{Bench1A-500\si{\kilo~it}}
		\end{subfigure}
		\begin{subfigure}[t]{0.19\textwidth} 
			\includegraphics[scale=0.153]{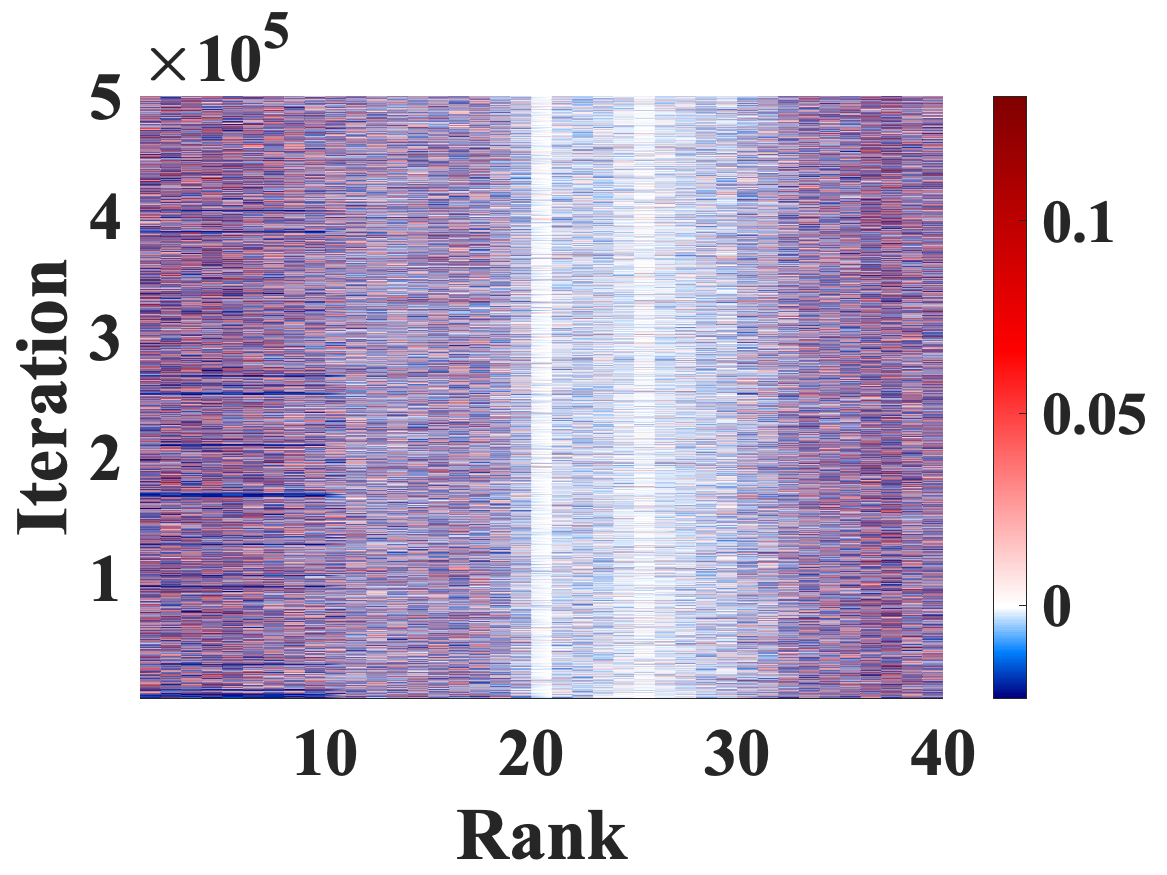}
			\caption{Bench2A-500\si{\kilo~it}}
		\end{subfigure}
		\begin{subfigure}[t]{0.19\textwidth} 
			\includegraphics[scale=0.153]{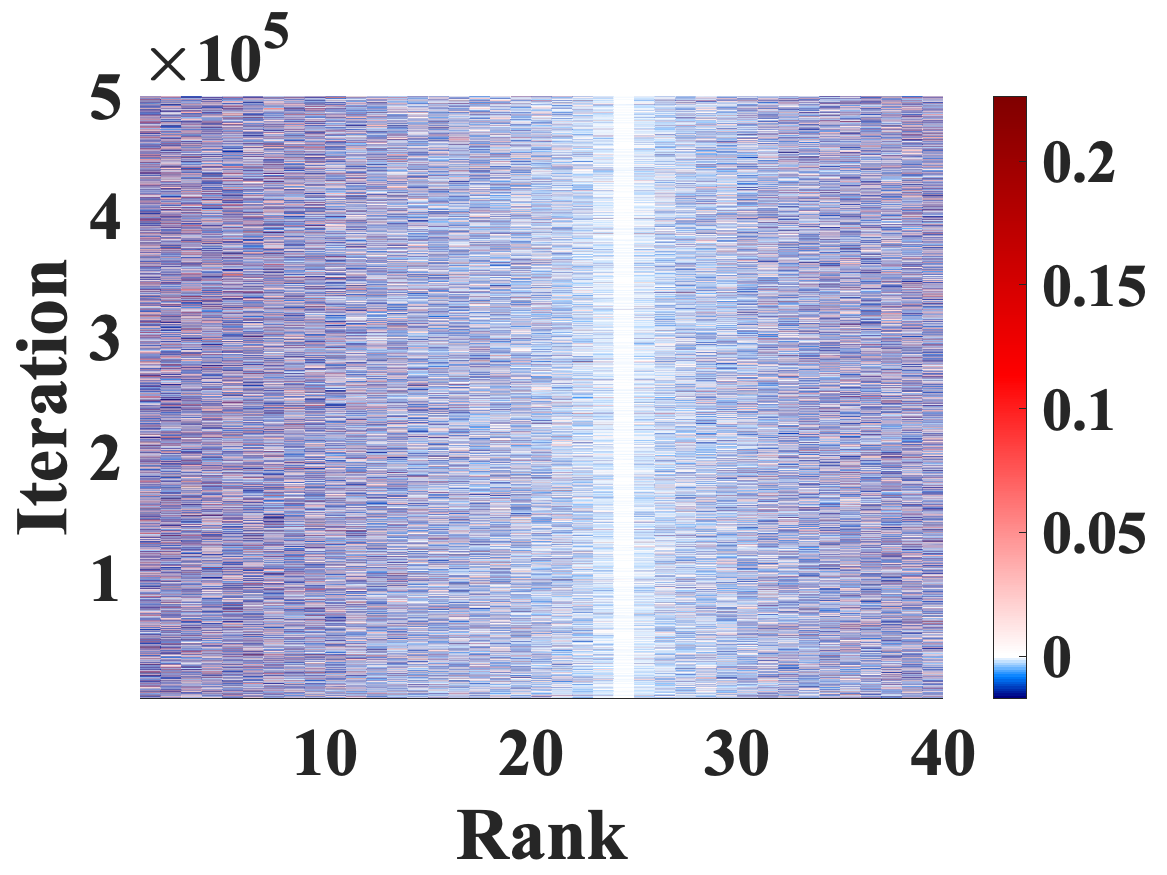}
		    \caption{Bench3A-500\si{\kilo~it}}
		\end{subfigure}
		\begin{subfigure}[t]{0.19\textwidth} 
			\includegraphics[scale=0.153]{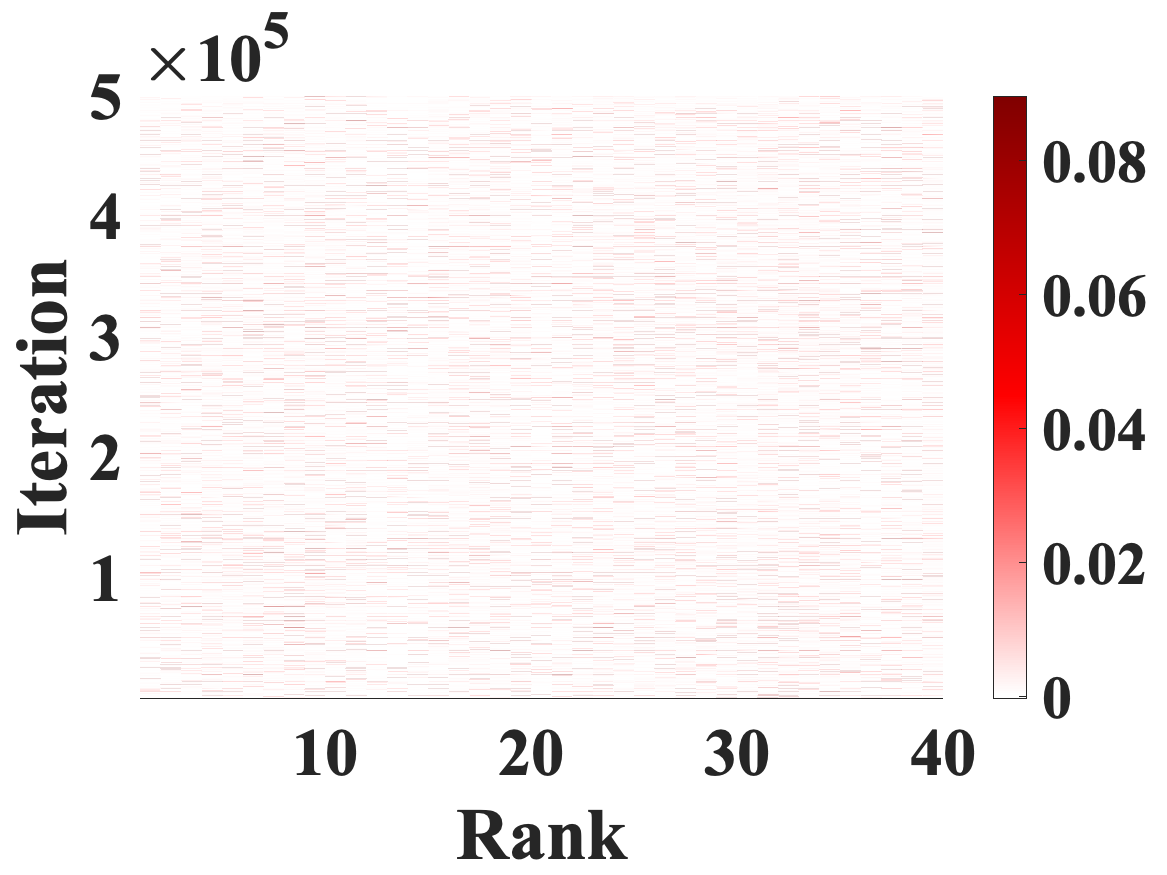}
		    \caption{Bench4A-500\si{\kilo~it}}
		\end{subfigure}
	\end{minipage}
	\begin{minipage}{\textwidth}
	    \begin{subfigure}[t]{0.19\textwidth} 
			\includegraphics[scale=0.153]{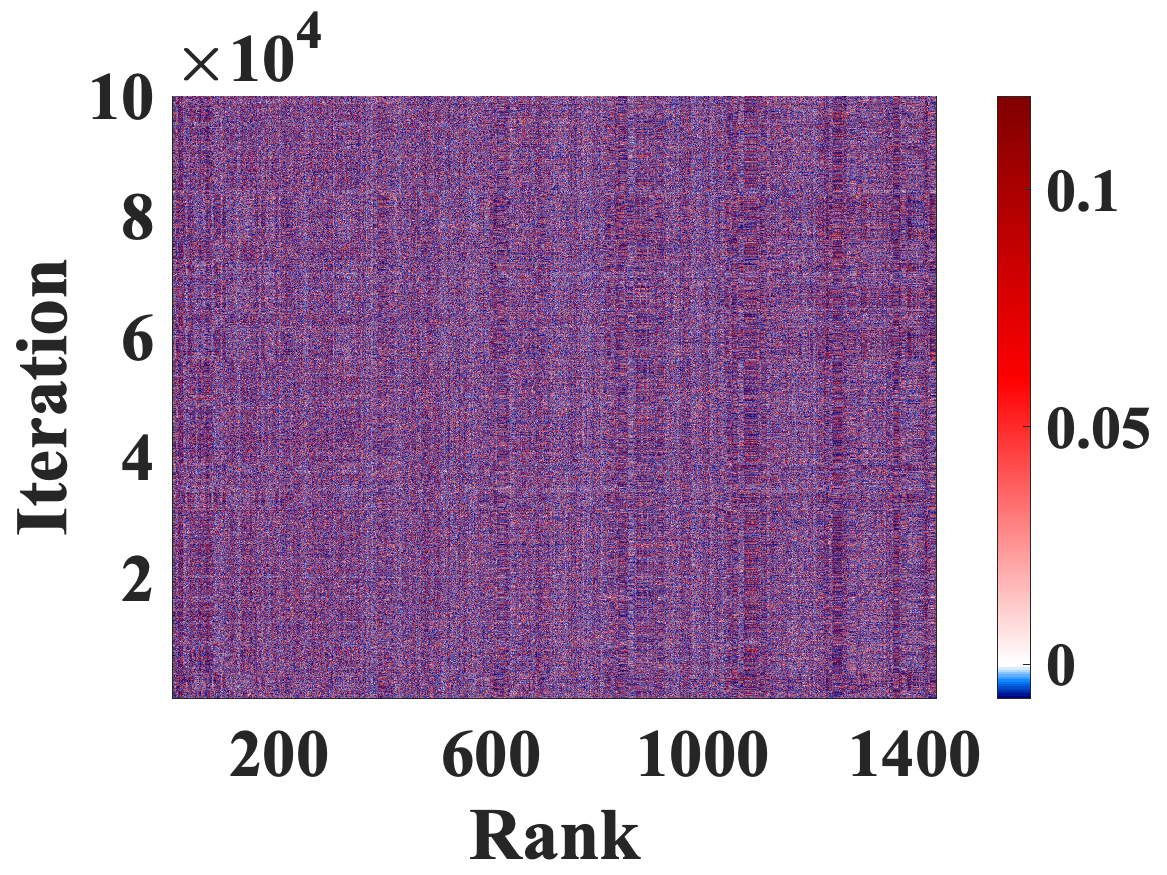}
		    \caption{spMVM-100\si{\kilo~it}}
		\end{subfigure}
		\begin{subfigure}[t]{0.19\textwidth} 
			\includegraphics[scale=0.153]{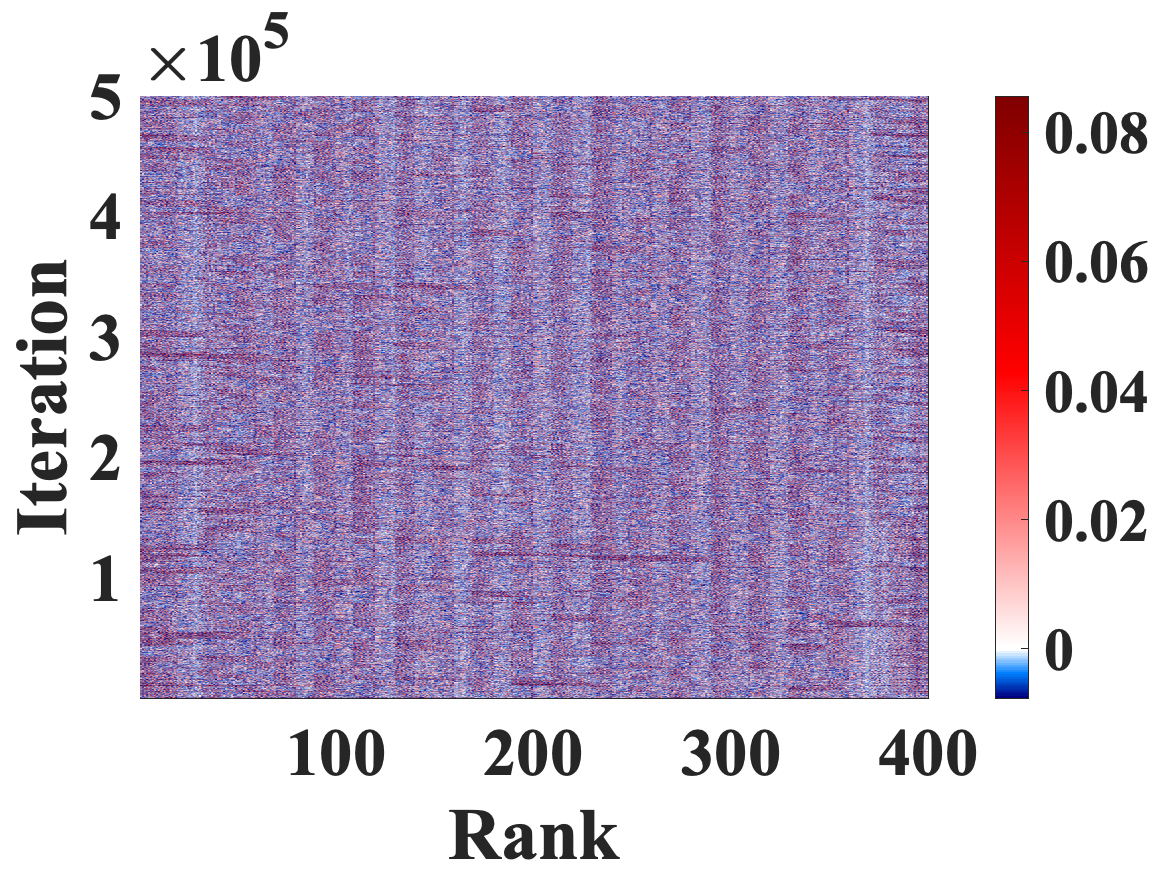}
		    \caption{Bench1B-500\si{\kilo~it}}
		\end{subfigure}
		\begin{subfigure}[t]{0.19\textwidth} 
			\includegraphics[scale=0.153]{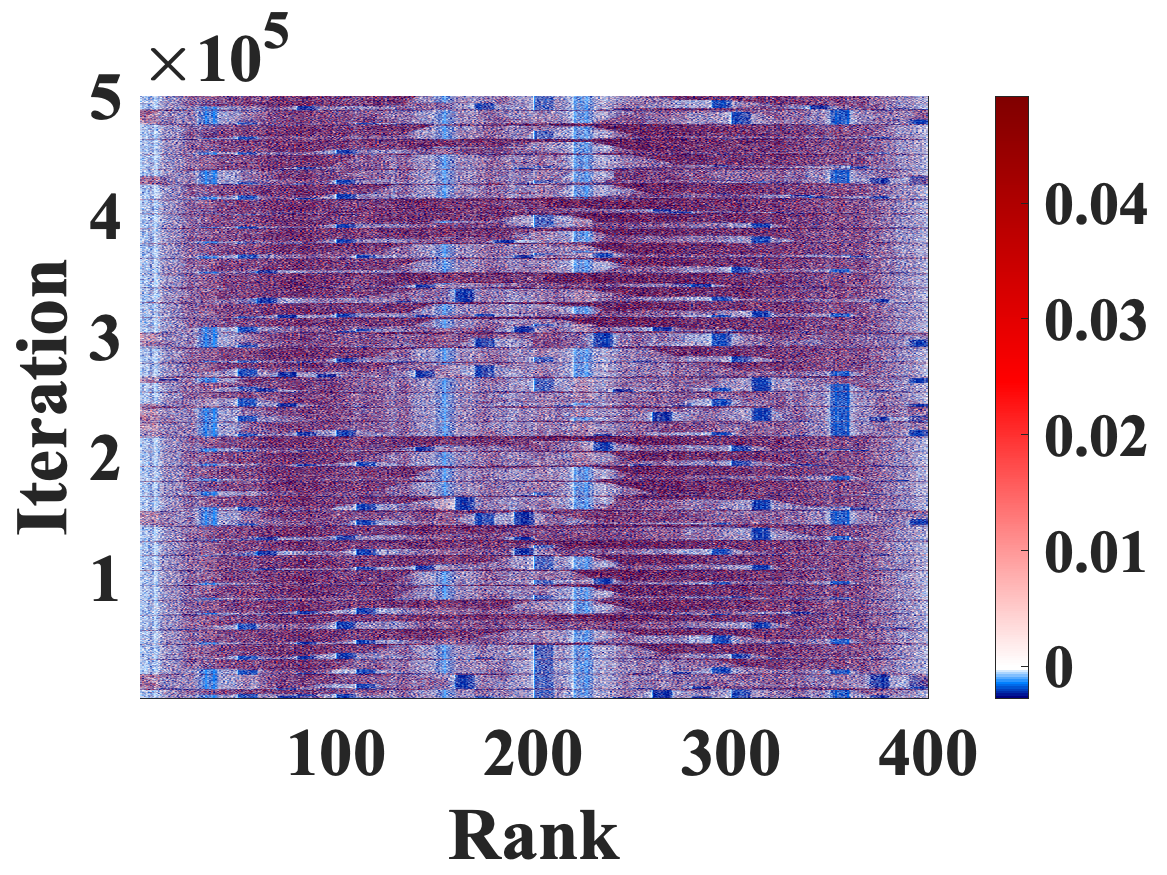}
		    \caption{Bench2B-500\si{\kilo~it}}
		\end{subfigure}
		\begin{subfigure}[t]{0.19\textwidth} 
			\includegraphics[scale=0.153]{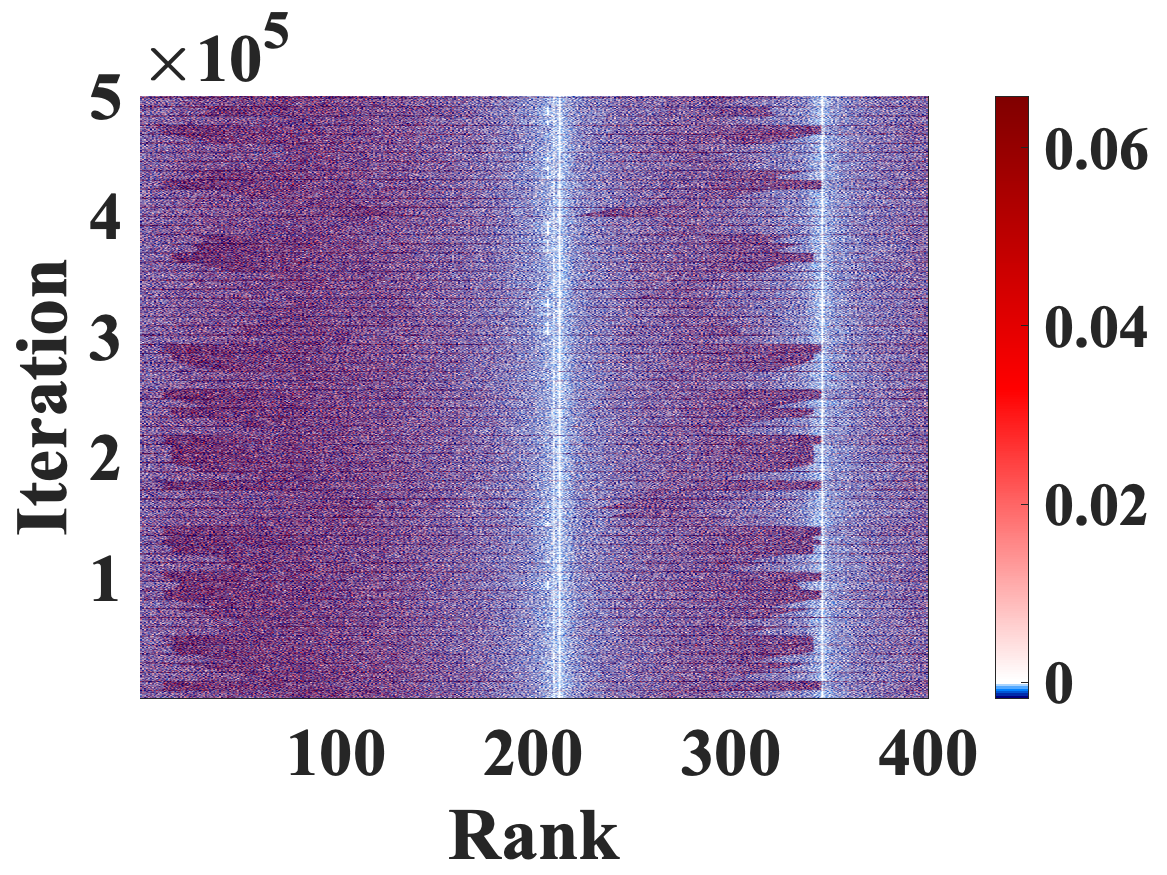}
		    \caption{Bench3B-500\si{\kilo~it}}
		\end{subfigure}
		\begin{subfigure}[t]{0.19\textwidth} 
			\includegraphics[scale=0.153]{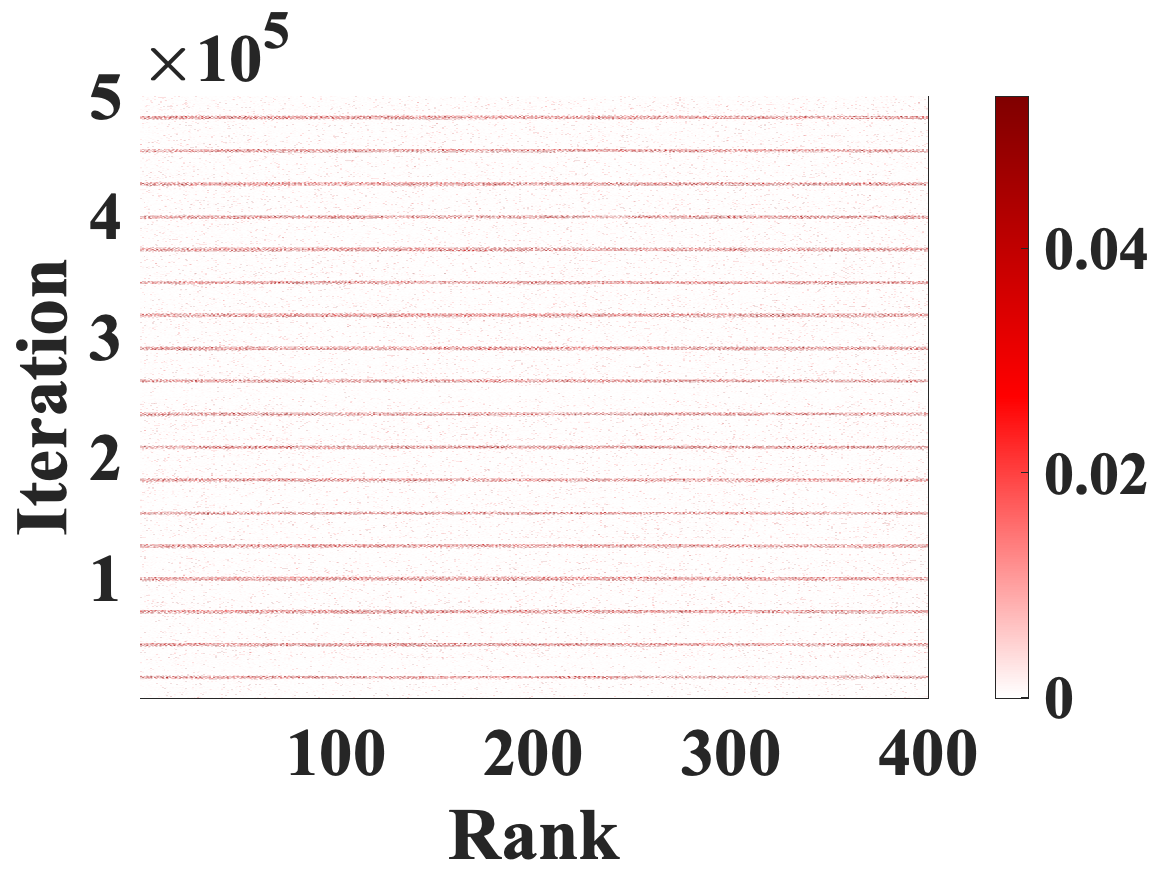}
		    \caption{Bench4B-500\si{\kilo~it}}
		\end{subfigure}
	\end{minipage}
	\caption{Compact rank($x$-axis)-iteration($y$-axis) timelines of MPI waiting times [\si{\sec}] for all benchmarks and full end-to-end runs.}
    \label{fig:compacttimelines}
\end{figure*}
    %%%%%%%%%%%%%%%%%%%
    \subsection{Timeline in compact representation} % pcolor((MPItime(:,:)-mean(MPItime(:,:)))) % shading interp
    MPI waiting times facilitate a compact representation of the timeline of a parallel program. Figure~\ref{fig:compacttimelines} show ranks on the $x$-axis and time steps on the $y$-axis, with the normalized MPI waiting time for each rank and time step color coded. For this representation, the mean value of MPI times across all processes and time steps is shown in white while red (positive value) and blue (negative value) represent values above and below the mean, respectively.
    This makes it possible to distinguish between synchronized and desynchronized groups of processes: Strongly desynchronized processes spend more time in MPI (red), while white 
    %\GHcomm{and blue?}
    color marks synchronized processes (Ranks 1-10 and 20-30 in Figure~\ref{fig:compacttimelines}(b) and (c), respectively).
    % or blue? Absolute waiting time can go down with desynchronization in case of network contention.
    This visualization is similar to what tracing tools like ITAC or Vampir display; however, these tools often encompass too much information, and depending on the chosen resolution one can easily get lost in the data. 
    In contrast, compact timelines of the waiting time per time step deliver a condensed view on this information and help to better visualize certain phenomena.
    For instance, the weaker saturation cases collect lower idle times which can be seen when comparing Figures~\ref{fig:compacttimelines}(c)--(e) and (h)--(j)).
    Asymptotic behavior in longer runs can be observed at the top part of the plot in all cases. % (Figure~\ref{fig:compacttimelines}(a)-(j)).
    Idle waves are prominently visible as dark-blue stripes in the LBM benchmark (Figure~\ref{fig:compacttimelines}(a)).
    %One strong idle wave switched the synchronized socket to the neighboring socket for a certain phase and then it came back for another strong idle wave.
    %% GHa: Not sure about this interpretation. Did you mean computational wavefronts?
%%%%%%%%%%%%%%%%%%%%%%%%%%%%%%%%%%%%%%%%%%%%%%%%%%%%%%%%%%%%%%%%%%%%%  
\section{Advanced metrics for analysis} \label{sec:additionalmetrics}
    %%%%%%%%%%%%%%%%%%%
    \begin{figure*}[t]
	\centering
	\begin{minipage}{\textwidth}
	    \begin{subfigure}[t]{0.19\textwidth} 
			\includegraphics[scale=0.0765]{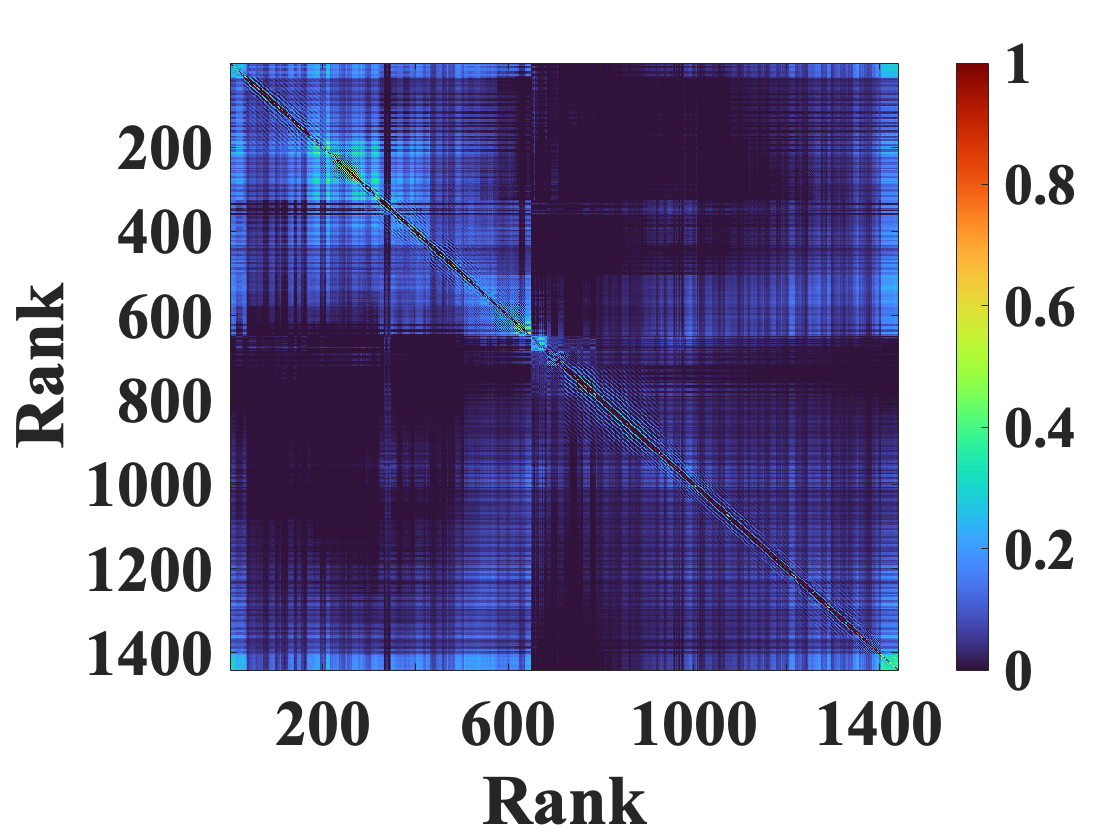}
		    \caption{LBM-100\si{\kilo~it}}
		\end{subfigure}
		\begin{subfigure}[t]{0.19\textwidth} 
			\includegraphics[scale=0.153]{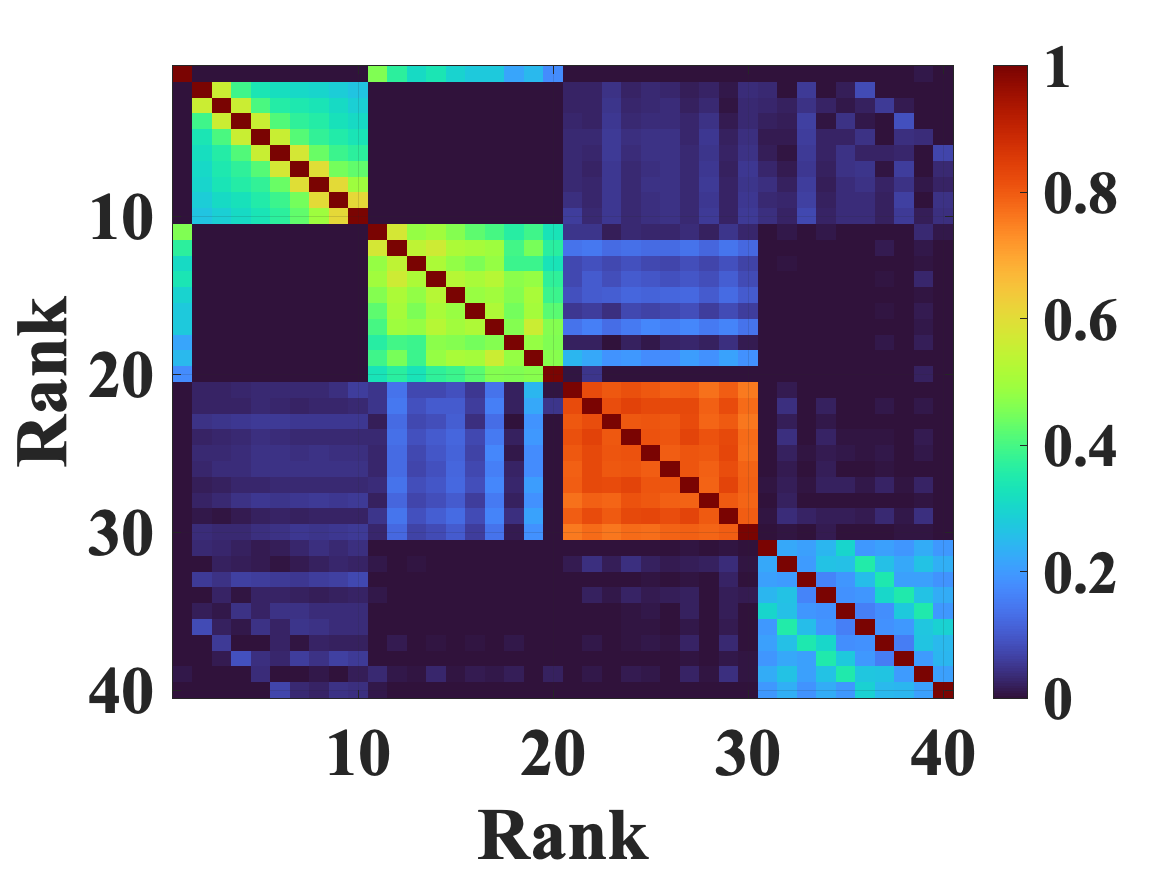}
			\caption{Bench1A-500\si{\kilo~it}}
		\end{subfigure}
		\begin{subfigure}[t]{0.19\textwidth} 
			\includegraphics[scale=0.153]{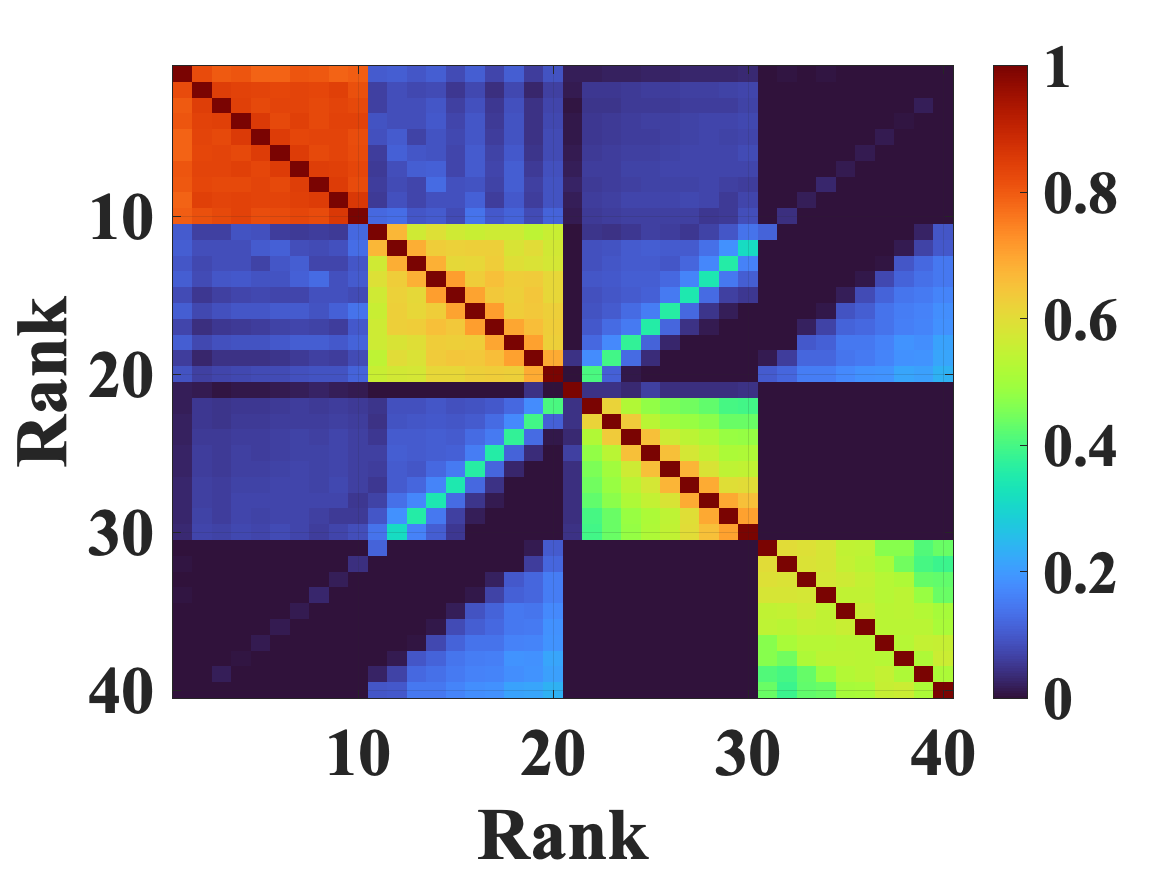}
			\caption{Bench2A-500\si{\kilo~it}}
		\end{subfigure}
		\begin{subfigure}[t]{0.19\textwidth} 
			\includegraphics[scale=0.153]{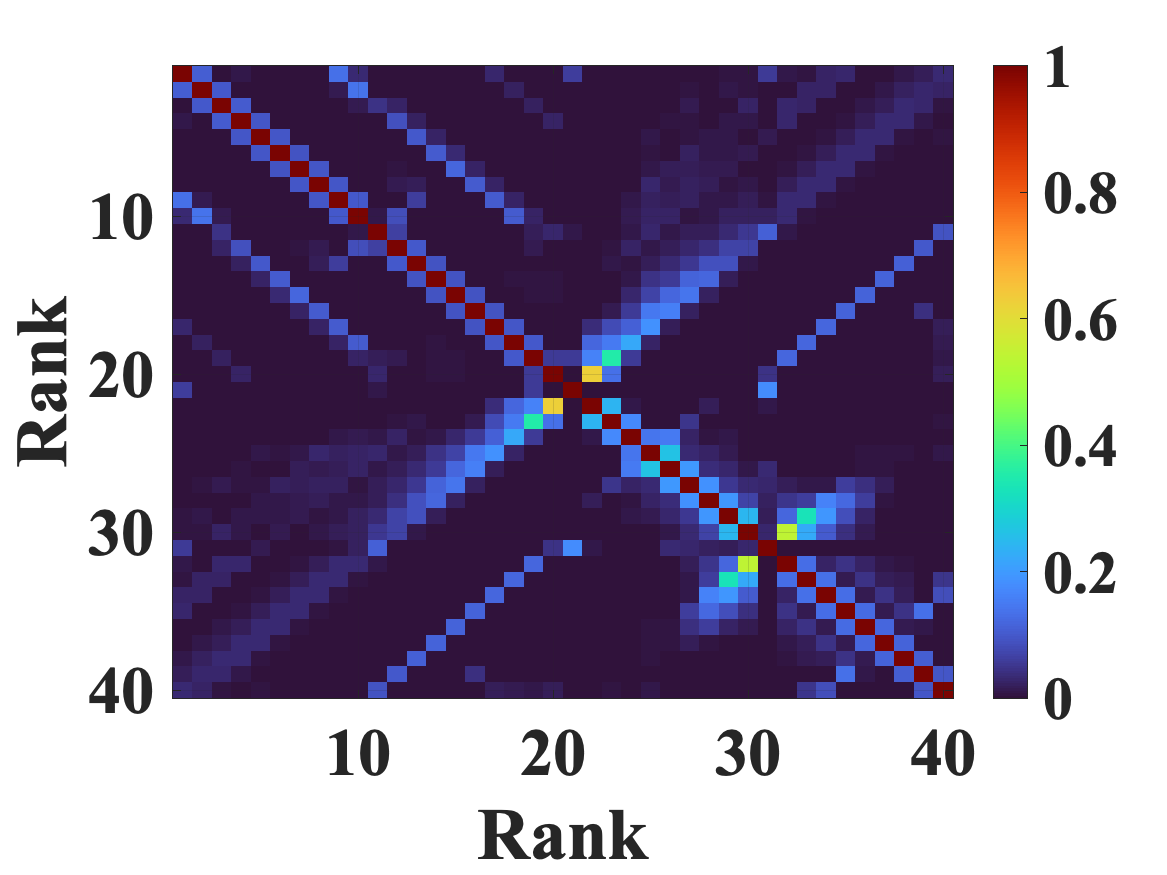}
		    \caption{Bench3A-500\si{\kilo~it}}
		\end{subfigure}
		\begin{subfigure}[t]{0.19\textwidth} 
			\includegraphics[scale=0.153]{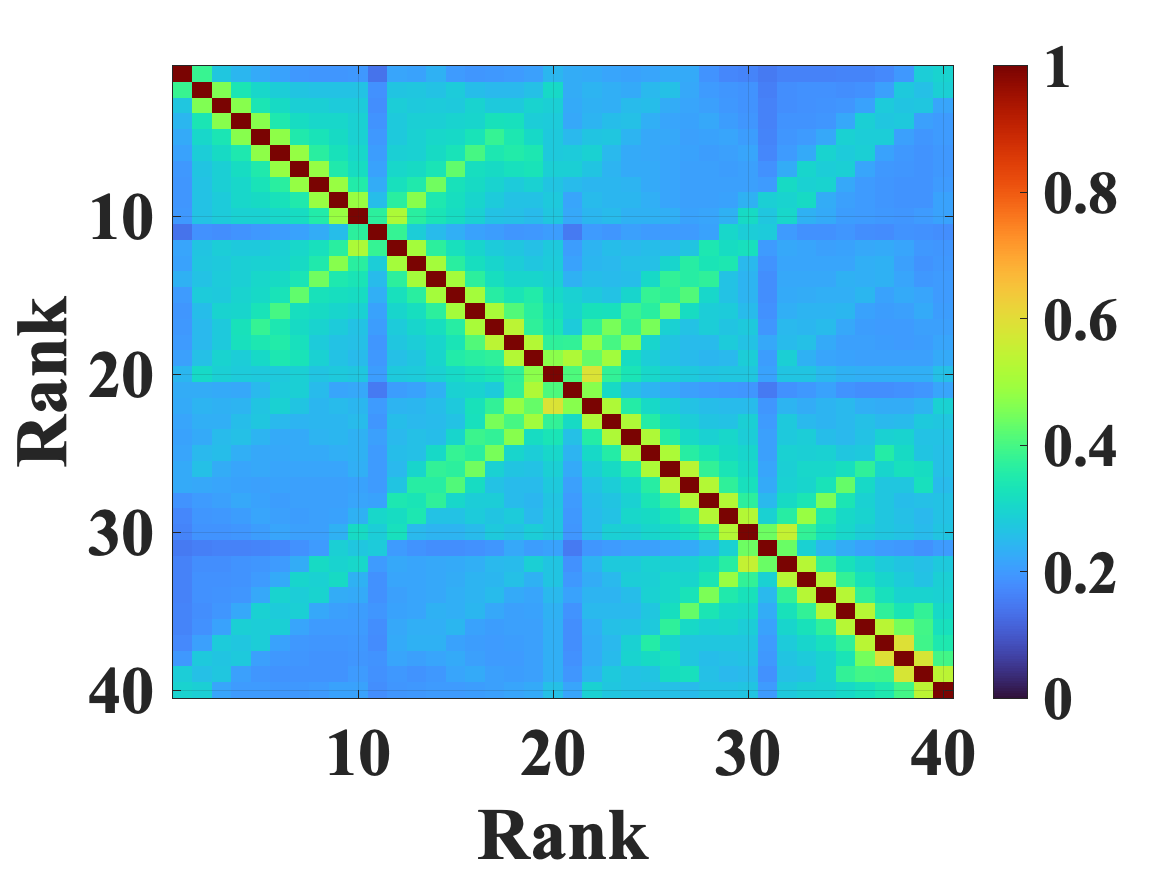}
		    \caption{Bench4A-500\si{\kilo~it}}
		\end{subfigure}
	\end{minipage}
	\begin{minipage}{\textwidth}
	    \begin{subfigure}[t]{0.19\textwidth} 
			\includegraphics[scale=0.153]{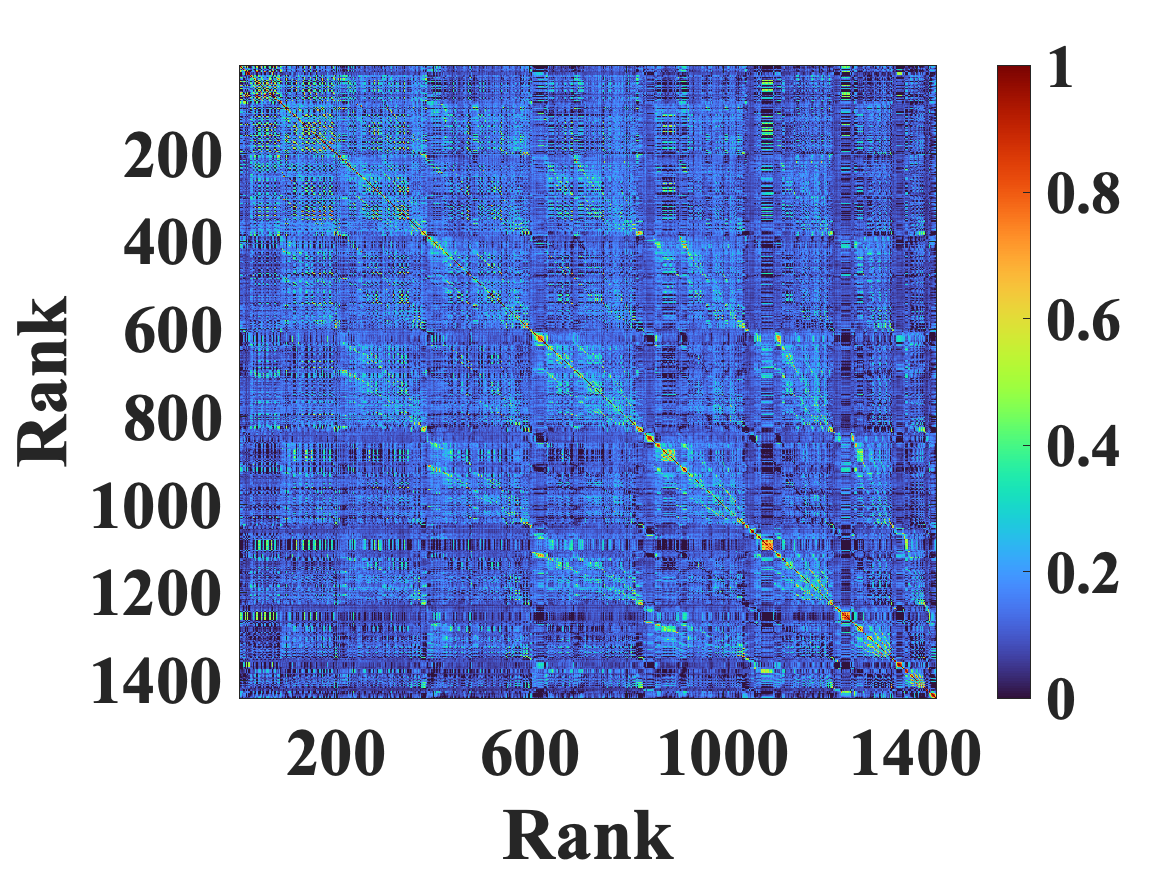}
		    \caption{spMVM-500\si{\kilo~it}}
		\end{subfigure}
		\begin{subfigure}[t]{0.19\textwidth} 
			\includegraphics[scale=0.153]{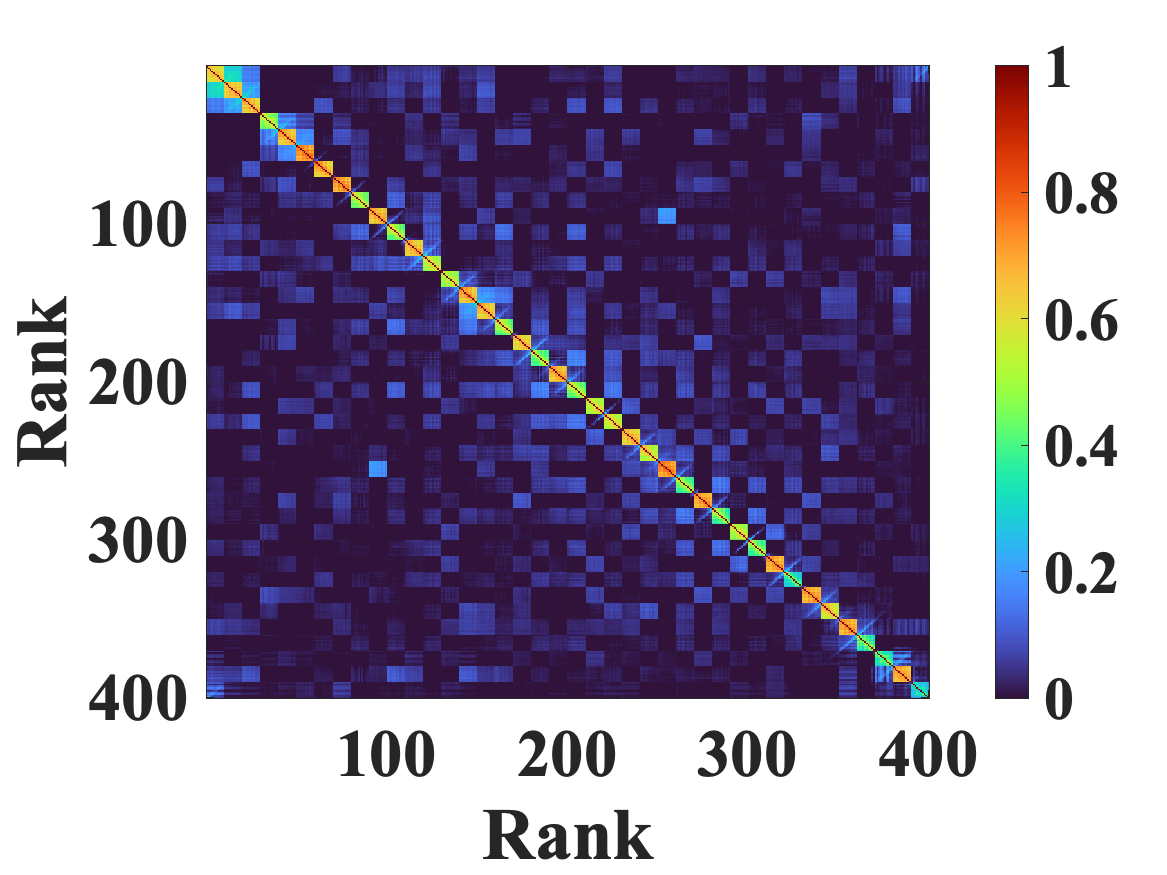}
		    \caption{Bench1B-500\si{\kilo~it}}
		\end{subfigure}
		\begin{subfigure}[t]{0.19\textwidth} 
			\includegraphics[scale=0.153]{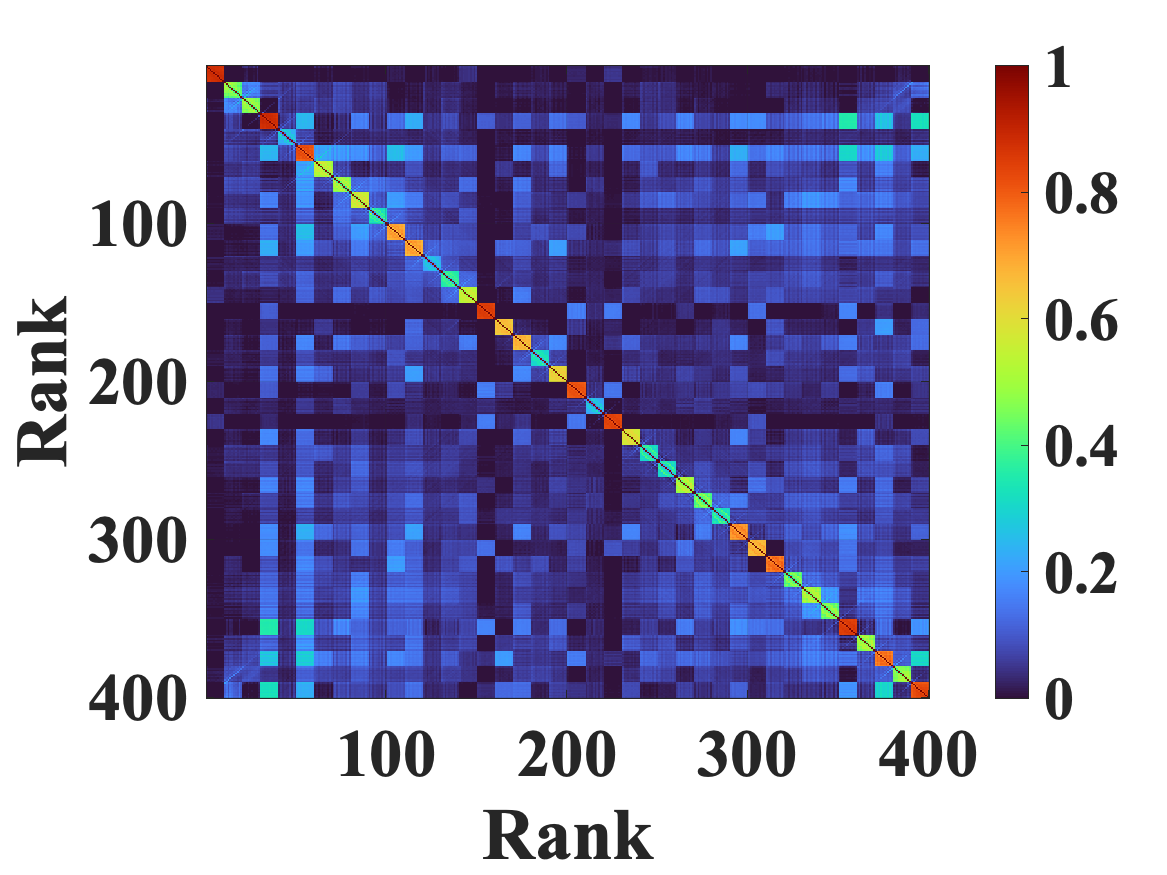}
		    \caption{Bench2B-500\si{\kilo~it}}
		\end{subfigure}
		\begin{subfigure}[t]{0.19\textwidth} 
			\includegraphics[scale=0.153]{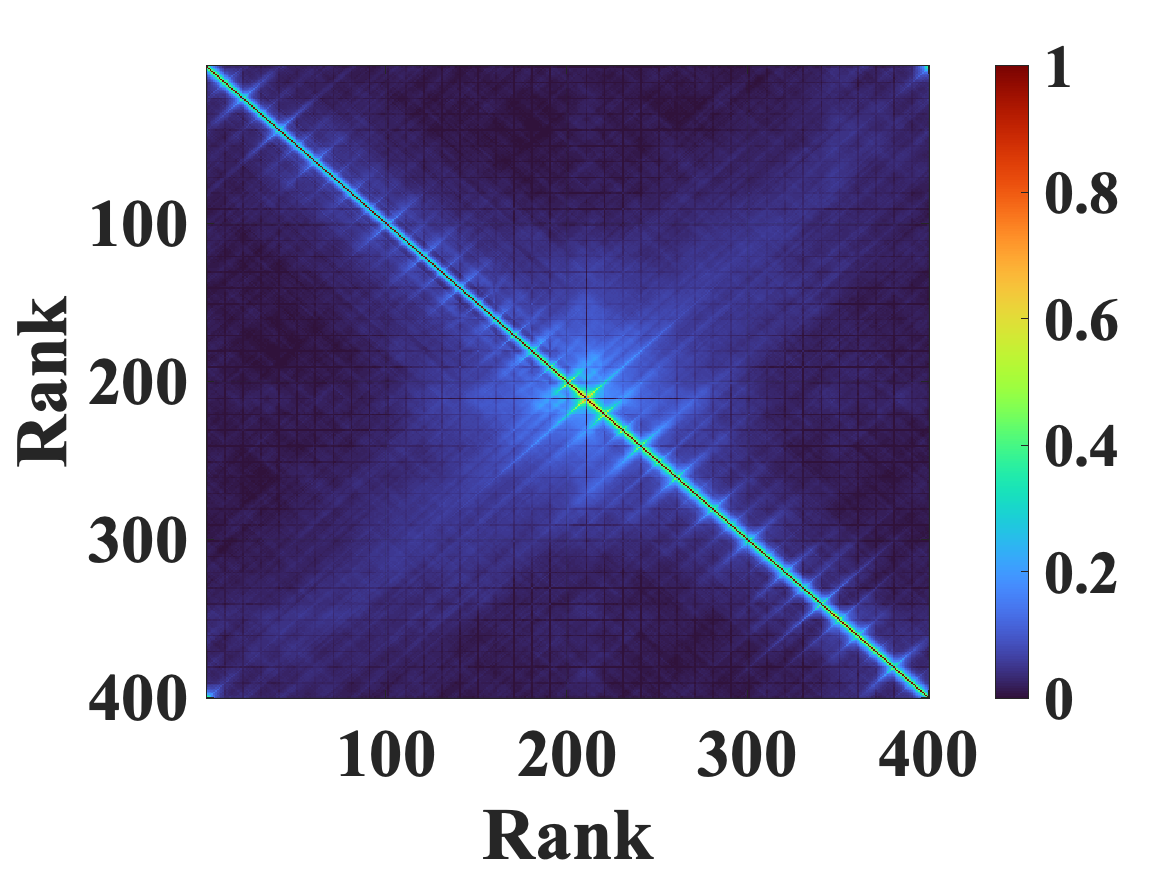}
		    \caption{Bench3B-500\si{\kilo~it}}
		\end{subfigure}
		\begin{subfigure}[t]{0.19\textwidth} 
			\includegraphics[scale=0.153]{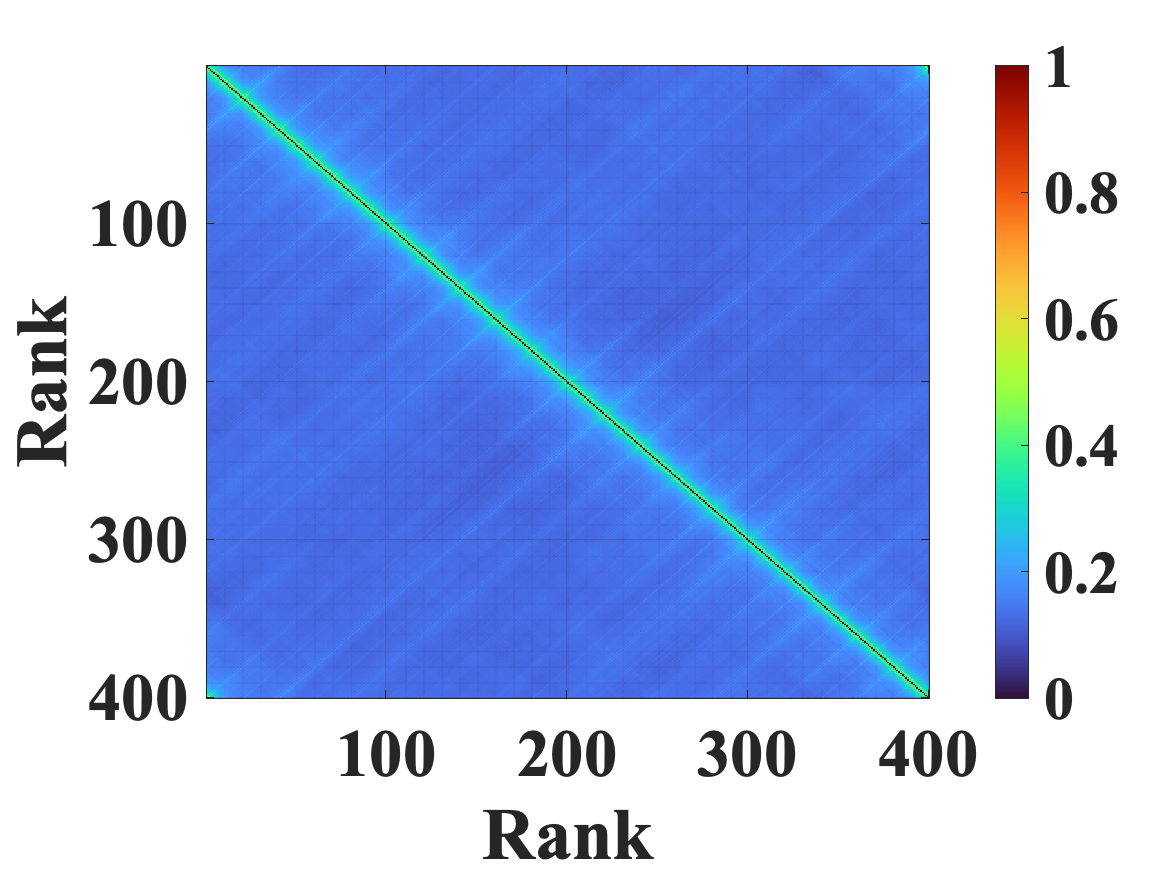}
		    \caption{Bench4B-500\si{\kilo~it}}
		\end{subfigure}
	\end{minipage}
	\caption{Correlation coefficiens of MPI times [\si{\sec}] between process pairs for all benchmarks.}
    \label{fig:CCMatrix}
\end{figure*}
    %%%%%%%%%%%%%%%%%%%
    Beyond timeline visualization and statistics, a plethora of advanced data analysis methods exist which can lead to deeper insights into the desynchronization process. Here we pick the \emph{correlation coefficient} and the \emph{phase space plot}.
    
    \subsection{Correlation coefficient} % imagesc(corrcoef(MPItime(:,:)))
    
    The correlation coefficient function~\cite{vetterling1992numerical} provides a simple way to uncover correlations between the timelines of two MPI processes. 
    %It normalizes the distribution of vectors in high-dimensional space by its mean and standard deviation.\GHcomm{I don't understand the previous sentence. Is it necessary?}\AAcomm{The correlation coefficients in MATLAB are calculated by taking the ratio of covariance to mean value. You can omit the description of the formula, if not required.}
    Figure~\ref{fig:CCMatrix} shows the color-coded correlation coefficients of rank pairs for all benchmarks, using the full end-to-end timelines.
    %It uses the full range of colors with the smallest coefficient value maps to the dark blue color and the largest value maps to the dark red color in the colormap.
    The matrices are obviously symmetric, and the diagonal entries (dark red) are set to one by convention.
    The correlation coefficients range from $-1$ to $1$, with $-1$ representing a direct, negative correlation, $0$ representing no correlation, and $1$ representing a direct, positive correlation.
    For the memory-bound applications, the ccNUMA domain structure is clearly visible in Figures~\ref{fig:CCMatrix}(a-c,f-h). This implies that processes within ccNUMA domains are strongly correlated, while they are less (or not) correlated across sockets.
    The data shows strong correlations within desynchronized sockets with a bi-modal distribution of MPI times since the socket already started to lose the sync pattern.
    In the open chain scenarios, processes on the last socket show a weaker correlation.
    In the SpMVM application, the sparse matrix structure is reflected in the correlation coefficients since the desynchronization process is strongly influenced by the communication structure (Figure~\ref{fig:CCMatrix}(f)).
    In weakly or non-saturated applications (Figures~\ref{fig:CCMatrix}(d-e, i-j)), correlations are generally weaker, as expected.
    %\GHcomm{But I do see the socket structure in (e) and also in (d); also, if all processes are equal, should the plots not be all red for PISOLVER?}\AAcomm{True, but all MPI processes will be equally correlated (and not very weakly correlated); its not red as correlations cannot be maximum because of random variations throughout the whole run of 500K iterations.}

    %%%%%%%%%%%%%%%%%%%
    \begin{figure*}[t]
	\centering
	\begin{minipage}{\textwidth}
% 	    \begin{subfigure}[t]{0.24\textwidth} 
% % 			\includegraphics[scale=0.11]{figures/phaseSpace/LBM-start.png}
%     		\includegraphics[scale=0.11]{figures/phaseSpace/LBM-start_redo.png}
% 		    \caption{LBM-initial}
% 		\end{subfigure}
% 		\begin{subfigure}[t]{0.24\textwidth} 
% % 			\includegraphics[scale=0.11]{figures/phaseSpace/LBM-end.png}
%  			\includegraphics[scale=0.11]{figures/phaseSpace/LBM-end_redo.png}
% 			\caption{LBM-mid}
% 		\end{subfigure}
		\begin{subfigure}[t]{0.24\textwidth} 
			\includegraphics[scale=0.11]{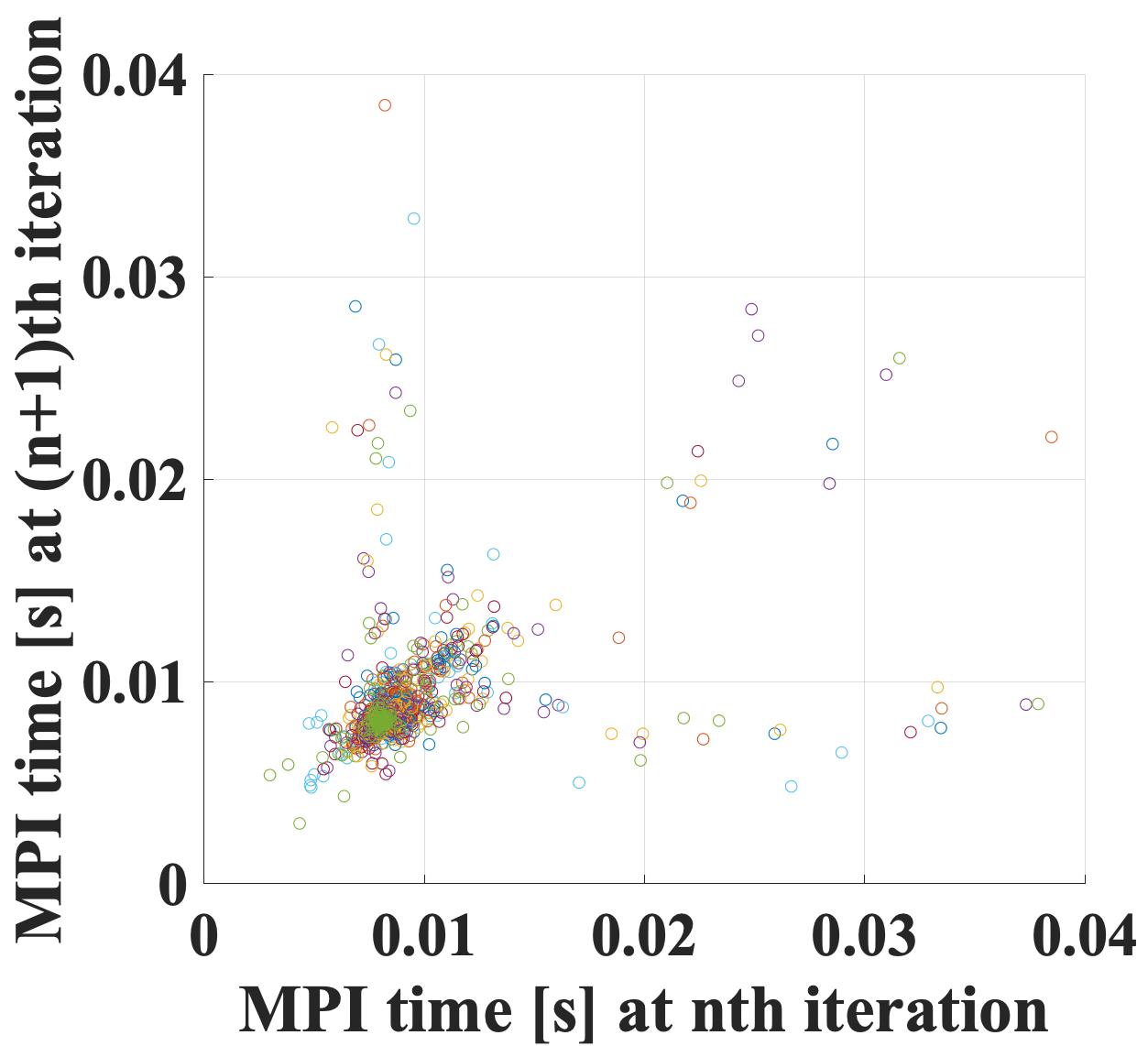}
			\caption{Bench1A-initial}
		\end{subfigure}
		\begin{subfigure}[t]{0.24\textwidth} 
			\includegraphics[scale=0.11]{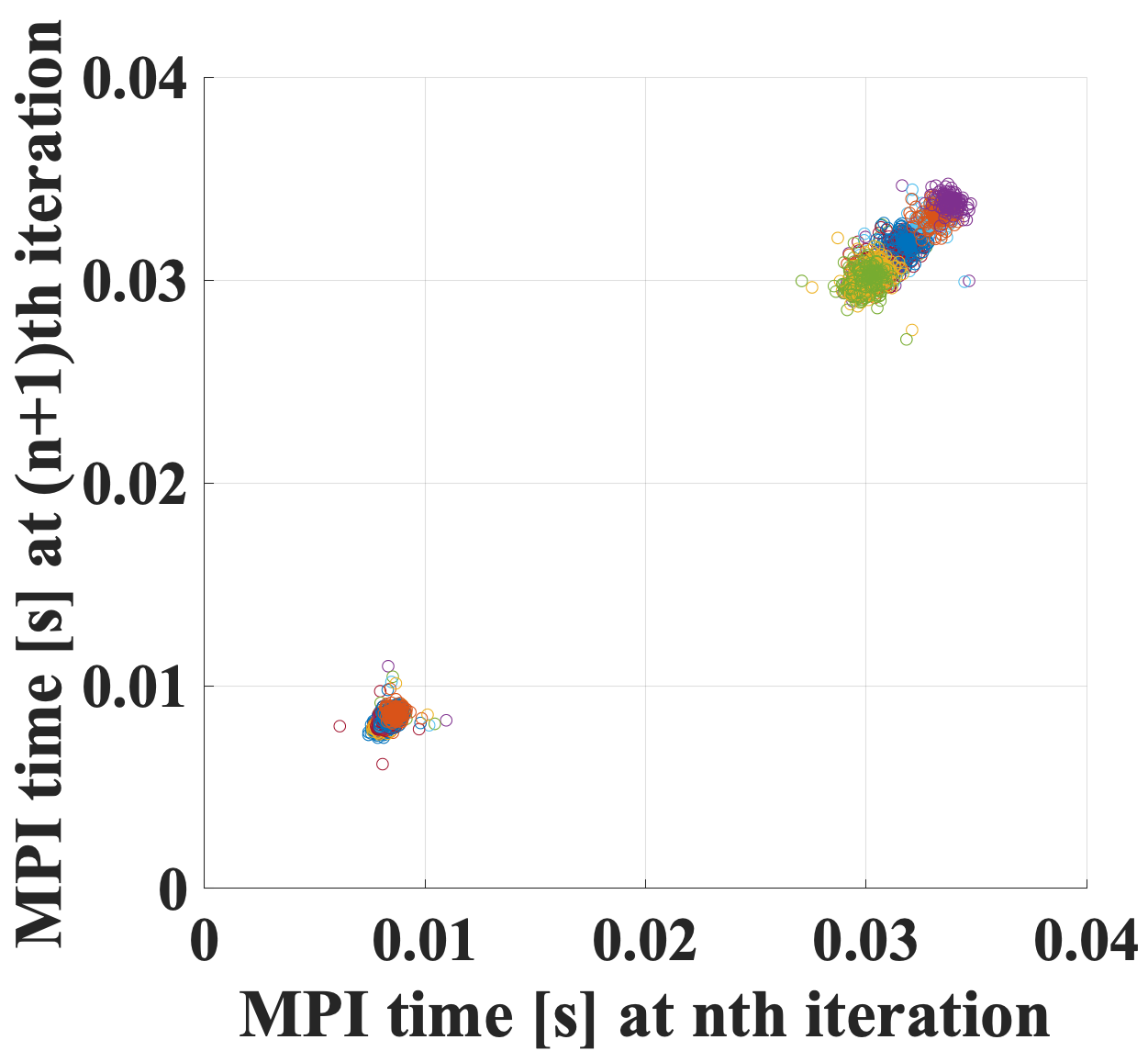}
		    \caption{Bench1A-mid}
		\end{subfigure}
		\begin{subfigure}[t]{0.24\textwidth} 
			\includegraphics[scale=0.11]{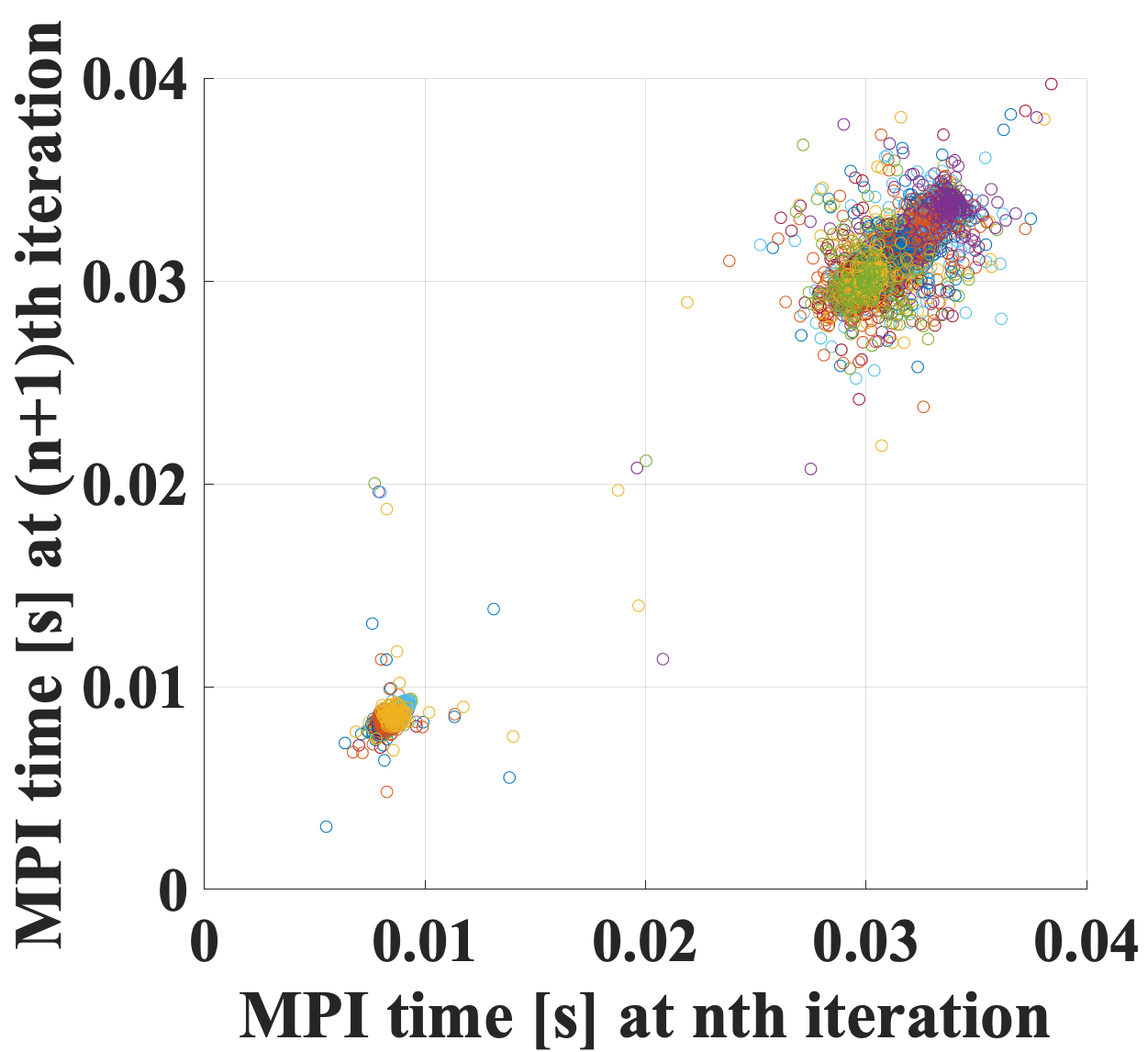}
		    \caption{Bench1A-end}
		\end{subfigure}
		\begin{subfigure}[t]{0.24\textwidth} 
			\includegraphics[scale=0.11]{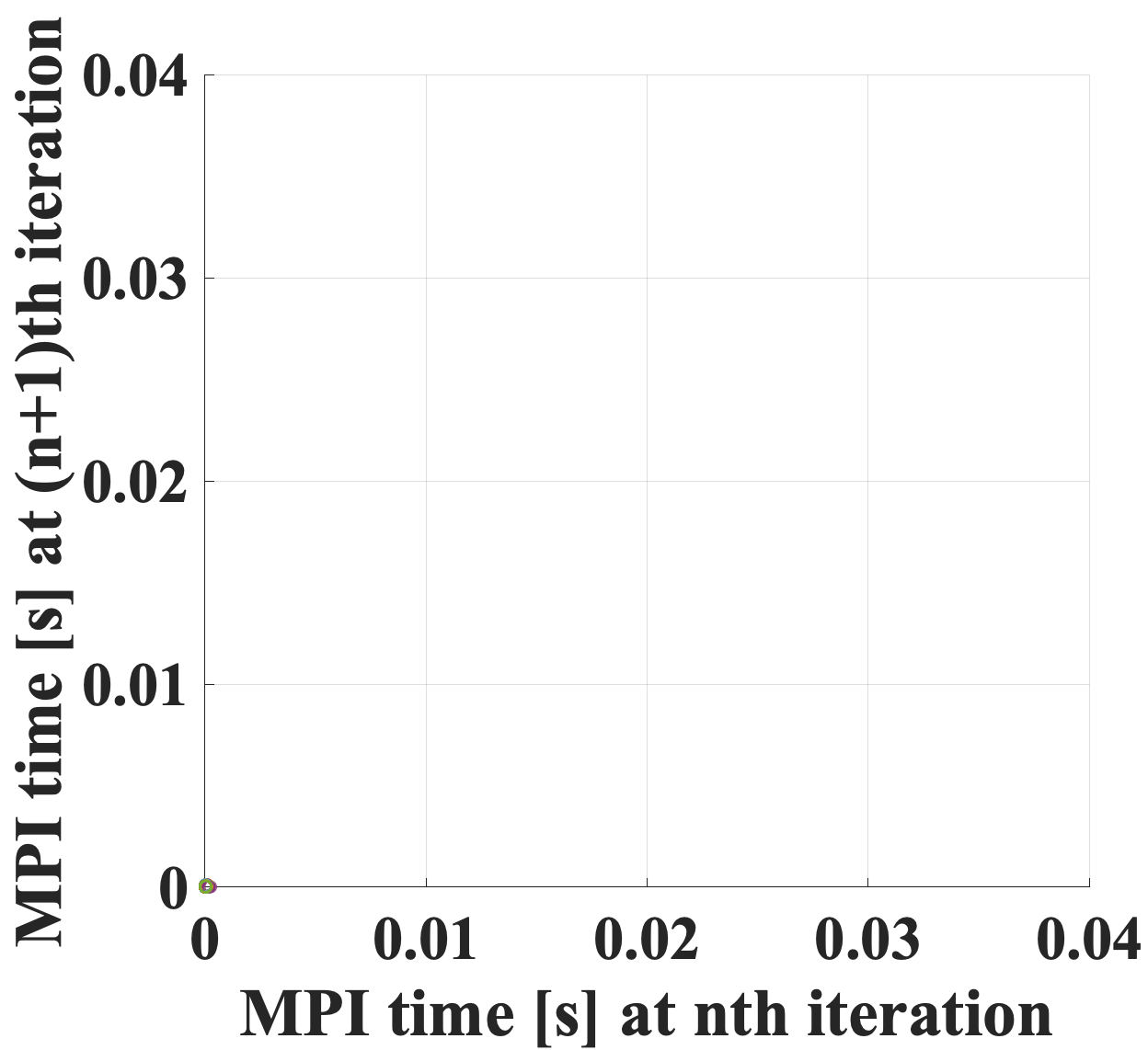}
		    \caption{Bench4A-initial}
		\end{subfigure}
	\end{minipage}
	\begin{minipage}{\textwidth}
		\begin{subfigure}[t]{0.24\textwidth} 
			\includegraphics[scale=0.11]{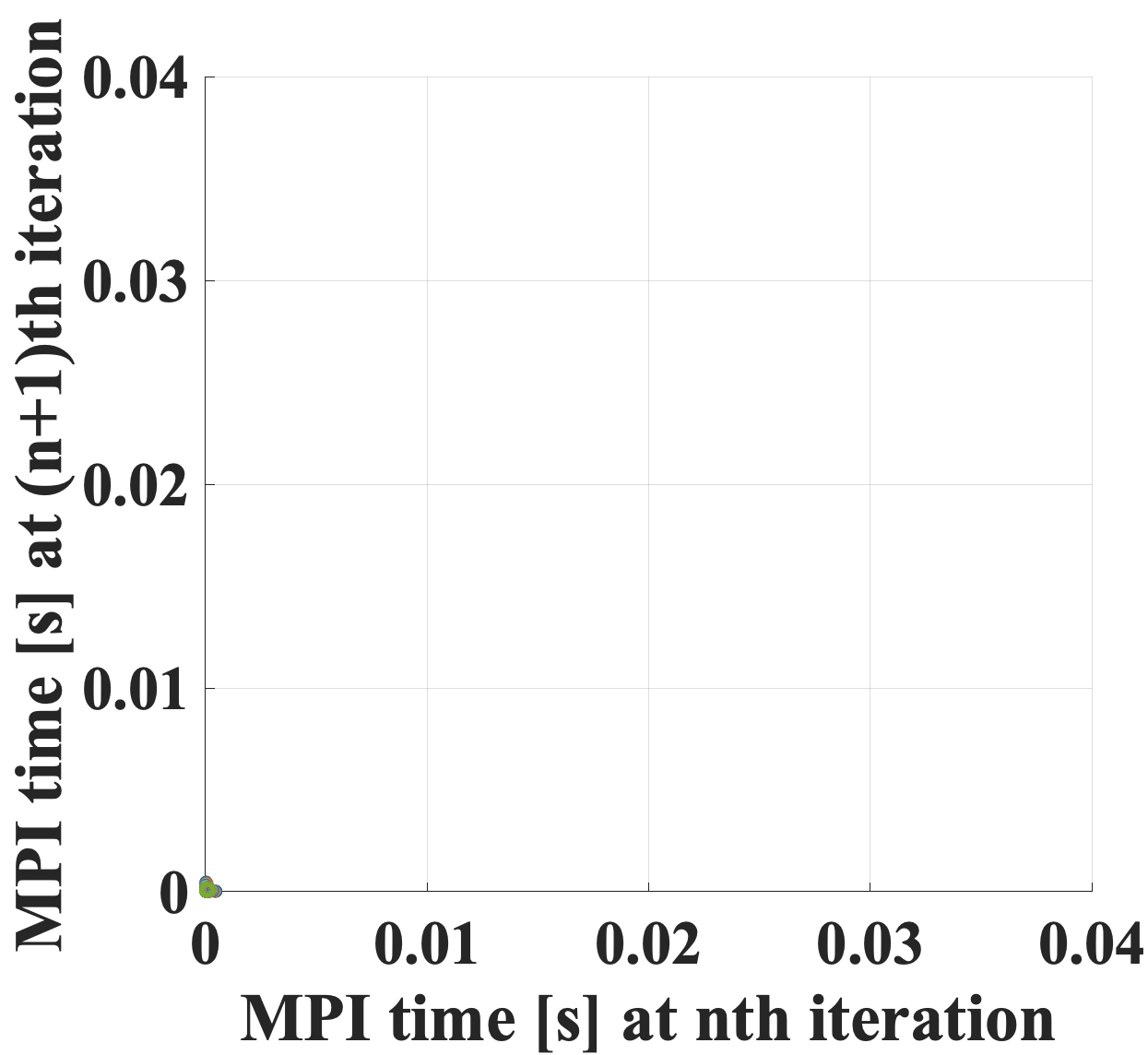}
		    \caption{Bench4A-end}
		\end{subfigure}
		\begin{subfigure}[t]{0.24\textwidth} 
			\includegraphics[scale=0.11]{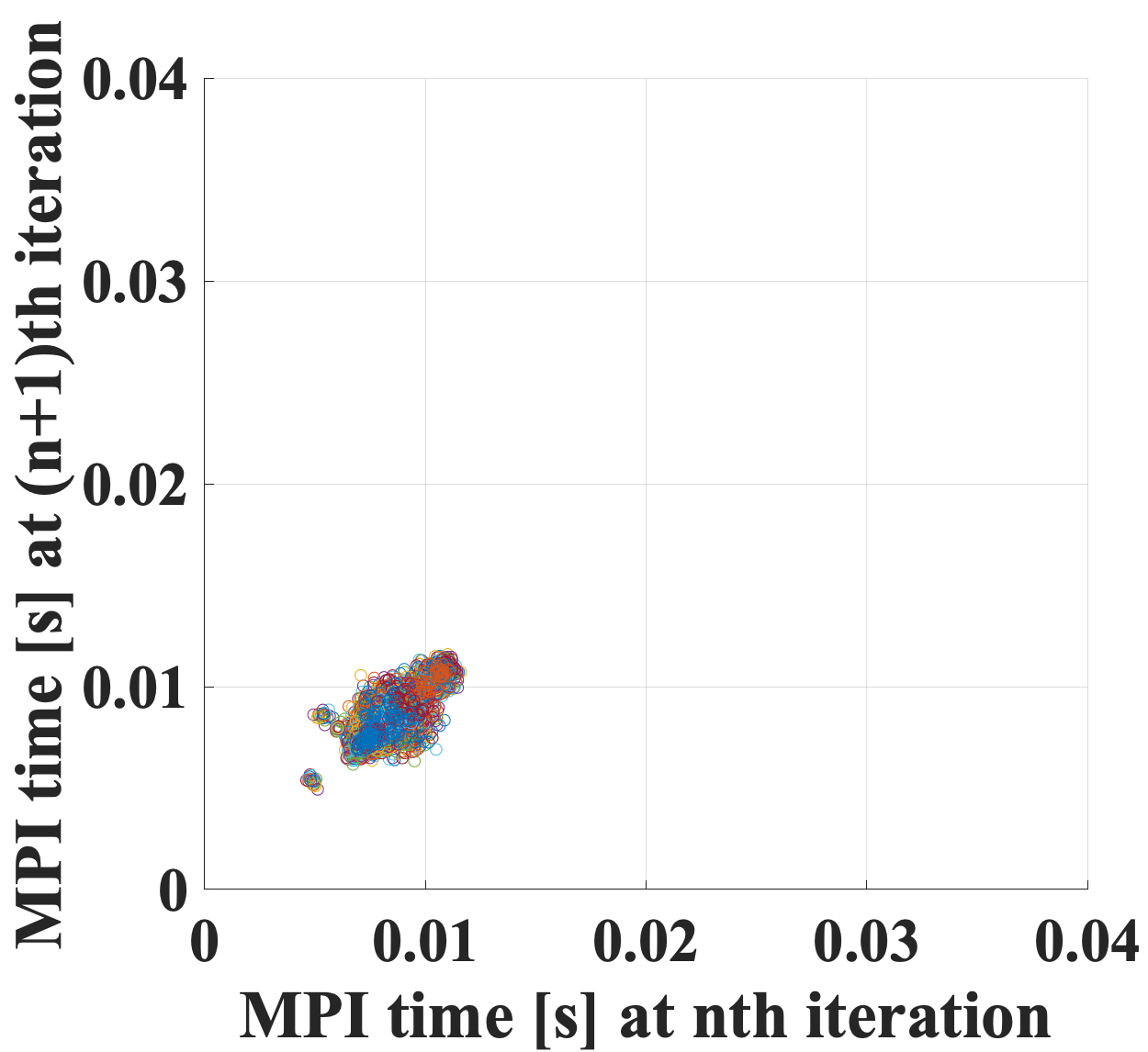}
		    \caption{Bench1B-initial}
		\end{subfigure}
		\begin{subfigure}[t]{0.24\textwidth} 
			\includegraphics[scale=0.11]{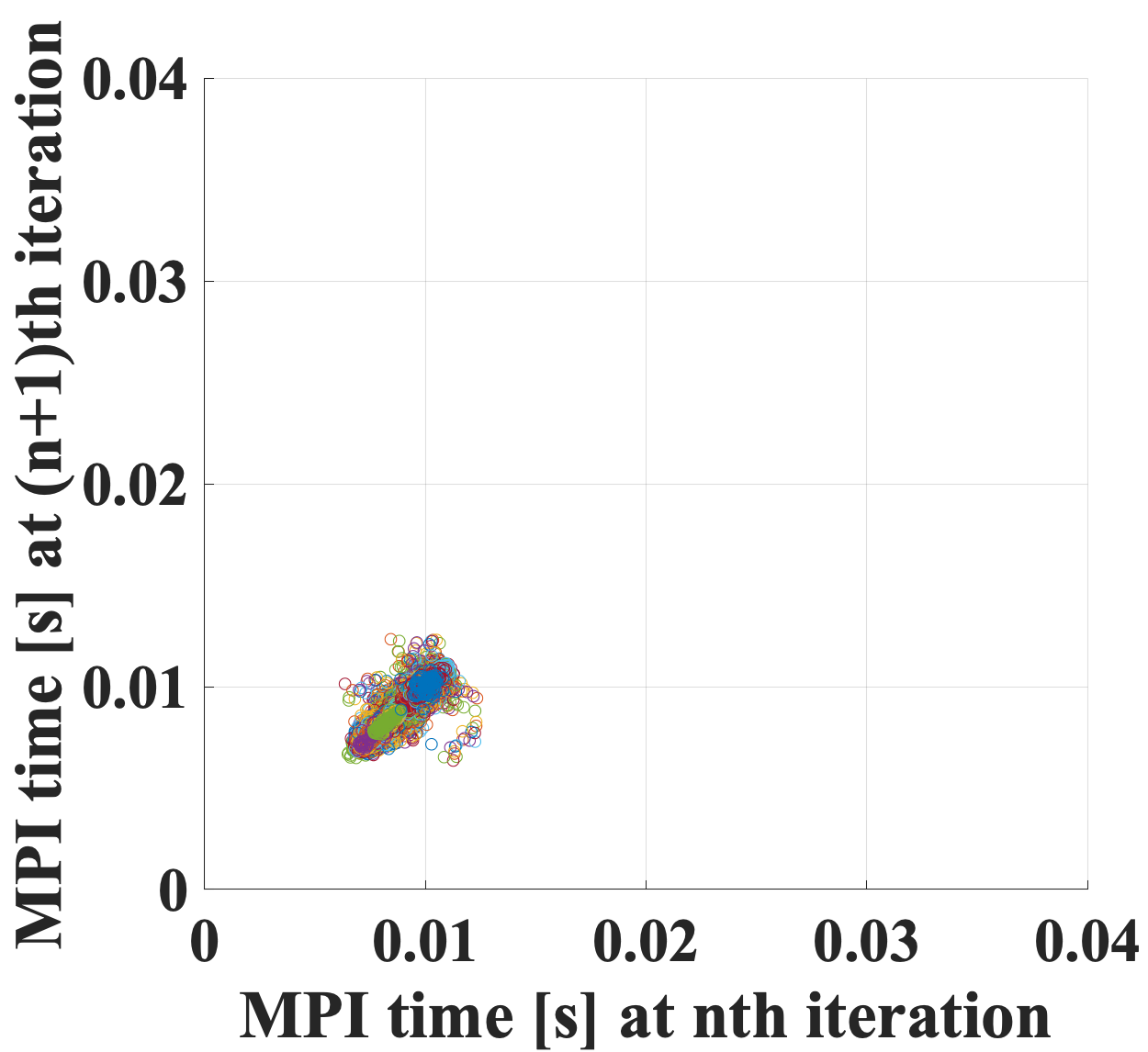}
		    \caption{Bench1B-mid}
		\end{subfigure}
		\begin{subfigure}[t]{0.24\textwidth} 
			\includegraphics[scale=0.11]{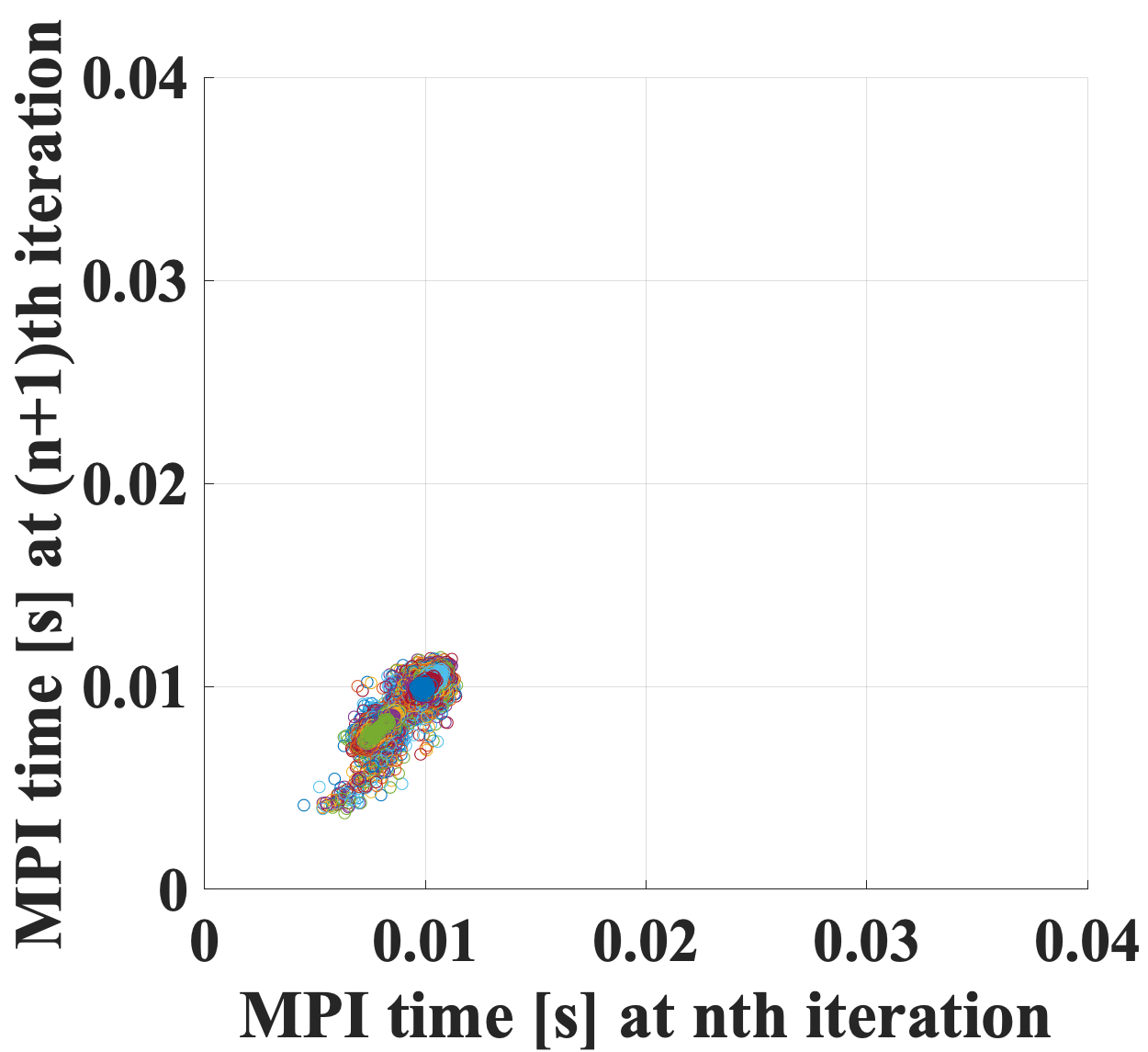}
		    \caption{Bench1B-end}
		\end{subfigure}
	\end{minipage}
	\caption{
	(a, c, f, h) \emph{Snippets view} of phase space of all MPI processes
	for \SI{100}{iterations} at the beginning.
	(b) Snapshot of \SI{1}{\kilo~iterations} (9.9-10 \si{\kilo~iterations}) for LBM in the middle state.
	(d, e, g, i, j) Snapshot of \SI{1}{\kilo~iterations}
	in the middle (1.9-2 \si{\kilo~iterations})
	and at the end evolved state (499.9-500 \si{\kilo~iterations}).}
    \label{fig:phaseSpace-snippet}
\end{figure*}
    
    \begin{figure*}[t]
	\centering
	\begin{minipage}{\textwidth}
	    \begin{subfigure}[t]{0.19\textwidth} 
			\includegraphics[scale=0.062]{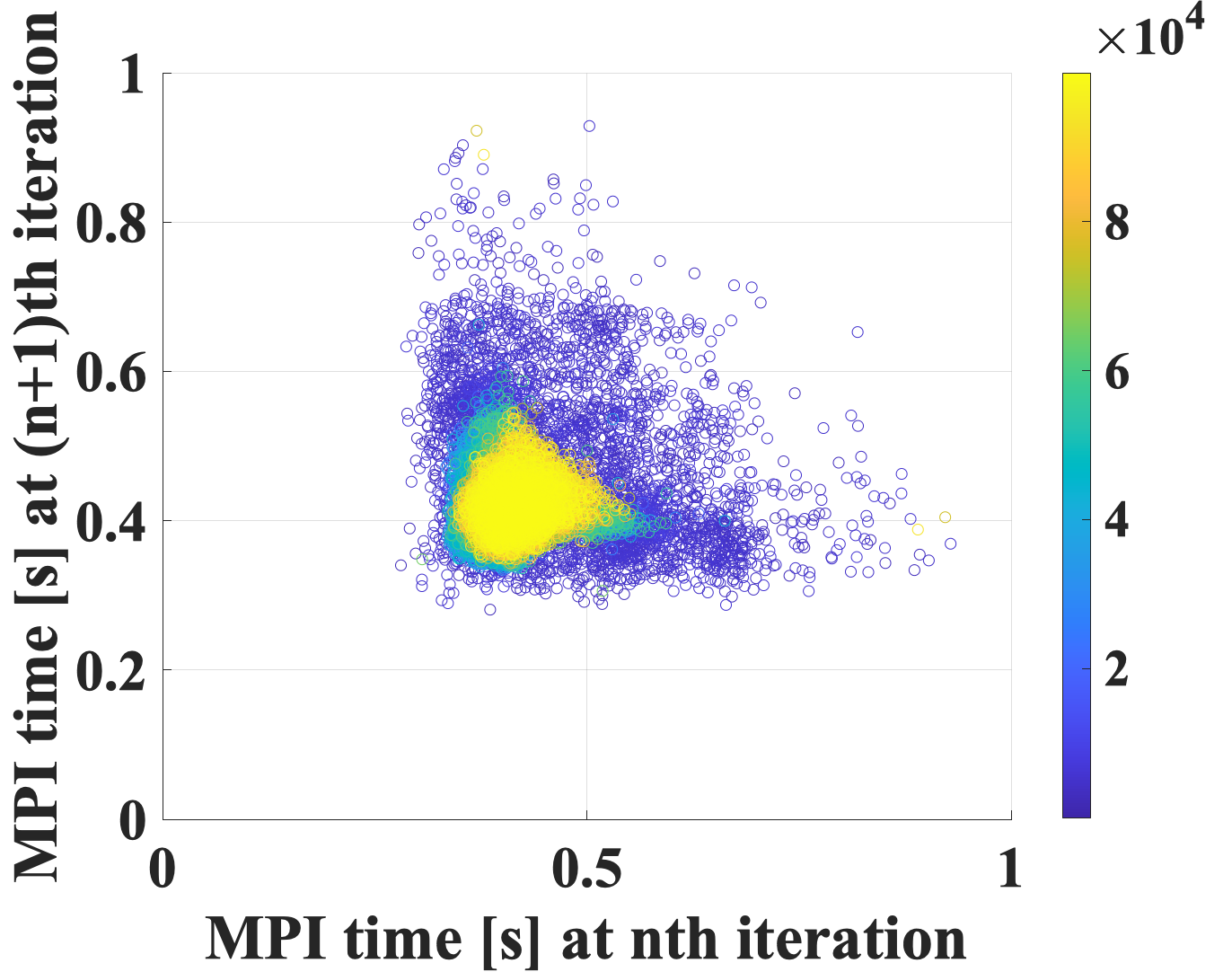}
		    \caption{LBM-100\si{\kilo~it}}
		\end{subfigure}
		\begin{subfigure}[t]{0.19\textwidth} 
			\includegraphics[scale=0.062]{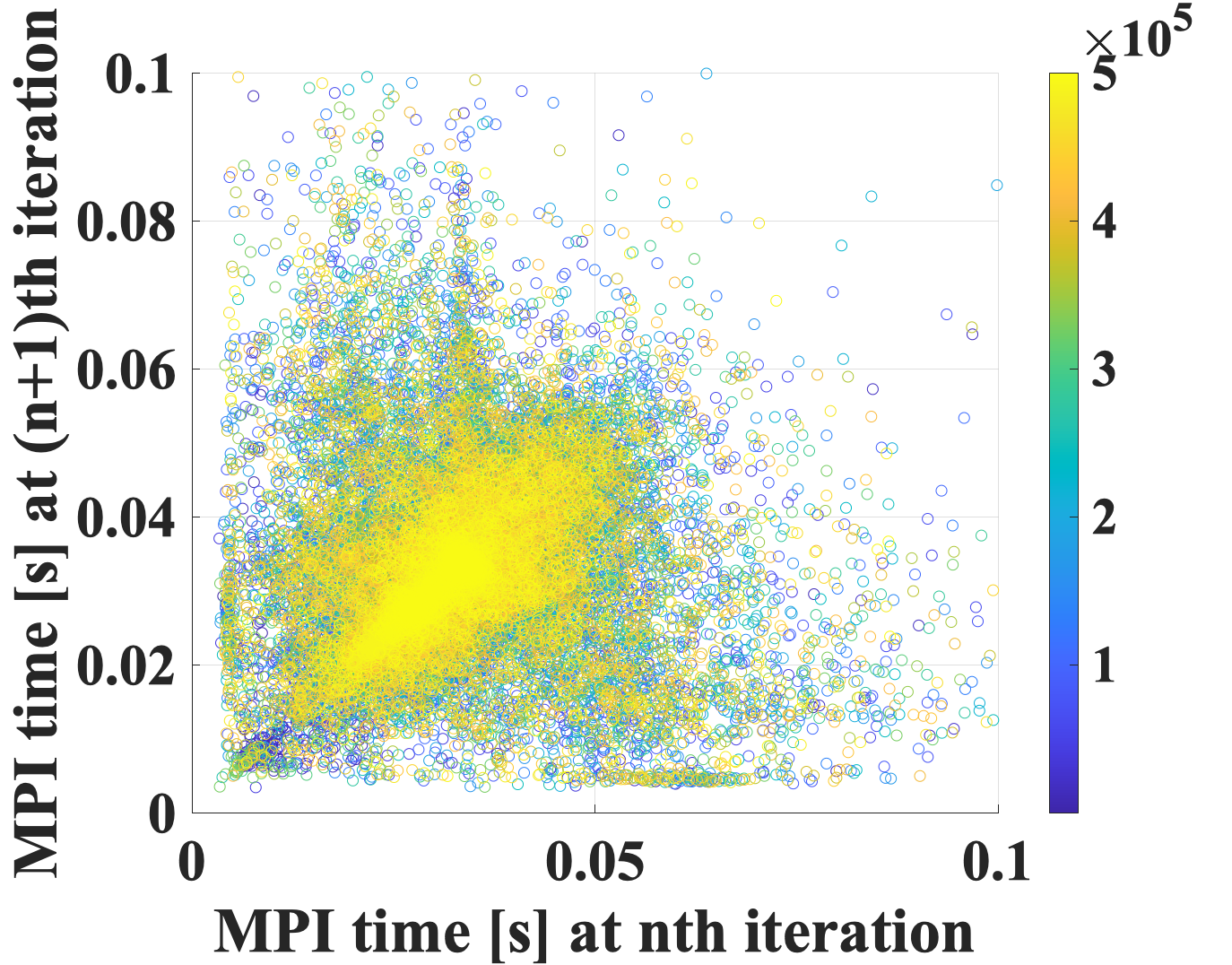}
			\caption{Bench1A-500\si{\kilo~it}}
		\end{subfigure}
		\begin{subfigure}[t]{0.19\textwidth} 
			\includegraphics[scale=0.062]{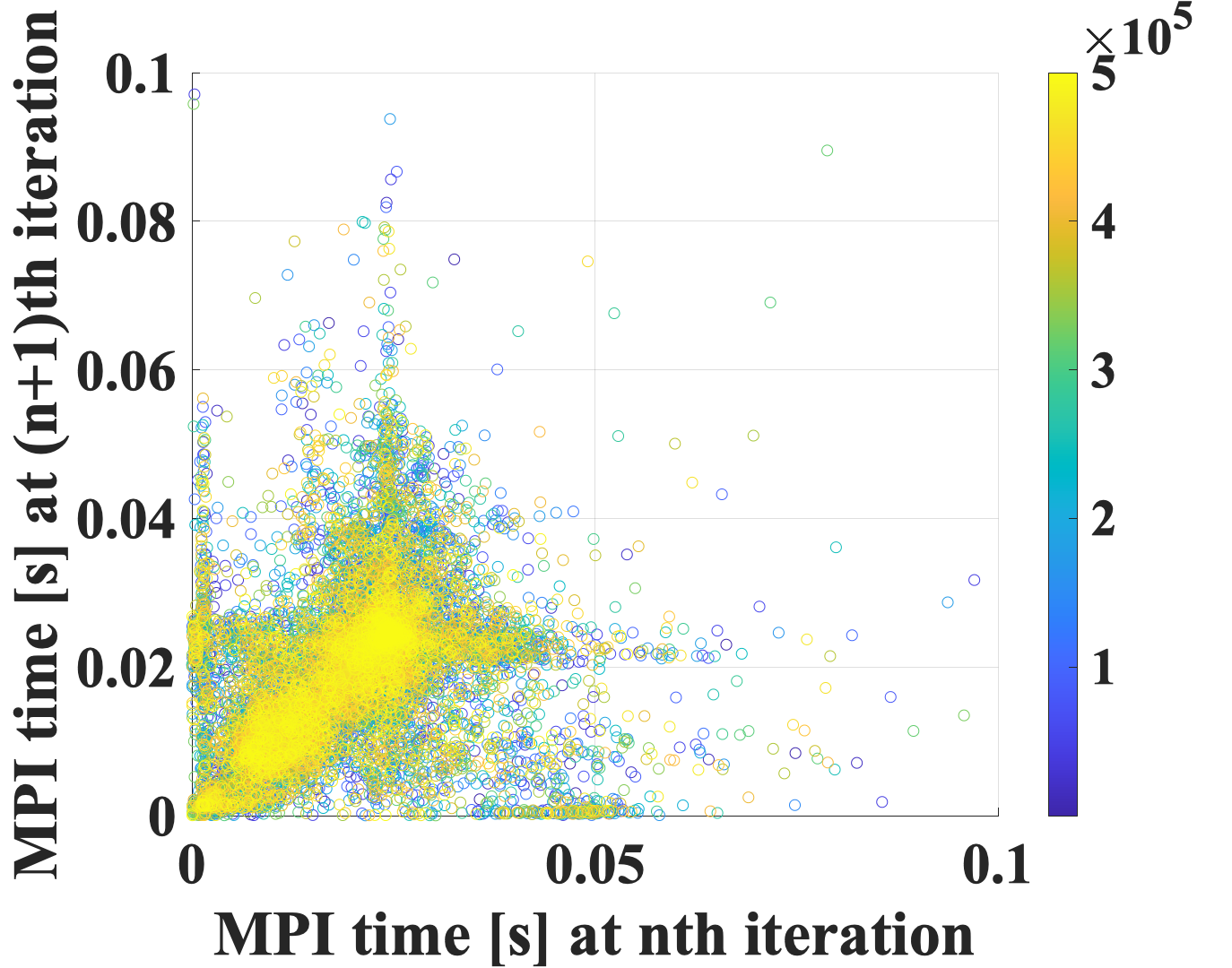}
			\caption{Bench2A-500\si{\kilo~it}}
		\end{subfigure}
		\begin{subfigure}[t]{0.19\textwidth} 
			\includegraphics[scale=0.062]{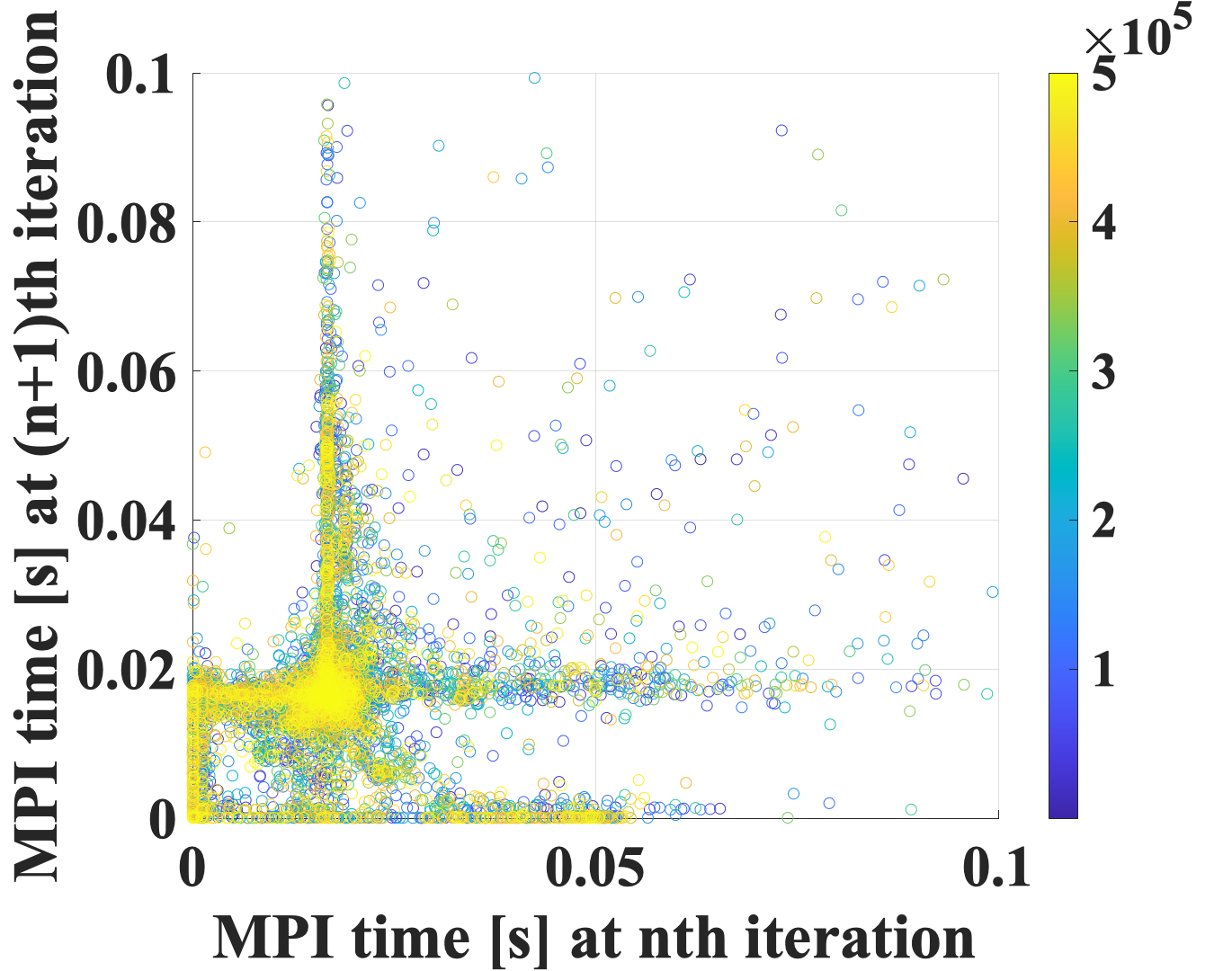}
		    \caption{Bench3A-500\si{\kilo~it}}
		\end{subfigure}
		\begin{subfigure}[t]{0.19\textwidth} 
			\includegraphics[scale=0.062]{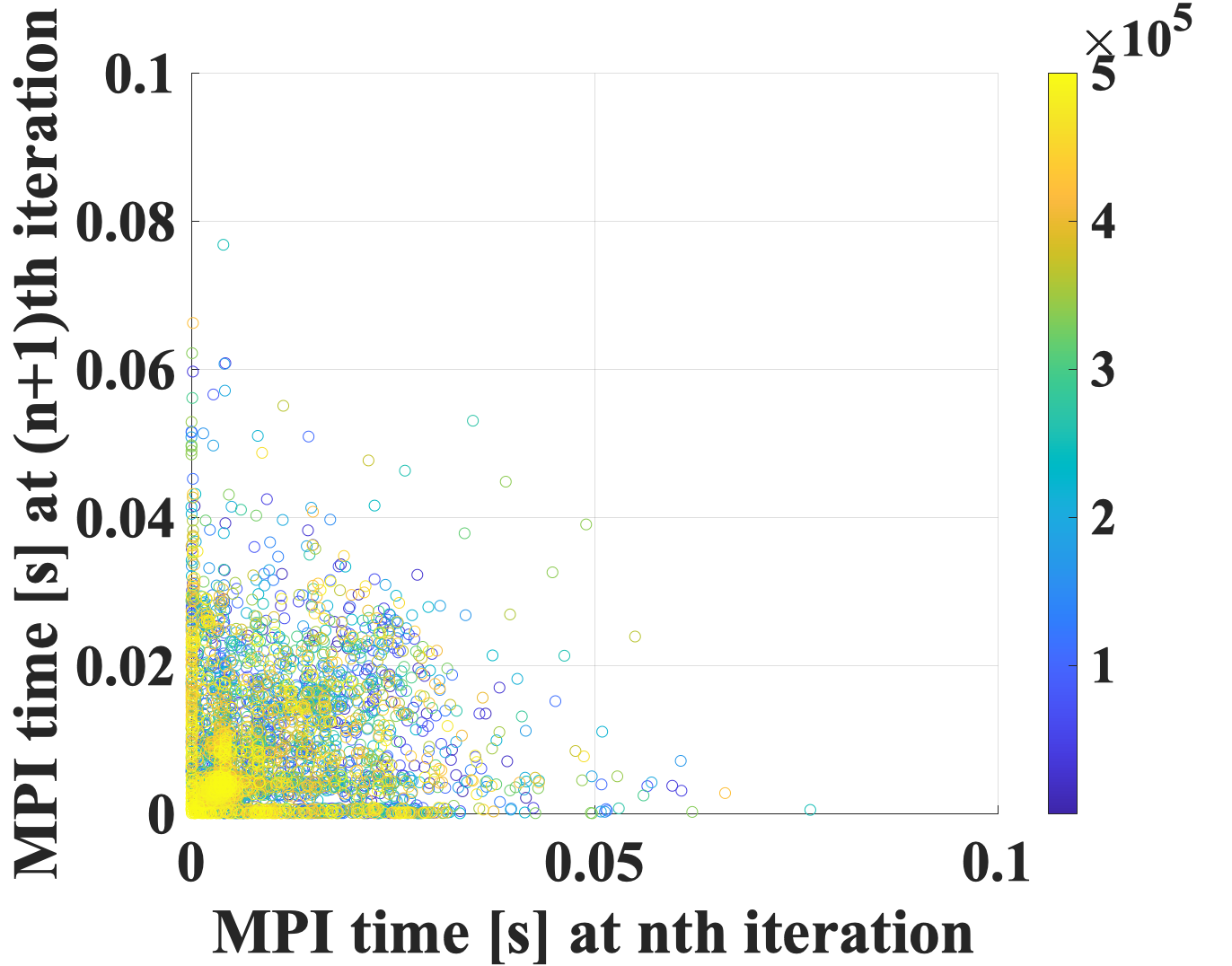}
		    \caption{Bench4A-500\si{\kilo~it}}
		\end{subfigure}
	\end{minipage}
	\begin{minipage}{\textwidth}
	    \begin{subfigure}[t]{0.19\textwidth} 
			\includegraphics[scale=0.062]{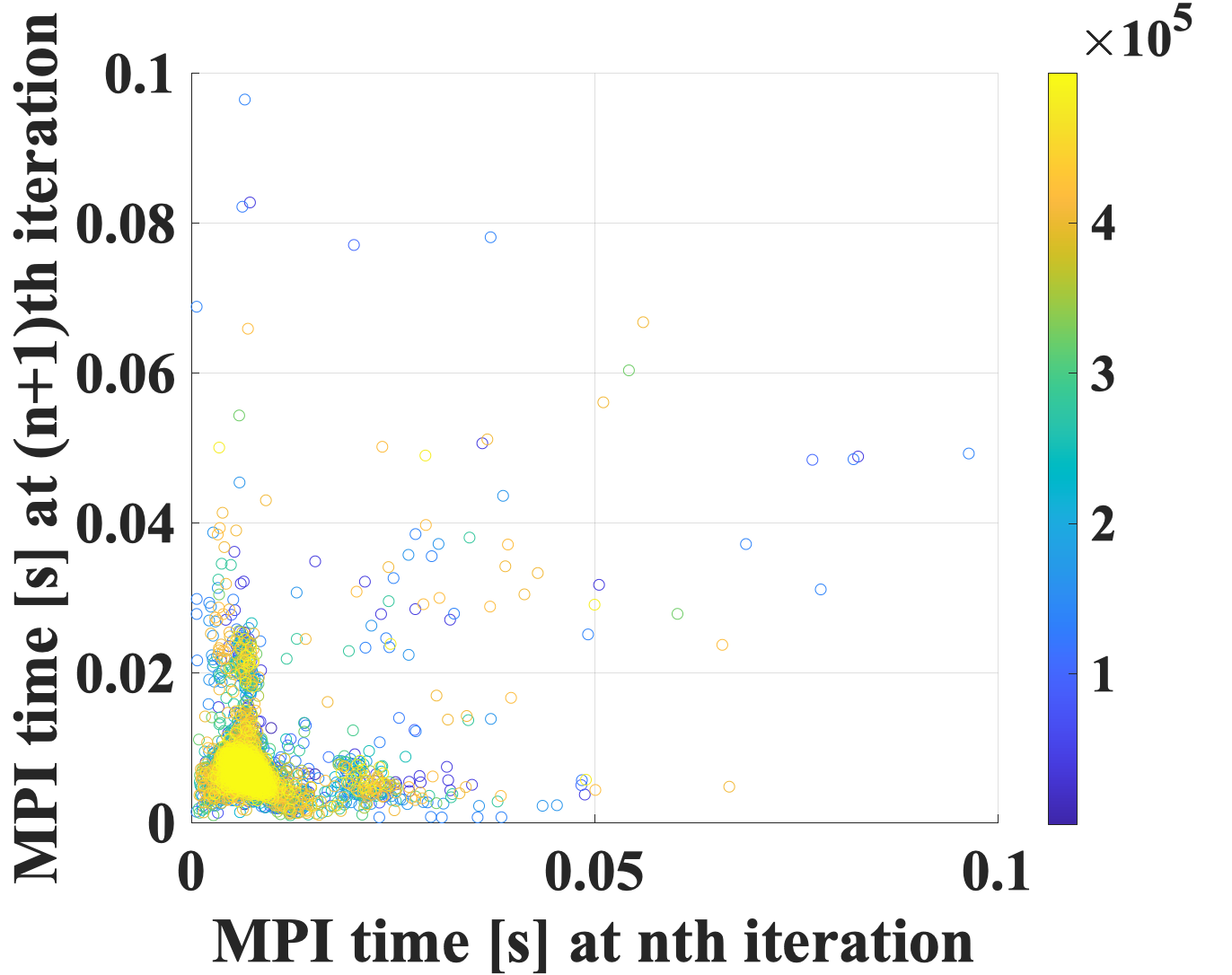}
		    \caption{spMVM-500\si{\kilo~it}}
		\end{subfigure}
		\begin{subfigure}[t]{0.19\textwidth} 
			\includegraphics[scale=0.062]{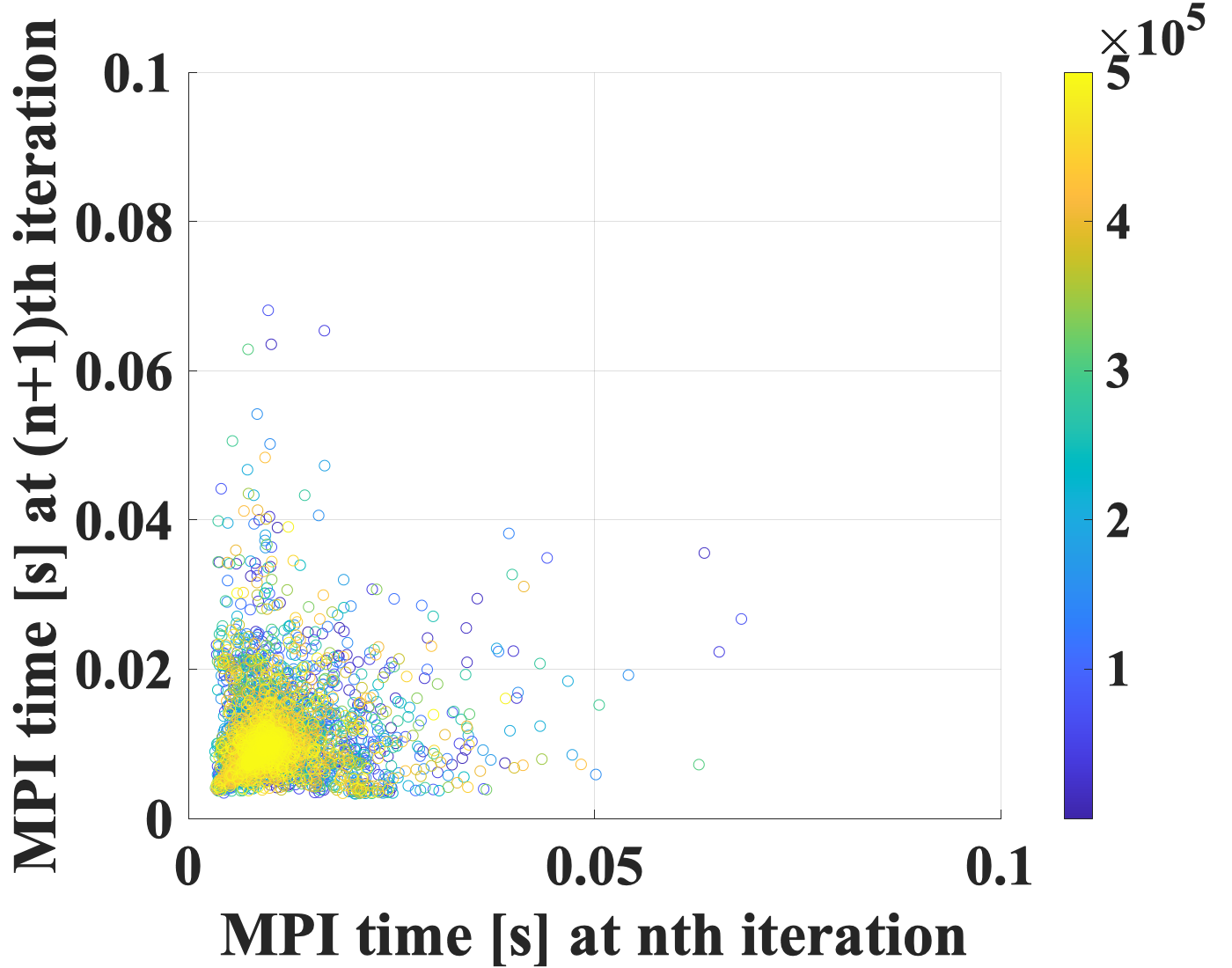}
		    \caption{Bench1B-500\si{\kilo~it}}
		\end{subfigure}
		\begin{subfigure}[t]{0.19\textwidth} 
			\includegraphics[scale=0.062]{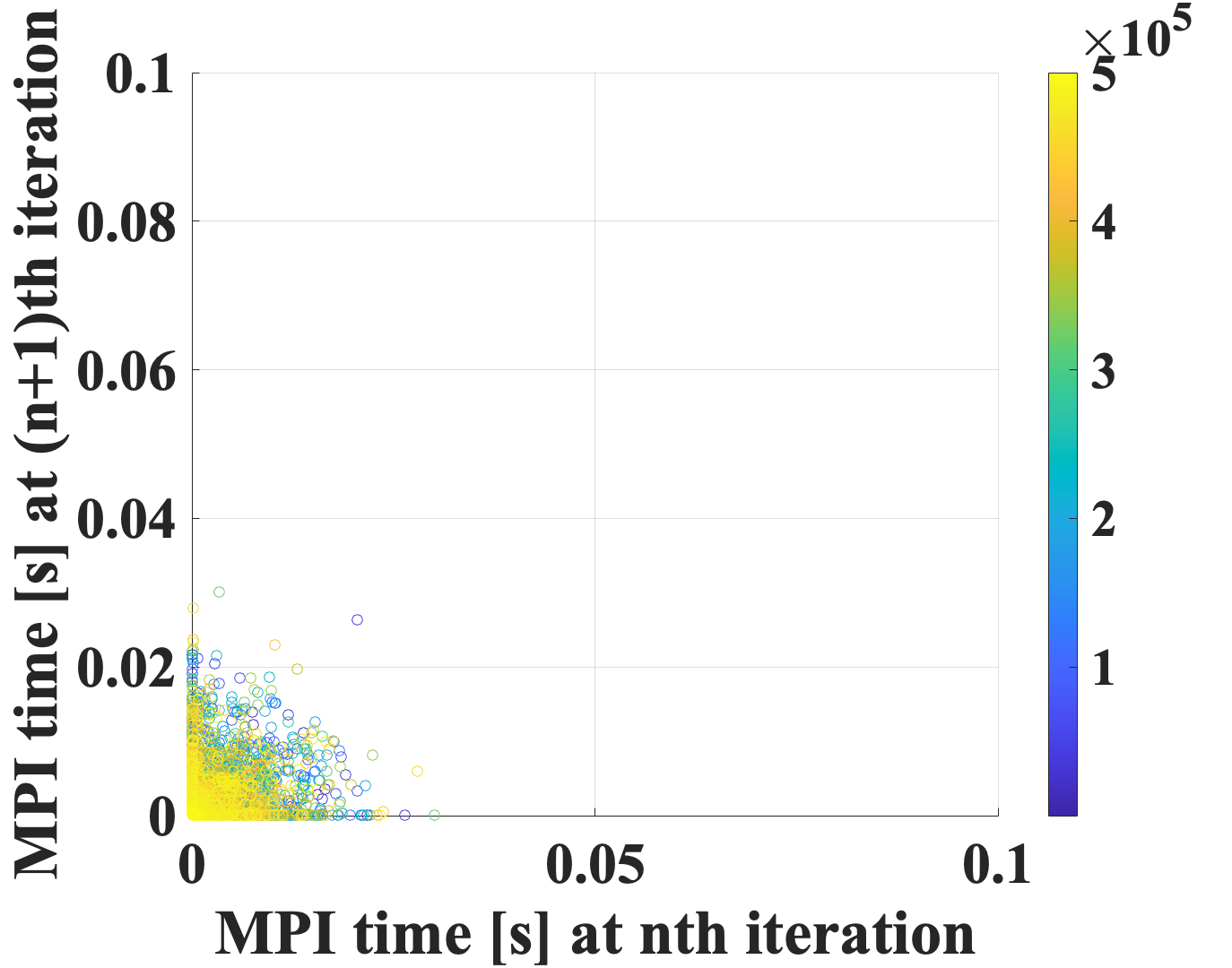}
		    \caption{Bench2B-500\si{\kilo~it}}
		\end{subfigure}
		\begin{subfigure}[t]{0.19\textwidth} 
			\includegraphics[scale=0.062]{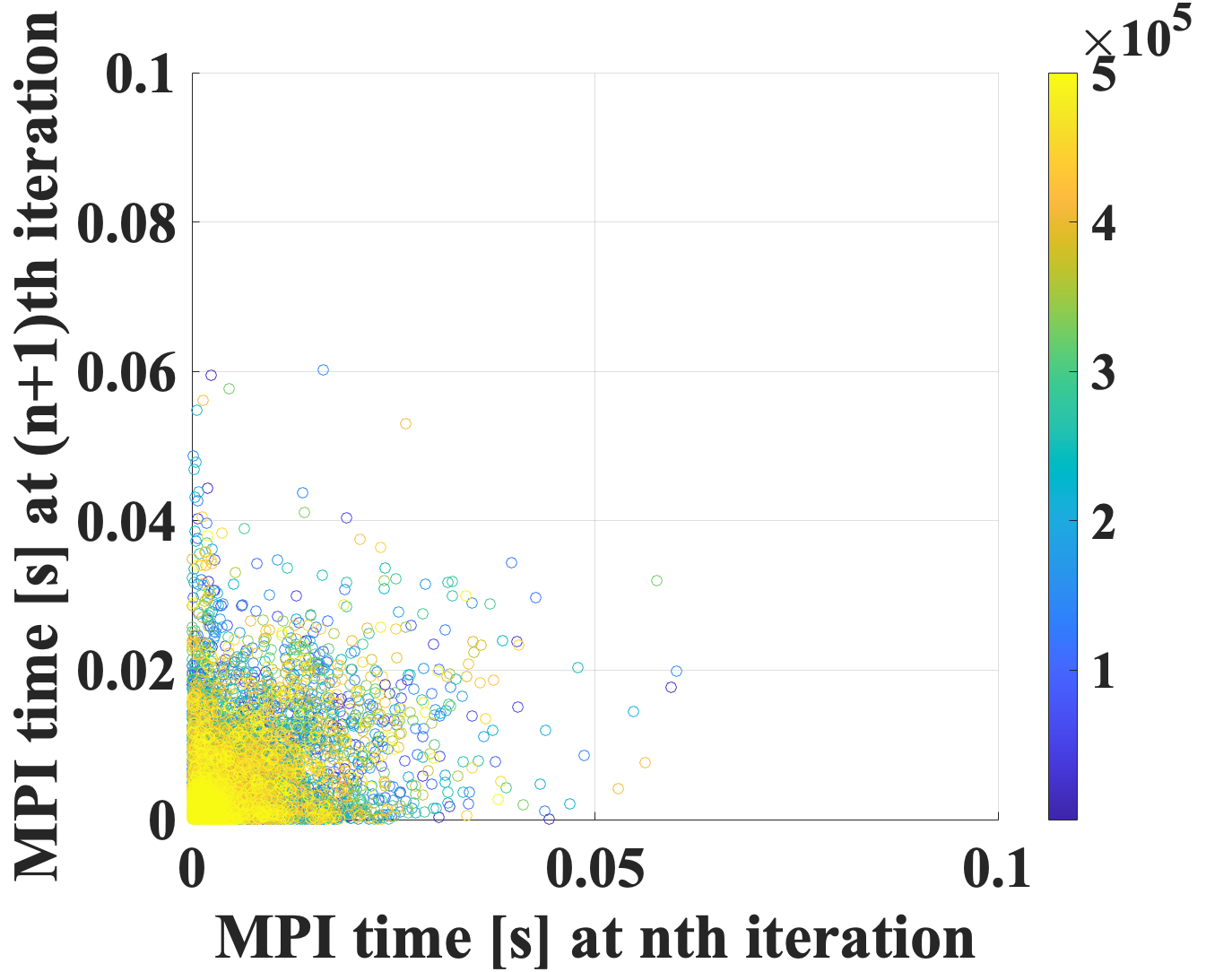}
		    \caption{Bench3B-500\si{\kilo~it}}
		\end{subfigure}
		\begin{subfigure}[t]{0.19\textwidth} 
			\includegraphics[scale=0.062]{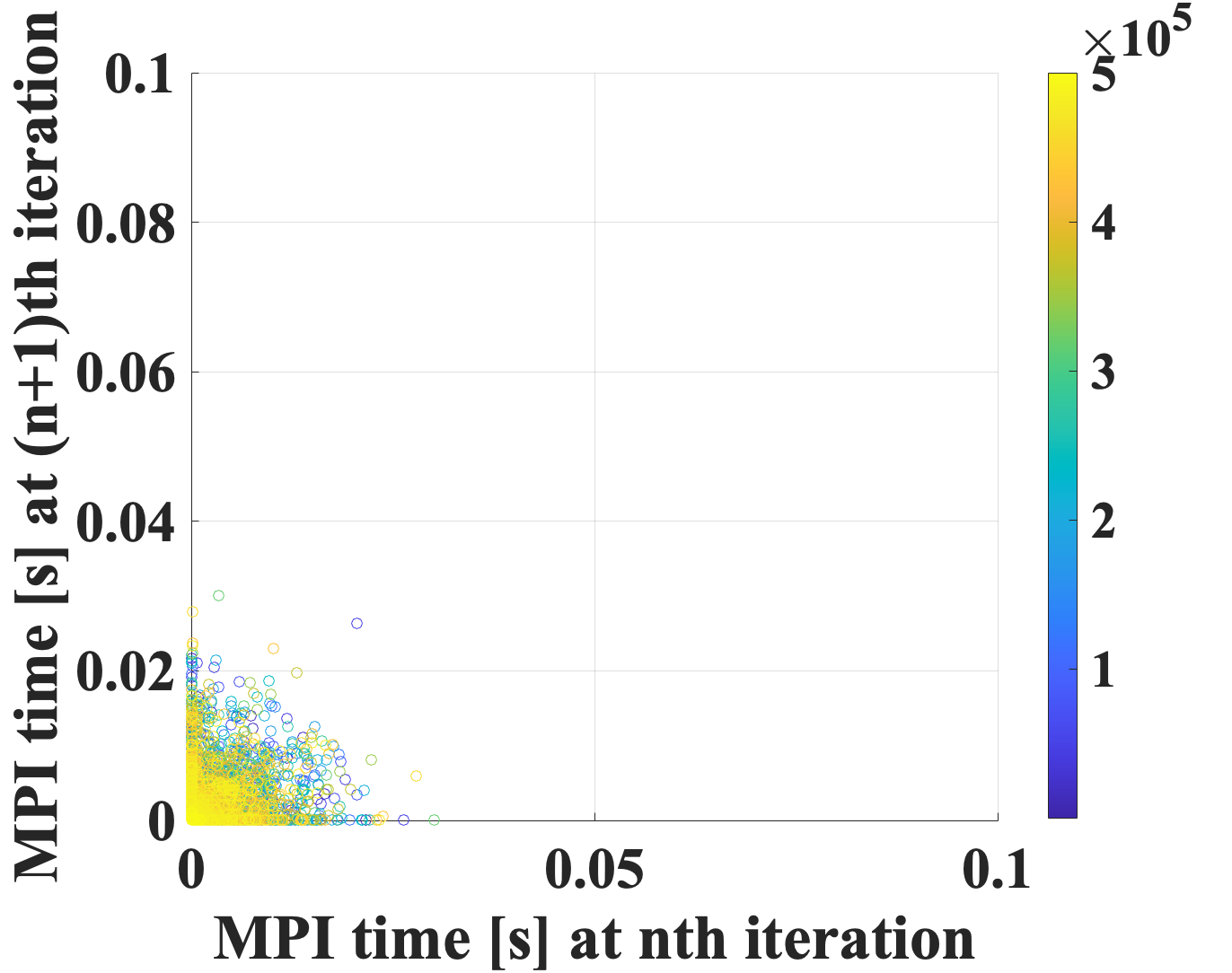}
		    \caption{Bench4B-500\si{\kilo~it}}
		\end{subfigure}
	\end{minipage}
	\caption{\emph{Overall view} of phase space for one MPI process (rank 32) of all benchmarks.
	The axes show the time spent in the MPI library at the $n$-th and $(n+1)$-th iteration, respectively.}
    \label{fig:phaseSpace}
\end{figure*}
    
    \begin{figure*}[tb]
	\centering
	\begin{minipage}{\textwidth}
% 	    \begin{subfigure}[t]{0.24\textwidth}
% 	    \vspace{-8em}
%         \begin{subfigure}[t]{0.55\textwidth} 
% 			\includegraphics[scale=0.05]{figures/PCA+Clustering/Case1b/PCA3_advance.png}
% 		\end{subfigure}
		
% 		\begin{subfigure}[t]{0.55\textwidth} 
% 		    \includegraphics[scale=0.05]{figures/PCA+Clustering/Case1b/PCA3_advance_end.png}
% 		\end{subfigure}
% 		 \caption{PCA and PCA-end}
% 		\end{subfigure}
        \begin{subfigure}[t]{0.24\textwidth} 
 			\includegraphics[scale=0.1]{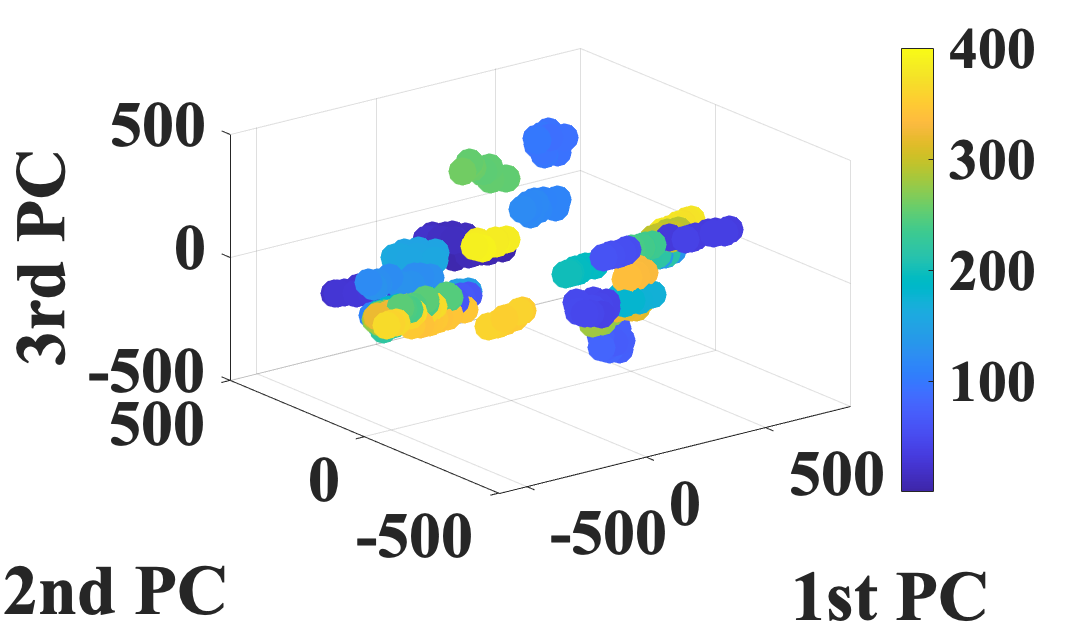}
		    \caption{PCA full run}
		\end{subfigure}
		\begin{subfigure}[t]{0.24\textwidth}
			\includegraphics[scale=0.095]{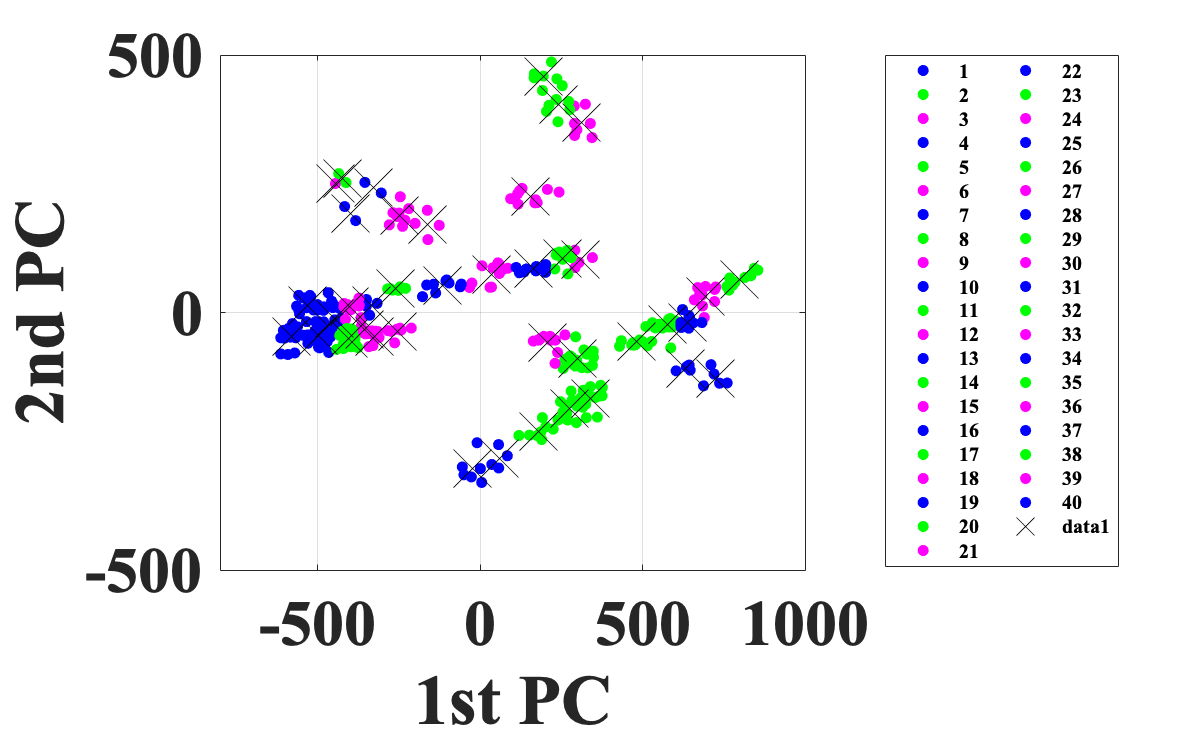}
		    \caption{K-mean full run}
		\end{subfigure}
		\begin{subfigure}[t]{0.24\textwidth} 
			\includegraphics[scale=0.04]{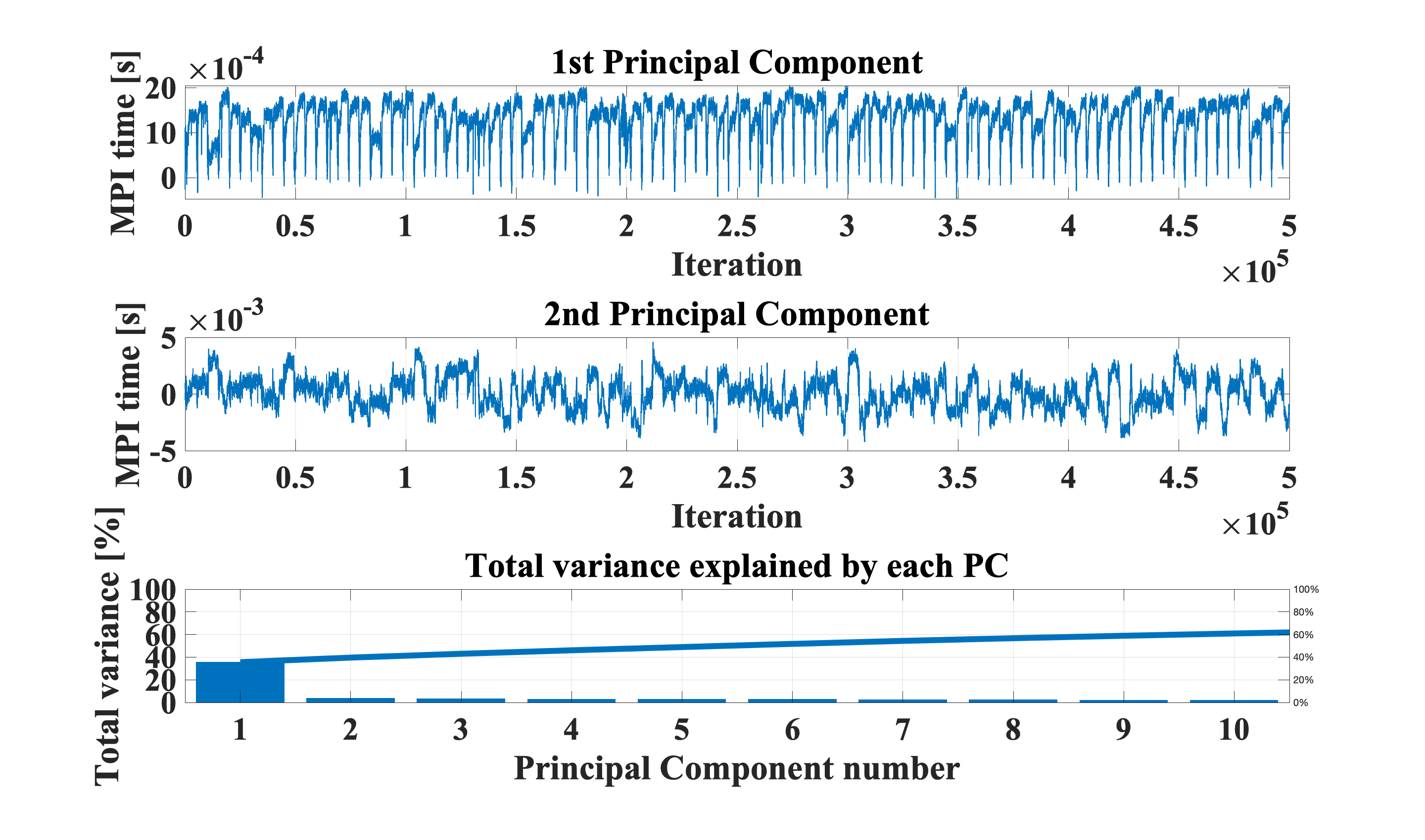}
			\caption{Eigenvectors full run}
		\end{subfigure}
 		\begin{subfigure}[t]{0.24\textwidth} 
 			\includegraphics[scale=0.04]{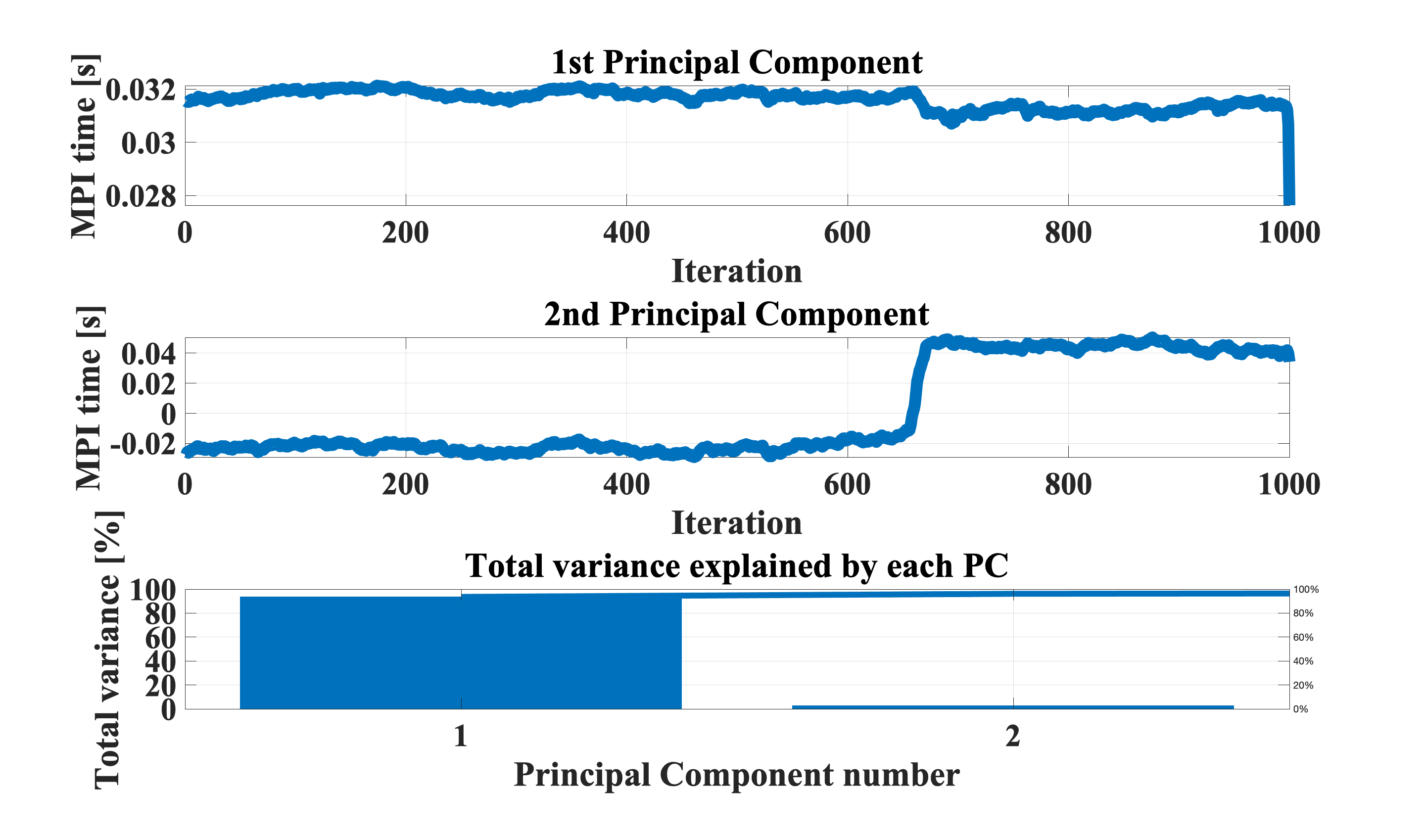}
 			\caption{Eigenvectors-end}
 		\end{subfigure}
	\end{minipage}
	\begin{minipage}{\textwidth}
	    \begin{subfigure}[t]{0.2\textwidth} 
 		    \includegraphics[scale=0.07]{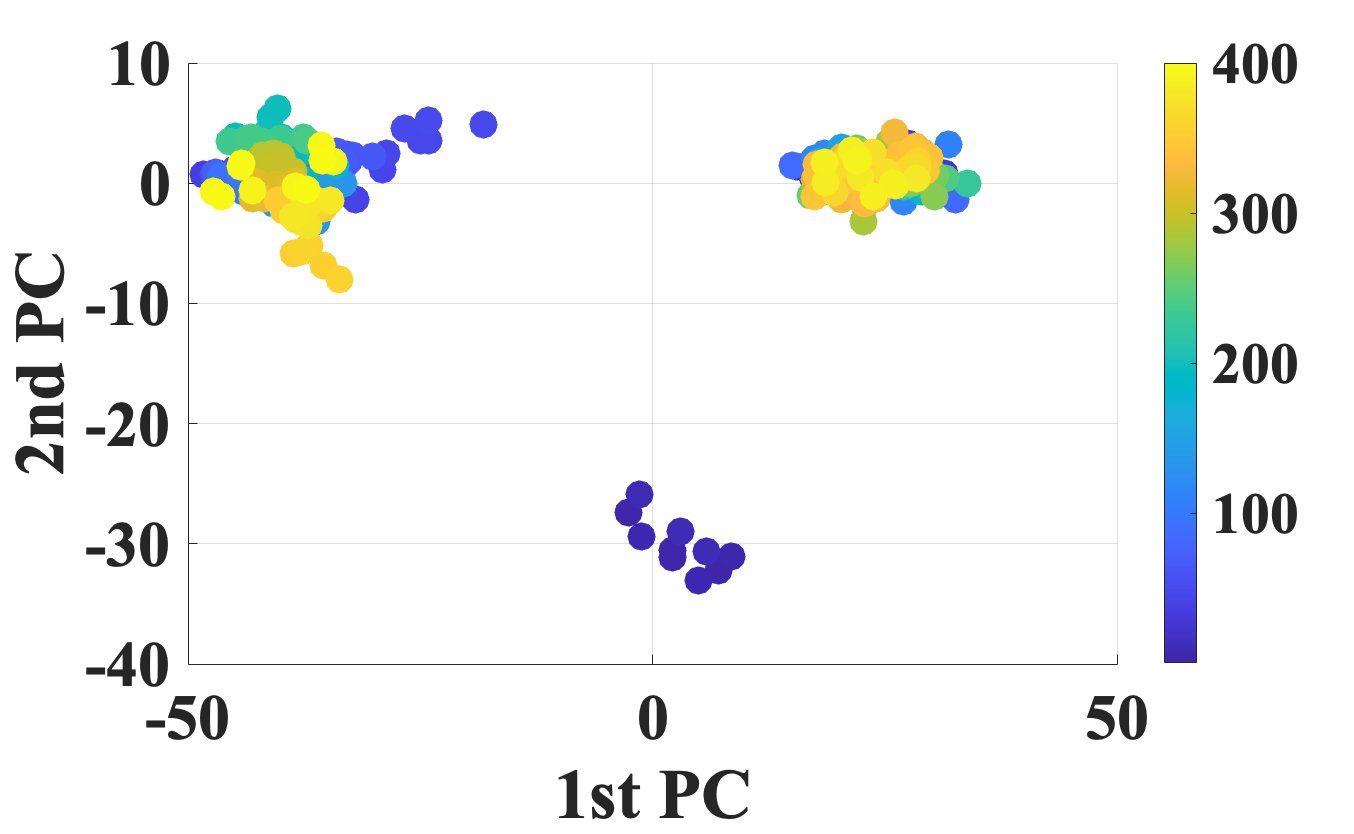}
			\caption{PCA-end}
		\end{subfigure}
		\begin{subfigure}[t]{0.17\textwidth} 
		    \includegraphics[scale=0.07]{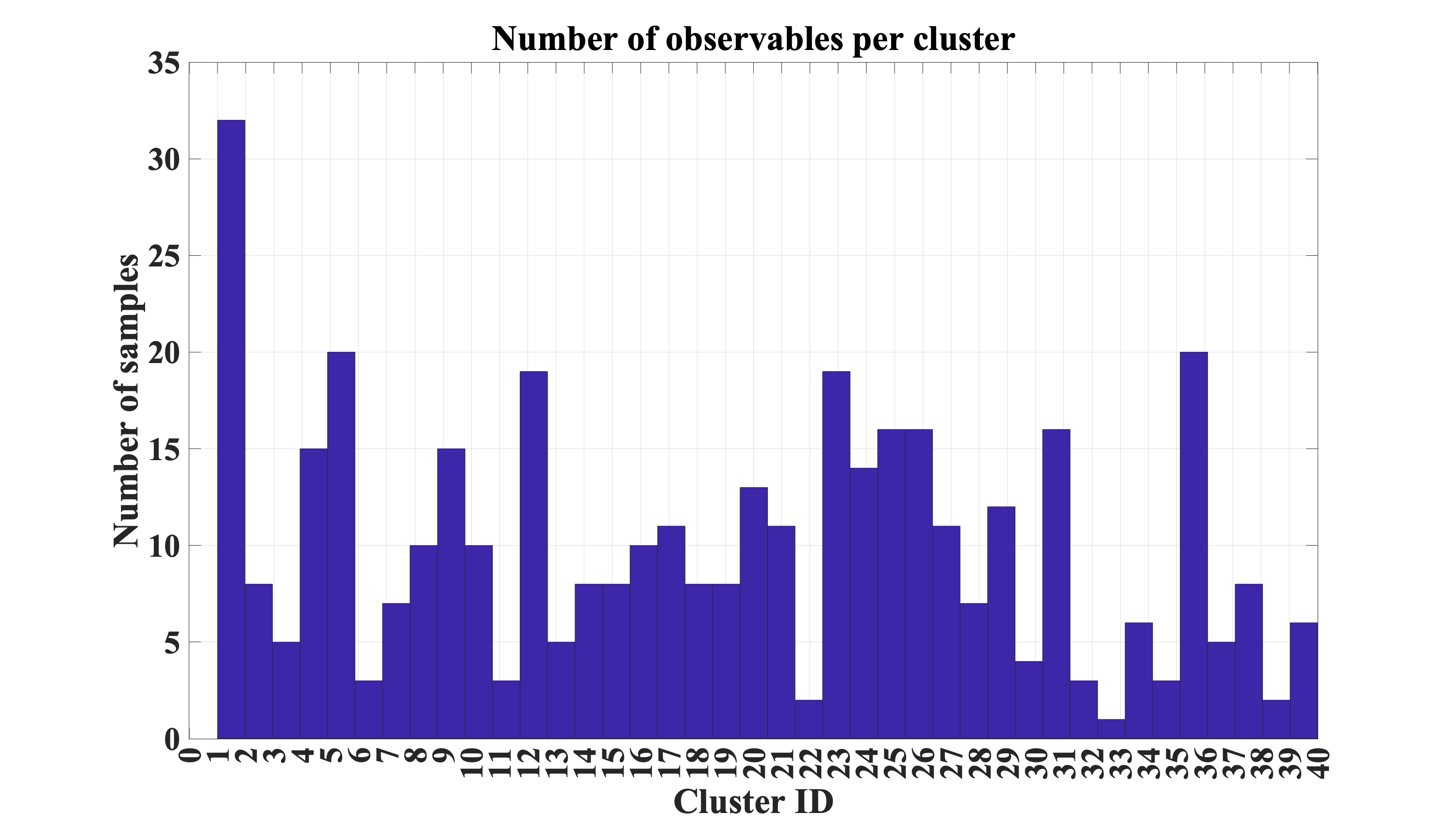}
		    \caption{Percentile}
		\end{subfigure}
		\hspace{2.2em}
		 \begin{subfigure}[t]{0.17\textwidth} 
		    \includegraphics[scale=0.12]{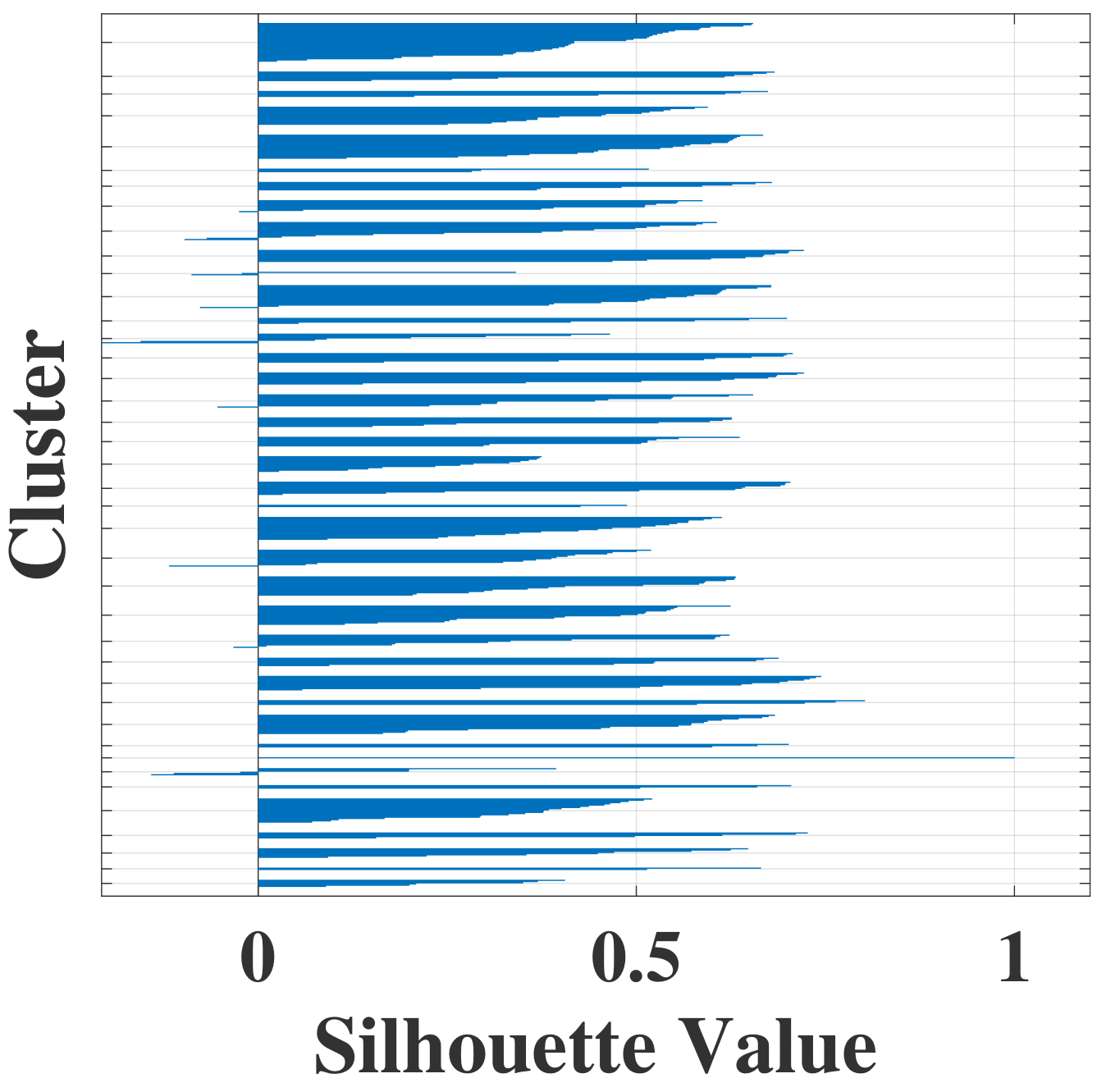}
 			\caption{Silhouette}
 		\end{subfigure}
  		\hspace{-0.5em}
		\begin{subfigure}[t]{0.17\textwidth} 
			\includegraphics[scale=0.07]{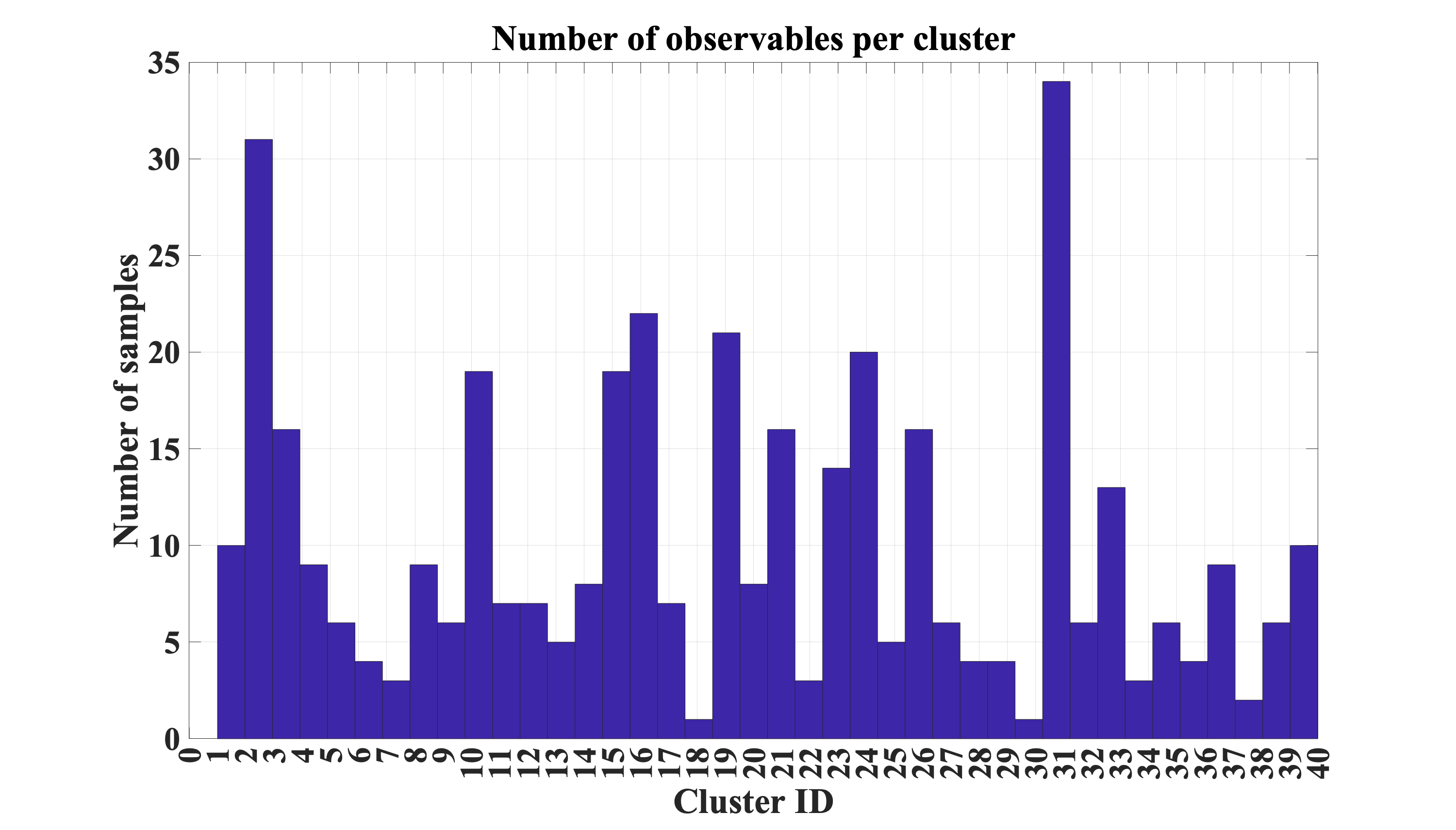}
			\caption{Percentile-end}
		\end{subfigure}
		\hspace{2.2em}
		\begin{subfigure}[t]{0.17\textwidth} 
			\includegraphics[scale=0.12]{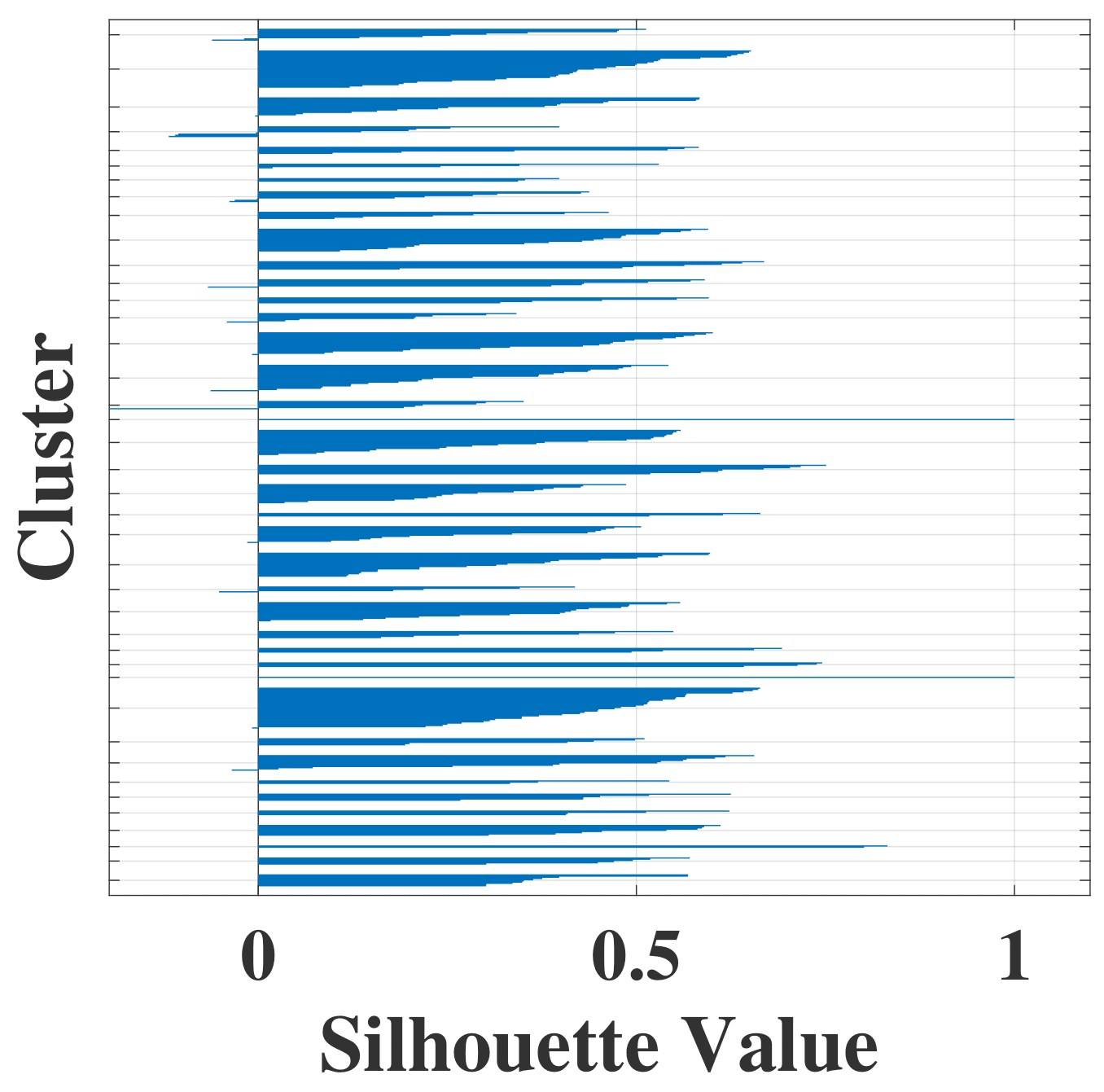}
		    \caption{Silhouette-end}
		\end{subfigure}
	\end{minipage}
	\caption{\PCA, k-mean clustering, eigenvectors, percentile and Silhouette analyses of Bench1B
	(a-c, f-g) for the whole run time of 500\si{\kilo~iterations} and
	(d-e, h-i) for snapshot of last \SI{1}{\kilo~iterations} only.
	%\GHcomm{change order so that full and snippet analysis are not interleaved}\AAcomm{analysis already in order and then the validation in order; this separation has been introduced to fit figures into two rows.}
	}
    \label{fig:pca}
\end{figure*}
    %%%%%%%%%%%%%%%%%%%
    \subsection{Phase space plots} %scatter(MPItime(1:end-1,:),MPItime(2:end,:))
    %In contrast to reducing dimensional space using PCA, 
    In order to capture the temporal evolution of MPI waiting time, we set up a scatter plot where each data point has coordinates $(\text{MPItime}(t_i,r),\text{MPItime}(t_i+1,r))$. For each process, a fixed point in this ``phase space'' is a point on the slope-1 line through the origin.  If the waiting time evolves, a process will move through the first quadrant; if waiting time increases over time (e.g., due to desynchronization), the path of a process will rise above the axis and move further up and to the right. 
    Color coding in the point cloud, from early (blue) to late (yellow), helps to visualize how processes move. 
    %At the few starting iterations of bulk-synchronous parallel programs, observables have almost the same values in each iteration and consequently all processes are part of one dot cloud. 
    We choose two different types of analysis.
    
    In the \emph{snippet view}, only a small part (e.g., 1000 iterations) of the data is visualized per plot; separate plots are used to show the long-term temporal evolution (\emph{initial}-\emph{mid}-\emph{end} in Figure~\ref{fig:phaseSpace-snippet}).
    In Figures~\ref{fig:phaseSpace-snippet}(a)--(c), after the initial in-sync phase, the cloud gets spread out.
    Asymptotically, we identify multiple weak and strong clusters (smaller and bigger attractors basin for observable).
    %where observables change in quantized intervals within inner structure.
    Stronger or weaker clustering along the diagonal line expresses how much the observable fluctuates around a ``steady-state'' value.
    In the example shown, all but one (blue points) sockets get desynchronized.
    This separation of sockets should go away as time progresses for the close chain scenario (see Figures~\ref{fig:phaseSpace-snippet}(f)--(h)), but obviously the progression is too slow to be discernible in this visualization. 
    %All MPI processes are plotted on top of each other (clustered at a tiny point) as they are all part of one cloud for all synchronized MPI processes in the start (Figure~\ref{fig:phaseSpace-snippet}(h)) and for all asynchronized sockets in the end (Figure~\ref{fig:phaseSpace-snippet}(j)).
    %The shape for LBM is more oblong at the beginning and later becomes more spherical (Figures~\ref{fig:phaseSpace-snippet}(a)--(b)), indicating fluctuations around a fixed point. 
    For PISOLVER (Figures~\ref{fig:phaseSpace-snippet}(d)--(e)), the point cloud starts around the origin and remains there since this scalable code is self-synchronizing. 
    
    In the \emph{overall view} (see Figure~\ref{fig:phaseSpace}), the full timeline is shown for one process in one plot (plotting all processes would not allow useful conclusions)).
    Here the gradual evolution of waiting time is hard to see since it is drowned in fluctuations. However, especially in the open-chain scenarios (Figures~\ref{fig:phaseSpace}(b)--(d)) we observe structures parallel to the axes, indicating singular long delays of a few preferred lengths. These are signatures of traveling idle waves, which for a single process manifest themselves as singular high waiting time in one time step. 
    % \GHcomm{We should have dome experiments with synthetic benchmarks to map the phase space patterns to stuff we already know.\AAcomm{I'm not sure if I understand your point.}}\GHcomm{We are fishing in the dark with this data. Why not run an experiment with a strong idle wave and then see how the phase diagram looks for that, then run a full developed computational wave etc. etc. so we learn how to identify these patterns. Most of the data in this paper is poorly understood, including the analysis methods!} 
    
    %we get a near-linear line in phase for memory-bound applications as MPI waiting times increase over time for the progression from sync to desync socket, i.e., leaving the initial point (origin or not) and progress to the upper right.
    %Larger MPI times are collected for open chain cases including a single synchronized socket (Figure~\ref{fig:phaseSpace}(b, c)).   
    %More irregular MPI times that spread out to get away from the main diagonal implies that in one time step it has small waiting times, whereas in the next time step it has large waiting times (idle periods) and vice versa.
    %We observe an echo, i.e., hits more and more on these axes, which is stronger in \emph{Bench3A} or spMVM and implies that one of the delays in this sequence is preferred. 
    %For high communication data volume, e.g., \emph{Bench1}, every process will be along the diagonal already from the start. 

%%%%%%%%%%%%%%%%%%%%%%%%%%%%%%%%%%%%%%%%%%%%%%%%%%%%%%%%%%%%%%%%%%%%%  
\section{Machine learning techniques for analysis} \label{sec:MLTechniques}
    In order to prepare the timeline data for machine learning techniques, we subtract the mean values of MPI times across all time steps and processes of each experiment from the value at each step. This is one of many possible options for data normalization; better ones might exist.
    We then apply PCA~\cite{jolliffe2016principal} to the timelines of each run, using the MPI times of each process as feature vectors, and then classify the projections of the feature vectors on the first two principal component vectors using clustering techniques. 
    Finally, we validate the quality of the clustering for an accurate evaluation. To do that, we look at the reconstruction error that is generated using an essential number of \PC{}s only. 
    
    \subsection{Principal Component Analysis (PCA)} % [wcoeff,score,latent,~,explained,mu]=pca(zscore(((MPItime-mean(MPItime)))));
    \PCA projects the directions of high-dimensional data onto a lower-di\-men\-sional subspace while retaining most of the information.
    Ideally, the low-dimensional manifolds still retain almost all variance of the data set needed to identify and interpret generic behavior.
    Coarse features are captured by the first principal components while the highest-frequency features are captured by highest principal components.
    PCA centers the data and uses the \emph{singular value decomposition} algorithm on the non-square observable matrix.
    Rows and columns of the input matrix correspond to observations and variables, respectively.
    Each row vector is the timeline of MPI times in a process; the observable values in different iterations are the coordinates in a high-dimensional space. 
    %For instance, the space is $N_p$ dimensional for $N_p$ MPI processes in the timeline. 
    % The axes on high-dimensional space are iterations in the MPI process and accumulated observable value in multiple vectors at that point.

    \paragraph{Projection plot on the reduced Principal Components}
    % c = linspace(1,Ranks,length(score(:,1)));
    % scatter(score(:,1),score(:,2),[],c)
    % scatter3(score(:,1),score(:,2),score(:,3),[],c)
    The points in Figure~\ref{fig:pca}(a, e) indicate the score of each observation for the first three principal components in the Bench1B experiment.
    They show the PCA analysis on the full run and on the last 1000 iterations, respectively.
    %Three dimensional plot provides a better visualization of the first three \PC's, while the two dimensional plot will be still visible from the top of the 3-D plot.
    For the compute-bound PISOLVER (Bench4), all processes cluster around one point because of absence of contention on the sockets (data not shown).
    In contrast, for the memory-bound Triad variants, four or 40 clusters emerge at the start due to the presence of four or 40 ccNUMA domains, respectively. 
    As time progresses, all desynchronized sockets form weak clusters by collecting larger scores for PC1 and nonzero scores for PC2, while the in-sync domain forms a compact cluster due to lower scores for PC1 and zero scores for PC2.  
    Desynchronization is strongest for the processes on the top right of the plot.
    The negative values for projections on eigenvectors indicate an inverse relationship, but large (either positive or negative) values indicate that a vector has a strong overlap with that principal component.
    If all ccNUMA domains are eventually desynchronized, all processes cluster on the top right as shown in Figure~\ref{fig:pca}(e) for the last 1000 iterations.
    %All sockets in different places for the whole MPI timeline in (Figure~\ref{fig:pca}(a)) converge to one cluster in the vicinity of zero for the PCA of last \SI{1}{\kilo~iterations} (Figure~\ref{fig:pca}(e)).

    \paragraph{Principal Components (eigenvectors)} % plot(wcoeff(:,1))
    In order to get better insight into the governing characteristics of desynchronized execution, We analyze the essential eigenvectors.
    %The first PC captures the maximum characteristics behavior, i.e., high observable values; while the second PC captures second largest variance, non-correlating with the first PC, i.e., low observable values.
    Figures~\ref{fig:pca}(c, d) show the eigenvectors and how ranks contribute to the reduced number of principal components for the full run and the last 1000 iterations.
    %We analyze how the eigenfunctions differ via correlating with the actual behavior.
    In the full-run case (Fig.~\ref{fig:pca}(c)), the PC1 eigenvector characterizes desynchronizing processes
    %\GHcomm{You keen writing "desync socket" but it's actually processes within a ccNUMA domain that are desynchronized. You could use the term, but it's confusing to the reader.}\AAcomm{True.}
    and thus indicates a lot of waiting times with in-between downward spikes. The PC2 eigenvector characterizes in-sync processes and shows almost no waiting time, but upward spikes (idle periods) in between. It must be noted that the PCs for end-to-end runs encompass the entire evolution of the program, including initial in-sync phases, transient states, and final, stable states. Looking at the final 1000 iterations (Fig.~\ref{fig:pca}dc)), the signatures are much clearer; PC1 characterizes stable desynchronization while PC2 maps transient behavior where a noise-induced event between iteration 600 and 700 causes processes to change from a state with small waiting times to a state with large waiting times. One can interpret this as processes within a ccNUMA domain ``snapping'' out of sync.
    
    \paragraph{Total variance explained by each principal component}
    % pareto(explained)
    The percentage of the total variance explained by each principal component (Pareto plots in Figures~\ref{fig:pca}(c-bottom, d-bottom)) indicates for Bench1B how many PCs are required to reconstruct the original data sets using only the projections. In this particular case, one component is sufficient near the end of the run but many are required (with the first one still dominating) over the full run.
    Overall, the results show that more PCs are needed to explain the data variance for a more pronounced memory-bandwidth saturation on the ccNUMA domain. 
    %For instance, for memory-bound Triad \emph{BenchA} variants, two \PC's of the first two cases (\emph{Bench1A} and \emph{Bench2A}) and one \PC of the third case (\emph{Bench3A}) are enough to capture at least 98 \% variance.
    %It reduces the problem dependency from $40$ to two or one.
    In contrast, the compute-bound PISOLVER has much less variance as no typical structure exists except natural, noise-induced fluctuations. 
    Further, the more revealing socket behavior in short-runs is captured by the higher number of PCs compared to the asymptotic behavior in long runs.
    %\GHcomm{I don't understand: long runs do not capture asymptotic behavior well, because all the fluctuations along the way are part of the eigenvectors.\AAcomm{True} Also, if PISOLVER has less variance why are more PCs required?\AAcomm{Not true, there is not enough variance to be captured, and thereby \PCA doesn't make any sense.}}
    %% GHa I still don't get it but let's drop it for the time being
    
    \subsection{K-means clustering} %[idx,C]=kmeans(score(:,1:NPC),NSockets,'Distance','sqeuclidean')
    %gscatter(score(:,1),score(:,2),idx,'bgm')
    While PCA delivers insight into typical patterns, it does not allow for automatic grouping (clustering) of processes. This can be accomplished by partitioning the projection of observations on the principal components into $k$ clusters by using k-means. Rows of PC scores correspond to points and columns correspond to variables. We use the k-means++ algorithm~\cite{vassilvitskii2006k} for the cluster center initialization; it is strongly dependent on the distance metric used. % to partition the training data into different clusters.
    
    \paragraph{Distance types}
    Clustering quality was studied for four metrics.
    In the cluster, each centroid $c$ is either the
    mean ($(x-c)(x-c)'$) or component-wise median (($\sum^{p}_{j=1}|x_j-c_j|$) of the points in \textit{squared Euclidean} and \textit{city-block} metrics, respectively.
    Here, $x$ is a row of PC scores. 
    For the \textit{cosine} and \textit{correlation} metrics, each centroid $c$ is either the mean of the points which are already normalized to unit Euclidean length
    ($1-\dfrac{xc'}{\sqrt{(xx')(cc')}}$)
    or component-wise mean of the points which are already centered and normalized to zero mean and unit standard deviation
    ($1-\dfrac{(x-{\bar{x}})(c-{\bar{c}})'}{\sqrt{(x-{\bar{x}})(x-{\bar{x}})'}\sqrt{(c-{\bar{c}})(c-{\bar{c}})'}}$), with ${\bar{x}}=\dfrac{1}{p}(\sum^{p}_{j=1}x_j){1_p}$ and ${\bar{c}}=\dfrac{1}{p}(\sum^{p}_{j=1}c_j){1_p}$.

    The result is a matrix containing the $k$ cluster centroid locations and a vector containing cluster indices.
    Figure~\ref{fig:pca}(b) shows a scatter plot of essential PC scores grouped by the cluster indices of each observation in the \emph{Bench1B} case. K-means uses the squared Euclidean distance here.
    We expect one cluster if all processes are in a fully evolved desynchronized state.
    
    \begin{comment}
    %Excluded following analysis from this paper
    \paragraph{Classify cases 2-4 using cluster of \emph{Bench1A}} 
    %[wcoeff1,score1,latent1,~,explained1] = pca(zscore(((Idle2' -mean(Idle2')))));
    %[~,idx1] = pdist2(C,score1(:,1:NPC),'seuclidean','Smallest',1); 
    %gscatter(score1(:,1),score1(:,2),idx1','bgm','ooo')
    Instead of multiple eigenfunctions, we automatically use the PCA of first case-study as the basis for all other test cases and calculate the error.
    To create new clusters including cluster of \emph{Bench1}, we find the nearest centroid from the new case by using the one smallest pairwise distance between two cases.
    That is, \emph{the distance to the k cluster centroid locations of \emph{Bench1} for each reduced PC's scores of new case i (where i can be any number between 2-4)}.
    Four different distance metrics are used for minimization, as stated above.
    \begin{itemize}%[nosep]
        \item Figure~? \SMcomm{figure missing} shows a scatter plot of reduced PC’s scores grouped by the cluster indices of each observation corresponding to the smallest pairwise distances between new and baseline cases.
        \item Two identical clusters will have zero distances (one color) and the further away the data points get from baseline data points the more colors picked up and higher it gets in this scale.
        \item \emph{Bench2} should resemble more than \emph{Bench3} and \emph{Bench3} should resemble more than most dissimilar \emph{Bench4}. 
    \end{itemize}
    \end{comment}
    
    \paragraph{Number of observables per cluster} %hist(idx,NSockets)
    The number of clusters $k$ is chosen in a way that it assigns all unalike clusters, while the number of observables belonging to each cluster could be significantly different.   
    Figures~\ref{fig:pca}(f, h) show the histogram bar chart of the cluster indices in the vector, which are sorted into $k$ bins. We choose $k$ equal to the number of ccNUMA domains.
    The $x$-axis indicates the cluster IDs and the $y$-axis shows the number of samples.

    \paragraph{Validation of clustering quality}
    % [silh2,h] = silhouette(score(:,1:NPC),idx,'Euclidean');
    A potential application of \PCA is its evaluation by calculating the error between original and reconstructed signal from fewer PCs.
    To this end, one can reconstruct the signal by multiplying the scores with the first two PCs and then sum them up. This should be very close to the original signal if the reconstruction error (using the Euclidean norm) is less than some threshold value.
    In Figures~\ref{fig:pca}(g, i), we performed a Silhouette analysis~\cite{kaufman2009finding} to quantify the quality of the clustering.
    A highly representative clustering is associated with a large positive coefficient value close to one and indicates that the point is well matched to other points in its own cluster, and poorly matched to other clusters.
    On the other hand, a negative coefficient value represents a disqualified clustering.
    We get higher reconstruction error for the high-frequency signal of the PISOLVER case as expected.
    While exploring the influence of distance metrics, it turned out that \emph{cosine} is the best-suited and \emph{city-block} is the worst-suited distance metric.

%%%%%%%%%%%%%%%%%%%%%%%%%%%%%%%%%%%%%%%%%%%%%%%%%%%%%%%%%%%%%%%%%%%%%
\section{Summary and future work} \label{sec:conclusion}

    \paragraph{Key takeaways}
    We have presented expressive data analytics techniques for investigating the dynamics of MPI-parallel programs with regular compute-communicate cycles. 
    We consider MPI waiting time per time step and process as a good observable metric since it encompasses much of the relevant dynamics in a very condensed format.
    Our new ``phase space'' analysis based on this data provides an efficient, visual way to observe the evolution of a program from its initial, synchronized state into a desynchronized state. However, it is not strictly a data analytics technique since it involves manual inspection of the data (moving dot clouds). PCA and subsequent k-means clustering allow for a more automated analysis, providing feature extraction, i.e., typical timeline behavior, as well as  grouping of MPI processes into clusters with similar features. Hence, these methods could pave the way towards advanced job-specific monitoring of production jobs on clusters. We have also found that the analysis is more expressive when applied to snippets of the timeline in order to avoid mixing different characteristics. If one is interested in an evolved state only, the final iterations of a run are most relevant. 
    
%    All methods, based on MPI waiting time and performance, show the different representation of the data.
%   With applying methods like temporal evolution of MPI waiting times from one time step to the next in phase space diagram, PCA and k-mean clustering, and so forth, we get reasonable results according to our expectations of such phenomena. 
%    For choosing between analyzing a full waiting timeline or snapshots, we recommend using snapshots for analyzing whole dynamics to avoid mixed up effects at the beginning, at the mid and at the end. Since this selection is strongly dependent on the analysis type, if one is interested in evolved states then apply techniques only for a few last iterations.
%    The phase space analysis is better suited for the dynamics to see how a dot cloud evolves through the diagram at the beginning and at the end. However, the \PCA analysis is better suited for the evolved state to look at patterns for desynchronized ccNUMA domains. On the other hand, an evolution of the \PC could be interesting by itself.
    
    \paragraph{Future work}
    We are convinced that PCA applied to MPI waiting time data allows the investigation of unknown applications by mapping their temporal evolution to principal components found in prototypical benchmark runs. It is still an open question how to choose these benchmarks to extract relevant, distinguishable patterns that real application can be tested against. It will also be necessary to investigate how the waiting time metric should be normalized to be as generic as possible. Furthermore, we plan to apply the demonstrated techniques to a wider spectrum of real applications in order to fathom their true scope of applicability.
    
    %While considering the \PC evolution, a possible future path of analysis is how does an unknown application evolve and what are the best eigenvectors to choose to characterize different kinds of applications.
    %Can we find a set of vectors that somehow encompass different patterns?
    %The resulting eigenvectors from \PCA of a well-known prototype application can be used for analyzing an unknown application since it encompasses the basic condensed behavior of the asynchronized ccNUMA domains with a clear structure.
    %By projecting the MPI waiting times of an unknown application on different prototype eigenvectors, one can make a comparison to see whether this unknown application gradually evolved into the direction of the eigenvector or not or if it's very different. 
   % \AAcomm{I am not sure what should be provided as an additional helping material?
    
  %  (1) All scripts generating plots for all techniques
    
   % (2)Tracing data of all the cases which is in GB
    
  %  (3) Alot of additional plots that are not included in this paper. 
    
 %   For me, point 1 can make sense. 
%}\GHcomm{If that's easy to do, provide the scripts. If anything it shows goodwill.\AAcomm{okay}}
\ifblind
\else
\section*{Acknowledgments}
This research work is supported by KONWIHR, the Bavarian Competence Network for Scientific High Performance Computing in Bavaria, under project name ``OMI4papps.'' We are thankful to Dr.\ Thomas Zeiser and his admin team for excellent technical support on NHR@FAU systems. 
%at Erlangen National High-Performance Computing Center (NHR@FAU). 
\fi

\printbibliography

@InProceedings{AfzalHW19,
  author    = {Ayesha Afzal and Georg Hager and Gerhard Wellein},
  title     = {Propagation and Decay of Injected One-Off Delays on Clusters: {A} Case Study},
  booktitle = {2019 {IEEE} International Conference on Cluster Computing, {CLUSTER} 2019, Albuquerque, NM, USA, September 23-26, 2019},
  year      = {2019},
  pages     = {1--10},
  doi       = {10.1109/CLUSTER.2019.8890995},
}

@inproceedings{AfzalHW20,
  author = {Afzal, Ayesha and Hager, Georg and Wellein, Gerhard},
  booktitle = {Lecture Notes in Computer Science},% (including subseries Lecture Notes in Artificial Intelligence and Lecture Notes in Bioinformatics)}

@InProceedings{AfzalHW21,
	abstract = {Most distributed-memory bulk-synchronous parallel programs in HPC assume that compute resources are available continuously and homogeneously across the allocated set of compute nodes. However, long one-off delays on individual processes can cause global disturbances, so-called idle waves, by rippling through the system. This process is mainly governed by the communication topology of the underlying parallel code. This paper makes significant contributions to the understanding of idle wave dynamics. We study the propagation mechanisms of idle waves across the processes of MPI-parallel programs. We present a validated analytic model for their propagation velocity with respect to communication parameters and topology, with a special emphasis on sparse communication patterns. We study the interaction of idle waves with MPI collectives and show that, depending on the implementation, a collective may be permeable to the wave. Finally we analyze two mechanisms of idle wave decay: topological decay, which is rooted in differences in communication characteristics among parts of the system, and noise-induced decay, which is caused by system or application noise. We show that noise-induced decay is largely independent of noise characteristics but depends only on the overall noise power. An analytic expression for idle wave decay rate with respect to noise power is derived. For model validation we use microbenchmarks and stencil algorithms on three different supercomputing platforms.},
	author = {Afzal, Ayesha and Hager, Georg and Wellein, Gerhard},
	booktitle = {Lecture Notes in Computer Science},% (including subseries Lecture Notes in Artificial Intelligence and Lecture Notes in Bioinformatics)}

@misc{AfzalHW22role,
      title={The Role of Idle Waves, Desynchronization, and Bottleneck Evasion in the Performance of Parallel Programs}, 
      author={Ayesha Afzal and Georg Hager and Gerhard Wellein},
      year={2022},
      eprint={2205.04190},
      archivePrefix={arXiv},
      primaryClass={cs.DC}
}

@Article{markidis2015idle,
  author    = {Markidis, Stefano and Vencels, Juris and Peng, Ivy Bo and Akhmetova, Dana and Laure, Erwin and Henri, Pierre},
  title     = {Idle waves in high-performance computing},
  journal   = {Physical Review E},
  year      = {2015},
  volume    = {91},
  number    = {1},
  pages     = {013306},
  doi       = {10.1103/PhysRevE.91.013306},
  publisher = {APS},
}

@article{qian1992lattice,
	title={Lattice {BGK} models for {Navier}-{Stokes} equation},
	author={Qian, Yue-Hong and d'Humi{\`e}res, Dominique and Lallemand, Pierre},
	journal={EPL (Europhysics Letters)},
	volume={17},
	number={6},
	pages={479},
	year={1992},
	publisher={IOP Publishing}
}

@article{bhatnagar1954model,
	title={A model for collision processes in gases. {I}. {S}mall amplitude processes in charged and neutral one-component systems},
	author={Bhatnagar, Prabhu Lal and Gross, Eugene P and Krook, Max},
	journal={Physical review},
	volume={94},
	number={3},
	pages={511},
	year={1954},
	publisher={APS}
}

@Article{mccalpin1995memory,
  Title                    = {Memory bandwidth and machine balance in current high performance computers},
  Author                   = {McCalpin, John D and others},
  Journal                  = {IEEE computer society technical committee on computer architecture (TCCA) newsletter},
  Year                     = {1995},
  Number                   = {19--25},
  Volume                   = {2}
}

@article{fehske2004quantum,
	title={Quantum lattice dynamical effects on single-particle excitations in one-dimensional {M}ott and {P}eierls insulators},
	author={Fehske, H and Wellein, G and Hager, G and Wei{\ss}e, A and Bishop, AR},
	journal={Physical Review B},
	volume={69},
	number={16},
	pages={165115},
	year={2004},
	publisher={APS}
}

@book{vetterling1992numerical,
  title={Numerical Recipes: Example book C},
  author={Vetterling, William T and Vetterling, William T and Press, William H and Press, William H and Teukolsky, Saul A and Flannery, Brian P and Flannery, Brian P},
  year={1992},
  publisher={Cambridge University Press}
}

@article{jolliffe2016principal,
  title={Principal component analysis: a review and recent developments},
  author={Jolliffe, Ian T and Cadima, Jorge},
  journal={Philosophical Transactions of the Royal Society A: Mathematical, Physical and Engineering Sciences},
  volume={374},
  number={2065},
  pages={20150202},
  year={2016},
  publisher={The Royal Society Publishing}
}

@inproceedings{vassilvitskii2006k,
  title={k-means++: The advantages of careful seeding},
  author={Vassilvitskii, Sergei and Arthur, David},
  booktitle={Proceedings of the eighteenth annual ACM-SIAM symposium on Discrete algorithms},
  pages={1027--1035},
  year={2006}
}

@book{kaufman2009finding,
  title={Finding groups in data: an introduction to cluster analysis},
  author={Kaufman, Leonard and Rousseeuw, Peter J},
  year={2009},
  publisher={John Wiley \& Sons}
}

\end{document}